\documentclass[12pt]{article}

\usepackage{amsmath}
\usepackage{amsthm}
\usepackage{amscd}
\usepackage{amsfonts}
\usepackage[psamsfonts]{amssymb}

\DeclareMathOperator{\tr}{tr} 
\DeclareMathOperator{\Imx}{Im} 
\DeclareMathOperator{\Rex}{Re} 
\DeclareMathOperator{\sgn}{sgn} 

\DeclareMathOperator{\Spin}{Spin} 

\date{\today}
\newcommand\email[1]{{\tt {#1}}}
\def\acknowledgments{\centerline{{\bf Acknowledgments}}}
\begin{document}
\setlength{\baselineskip}{20pt}

\begin{titlepage}
\rightline{hep-th/0502076}
\rightline{UCB-PTH-04/37}
\rightline{LBNL-56763}
\vskip 1cm
\begin{center}
\Huge
Massless and Massive\\
Three Dimensional Super Yang-Mills Theory \\
and Mini-Twistor String Theory
\\
\end{center}
\smallskip
\centerline{ Dah-Wei Chiou, Ori J. Ganor, Yoon Pyo Hong, Bom Soo
Kim and Indrajit Mitra }
\begin{center}
{\it Department of Physics,
             University of California, Berkeley, CA 94720} \\
  and\\
{\it Theoretical Physics Group,
             Lawrence Berkeley National Laboratory, Berkeley, CA 94720}\\
E-mails: \email{dwchiou,origa,yph,bskim,imitra@socrates.berkeley.edu},\\
\end{center}
\bigskip
We propose various ways of adding mass terms to three-dimensional
twistor string theory. We begin with a review of mini-twistor
space---the reduction of $D=4$ twistor space to $D=3.$ We adapt
the two proposals for twistor string theory, Witten's and
Berkovits's, to $D=3$ super Yang-Mills theory. In Berkovits's
model, we identify the enhanced R-symmetry. We then construct
B-model topological string theories that, we propose, correspond
to $D=3$ Yang-Mills theory with massive spinors and massive and
massless scalars in the adjoint representation of the gauge group.
We also analyze the counterparts of these constructions in
Berkovits's model.
Some of our
constructions can be lifted to $D=4$, where infinitesimal mass
terms correspond to VEVs of certain superconformal gravity fields.
\end{titlepage}


\def\be{\begin{equation}}
\def\ee{\end{equation}}
\def\bear{\begin{eqnarray}}
\def\eear{\end{eqnarray}}
\def\nn{\nonumber}

\newcommand\belabel[1]{\begin{equation}\label{#1}}
\newcommand\bearlabel[1]{\begin{eqnarray}\label{#1}}

\newcommand\bra[1]{{\langle {#1}|}}  
\newcommand\ket[1]{{|{#1}\rangle}}  

\def\defineas{{:=}}

\def\a{\alpha}
\def\b{\beta}
\def\g{\gamma}
\def\r{\rho}
\def\th{{\theta}}
\def\lam{{\lambda}}

\def\dta{{\dot{\a}}}
\def\dtb{{\dot{\b}}}

\def\oa{{\overline{a}}} 

\def\bx{{\overline{x}}}
\def\bz{{\overline{z}}}
\def\bw{{\overline{w}}}

\def\bth{{\overline{\theta}}}
\def\blam{{\overline{\lambda}}}
\def\bpsi{{\overline{\psi}}}
\def\bsig{{\overline{\sigma}}}
\def\Dslash{{\relax{\not\kern-.18em \partial}}} 
\def\Spin{{{\mbox{\rm Spin}}}} 
\def\SL{{{\mbox{\rm SL}}}} 
\def\GL{{{\mbox{\rm GL}}}} 
\def\rt{{\rightarrow}}
\def\cc{{\mbox{c.c.}}}

\newcommand\SUSY[1]{{${\cal N}={#1}$}}  
\newcommand\px[1]{{\partial_{#1}}}
\newcommand\qx[1]{{\partial^{#1}}}

\newcommand\ppx[1]{{\frac{\partial}{\partial {#1}}}}
\newcommand\pspxs[1]{{\frac{\partial^2}{\partial {#1}^2}}}
\newcommand\pspxpx[2]{{\frac{\partial^2}{\partial {#1}\partial {#2}}}}
\newcommand\pypx[2]{{\frac{\partial {#1}}{\partial {#2}}}}


\newcommand{\field}[1]{\mathbb{#1}}
\newcommand{\ring}[1]{\mathbb{#1}}
\newcommand{\C}{\field{C}}
\newcommand{\R}{\field{R}}
\newcommand{\Z}{\ring{Z}}
\newcommand{\N}{\ring{N}}
\newcommand{\CP}{\field{C}\mathbb{P}} 
\newcommand{\RP}{\field{R}\mathbb{P}} 


\providecommand{\abs}[1]{{\lvert#1\rvert}}
\providecommand{\norm}[1]{{\lVert#1\rVert}}
\providecommand{\divides}{{\vert}}
\providecommand{\suchthat}{{:\quad}}



\def\bz{{\overline{z}}}
\def\gYM{{g_{\textit{YM}}}} 
\def\gst{{g_s}} 
\def\Mst{{M_s}} 
\newcommand\rep[1]{{\bf {#1}}} 

\newcommand\inner[2]{{\langle {#1}, {#2} \rangle}} 
\def\half{{\tfrac{1}{2}}} 


\def\Horava{{Ho\v{r}ava\ }}
\def\Cech{{\v{C}ech\ }}


\newtheorem{thm}{Theorem}[section]
\newtheorem{cor}[thm]{Corollary}
\newtheorem{lem}[thm]{Lemma}
\newtheorem{prop}[thm]{Proposition}
\newtheorem{ax}{Axiom}
\newtheorem{conj}[thm]{Conjecture}

\newtheorem{ex}{Exercise}[section]

\theoremstyle{definition}
\newtheorem{defn}{Definition}[section]
\theoremstyle{remark}
\newtheorem{rem}{Remark}[section]
\newtheorem*{notation}{Notation}
\newtheorem*{example}{Example}

\numberwithin{equation}{section}

\newcommand{\thmref}[1]{Theorem~\ref{#1}}
\newcommand{\secref}[1]{\S\ref{#1}}
\newcommand{\lemref}[1]{Lemma~\ref{#1}}

\newcommand{\propref}[1]{Proposition~\ref{#1}}
\newcommand{\exref}[1]{Exercise~\ref{#1}}
\newcommand{\figref}[1]{Figure~\ref{#1}}

\newcommand{\appref}[1]{Appendix~\ref{#1}}

\def\XD{{\mathbb{D}}} 
\def\Ol{{\cal O}} 

\def\Mf{{\cal M}} 
\def\Tw{{{\mathbb T}}} 
\def\MfC{{\Mf_\C}} 
\def\CxSt{{\xi}} 
\def\aPl{{\mathbb D}} 

\def\Kah{K\"ahler}
\def\HK{Hyper-\Kah}
\def\hK{hyper-\Kah}

\def\wPhi{{\widetilde{\Phi}}} 
\def\Action{{I}} 
\def\wAction{{\widetilde{I}}} 

\def\wG{{\widetilde{G}}} 
\def\IntPath{{C}} 

\def\cD{{\cal D}} 
\def\IntPhPath{{\Upsilon}} 

\def\wG{{\widetilde{G}}} 
\def\bq{{\overline{q}}}
\def\vphi{{\varphi}}
\def\fvphi{{\widetilde{\varphi}}}
\def\vx{{\vec{x}}}
\def\vy{{\vec{y}}}
\def\cO{{\cal O}} 
\def\vn{{\vec{n}}} 
\def\vA{{\vec{A}}}
\def\tlam{{\tilde{\lam}}} 
\def\u{{\mu}}
\def\bQ{{\overline{Q}}} 

\def\tF{{\widetilde{F}}} 
\def\tG{{\widetilde{G}}} 

\def\semidirect{{\ltimes}} 
\def\vk{{\vec{k}}} 
\def\vn{{\vec{n}}} 

\def\wA{{\widetilde{A}}} 
\def\wt{{\widetilde{t}}} 

\def\twline{{\ell}} 
\def\Curve{{\Sigma}} 
\def\wCurve{{\widetilde{\Curve}}} 
\def\pbOmega{{\widetilde{\Omega}}} 
\def\cK{{\cal K}} 
\def\Focal{{\cal F}} 
\newcommand\Dxt[2]{{{\cal I}_{{#1}}({{#2}})}} 
\def\HoloD{{\Upsilon}}
\def\bHoloD{{\overline{\Upsilon}}}
\def\Vertex{{\cal V}} 
\def\hX{{\hat{X}}} 
\def\hY{{\hat{Y}}} 
\def\twv{{\mathfrak t}} 
\def\wtfd{{\tilde{\mathfrak t}}} 

\newcommand\rsprod[2]{{\langle {#1}, {#2}\rangle}} 
\def\Lag{{\cal L}} 
\def\oi{{\overline{i}}} 
\def\oj{{\overline{j}}} 
\def\bpar{{\overline{\partial}}} 

\def\ob{{\overline{b}}} 
\def\oeta{{\overline{\eta}}} 
\def\tW{{\tilde{W}}} 
\def\tZ{{\tilde{Z}}} 
\def\btW{{\overline{\tilde{W}}}} 
\def\btZ{{\overline{\tilde{Z}}}} 
\def\bX{{\overline{X}}} 
\def\oTheta{{\overline{\Theta}}} 
\def\varth{{\vartheta}} 
\def\OOs{{{\mathfrak s}}} 
\newcommand\Vpch[1]{{V_{\{{#1}\}}}} 

\def\Br{{\widetilde{B}}} 
\def\rY{{\widetilde{Y}}} 
\def\rUpsilon{{\widetilde{\Upsilon}}} 
\def\rU{{U}} 
\def\rTheta{{\widetilde{\Theta}}} 
\def\Vo{{{\cal V}}} 
\def\Jc{{{\cal J}}} 
\def\Qc{{{\cal Q}}} 
\def\bQc{{\overline{\cal Q}}} 
\def\Pc{{{\cal P}}} 

\def\cA{{{\cal A}}} 
\def\wchi{{\widetilde{\chi}}} 
\def\wvrho{{\widetilde{\varrho}}} 
\def\wvrho{{\widetilde{\varrho}}} 
\def\hwvrho{{\widehat{\widetilde{\varrho}}}} 
\def\hvarrho{{\widehat{\varrho}}} 
\def\hzeta{{\widehat{\zeta}}}

\def\zws{{{\mathfrak z}}} 
\def\bzws{{\overline{\mathfrak z}}} 

\def\wVo{{\tilde{\cal V}}} 
\def\ZZo{{\Xi}} 

\def\chA{{\widetilde{\cal A}}} 

\centerline{\bf DISCLAIMER}
This document was prepared as an account of work sponsored by the United States Government. While this document is believed to contain correct information, neither the United States Government nor any agency thereof, nor The Regents of the University of California, nor any of their employees, makes any warranty, express or implied, or assumes any legal responsibility for the accuracy, completeness, or usefulness of any information, apparatus, product, or process disclosed, or represents that its use would not infringe privately owned rights. Reference herein to any specific commercial product, process, or service by its trade name, trademark, manufacturer, or otherwise, does not necessarily constitute or imply its endorsement, recommendation, or favoring by the United States Government or any agency thereof, or The Regents of the University of California. The views and opinions of authors expressed herein do not necessarily state or reflect those of the United States Government or any agency thereof or The Regents of the University of California.

\newpage
\tableofcontents

\section{Introduction}\label{sec:intro}
Over the last twenty years following the work of Parke and Taylor
\cite{Parke:1986gb} (among others, see also \cite{Berends:1987me})
it has become clear that the scattering amplitudes of Yang-Mills
theory in four dimensions are much simpler than one would guess.
This simplicity was known to persist not only for tree-level
results, but also at one loop level (for a nice review, see
\cite{Bern:1996je}). Witten has recently shown
\cite{Witten:2003nn} that these amplitudes are most succinctly
expressed not in terms of the polarization and momenta of the
incoming and outgoing photons, but rather in terms of Penrose's
twistor variables \cite{Penrose:1967wn}(which of course encode
information about the polarization and momenta of the particles).
By applying the so-called twistor transform to the amplitudes
expressed in terms of spinor variables, Witten showed that the
results collapse to simple algebraic curves in twistor space.

Twistor theory uncovers holomorphic structure underlying massless
free field equations of motion. The ``twistor transform'' converts
harmonic functions on a manifold $\Mf$ to meromorphic functions on
its ``twistor space'' $\Tw\Mf.$ The twistor transform can be used
to convert scattering amplitudes of $n$ massless gluons in
perturbative Yang-Mills theory to a meromorphic function (i.e., a
section of a certain line bundle) of $n$ points on $\Tw(\R^4).$ It
was conjectured in \cite{Witten:2003nn} that the $l$-loop
contribution to the transformed amplitude is nonvanishing only if
the $n$ points on twistor space that correspond to the $n$ gluons
lie on a holomorphic curve whose degree and genus are determined
by the helicity of the particles and by $l.$ This led Witten
to conjecture that there exists a dual string
theory and that it is a topological B-model \cite{Witten:2003nn}.
The target space of this string theory
is the twistor space of $\R^4$, and Witten showed how the
D-instantons of this string theory can compute the scattering
amplitudes of the perturbative gauge theory. (The issue of whether one
should consider connected or disconnected instantons in order to
reproduce the gauge theory amplitudes was rather nicely resolved in
\cite{Gukov:2004ei}.) This surprising
duality is a ``weak-weak'' duality in the sense that a weakly
coupled string theory is dual to a weakly coupled gauge theory,
unlike the ``strong-weak'' duality
\cite{Maldacena:1997re}\cite{Gubser:1998bc}\cite{Witten:1998qj}
where a strongly coupled string theory was dual to a weakly
coupled gauge theory.

We wish to extend these results to three dimensions and
to understand the string dual of weakly
coupled super Yang-Mills theory in $D=3.$ The target
space of this string theory is the twistor space of $\R^3$. This
turns out to be the space of oriented lines in $\R^3$ (we mostly
follow \cite{Hitchin:1982gh} and the nice
review article \cite{Baird}).
It is called ``mini-twistor'' space \cite{Hitchin:1982gh},
and is related to the twistor space of
$\R^{3,1}$ by dimensional reduction.
This well-known elegant relation \cite{Hitchin:1982gh}
will guide us in developing an algorithm to obtain gauge theory
scattering amplitudes in $D=3$ from the corresponding ones in
$D=4$. We shall see that the $D=3$ amplitudes are still supported
on holomorphic curves.

Recall that the twistor space of Minkowski space $\R^{3,1}$ is
$\Tw(\R^{3,1})\simeq\CP^3\setminus\CP^1$ (i.e., $\CP^3$ with a
rational curve excised\footnote{ We use the operator $\setminus$
to denote the set-theory ``minus'' and it is not to be confused
with division.}); the twistor space of $\R^3$ is $T\CP^1$ (the
tangent space of $\CP^1$)
\cite{Hitchin:1982gh}\cite{Baird}.\footnote{ This space has also
been discussed in footnote 13 of \cite{Witten:2003nn}, and in
\cite{Bars:2004dg} it was derived by dimensional reduction with
constraints, from 2-Time physics.} The
mini-twistor space does not possess the full $D=3$ superconformal
symmetry $SO(3,2),$ but only its Poincar\'e subgroup
$SO(3)\semidirect \R^3.$ It can be obtained by dimensional
reduction as follows \cite{Hitchin:1982gh}. The
3(complex)-dimensional $D=4$ twistor space $\CP^3\setminus\CP^1$
can be written as a fiber bundle with the 2(complex)-dimensional
$D=3$ mini-twistor space $T\CP^1$ as the base, and the fiber is
$\C.$ The structure group is the additive translation group
$\sim\C$ (as opposed to the multiplicative group $\C^*$).
However, the fibration is not canonical;
it depends on a choice of
direction in the physical space $\R^4.$
This is the direction of the dimensional reduction.
 For a given choice of this direction
(which we will refer to as the ``$4^{th}$ direction'') there is a
natural projection from the $D=4$ twistor space
$\CP^3\setminus\CP^1$ onto the $D=3$ mini-twistor space $T\CP^1.$
We will use this fibration to calculate mini-twistor amplitudes of
$D=3$ Yang-Mills theory by taking the $D=4$ amplitudes and
integrating them over the $\C$ fibers of the above fibration. In
the worldsheet theory of the $D=4$ twistor string theory---the
B-model with target space $\CP^{3|4}$---we realize dimensional
reduction by gauging one of the four translation symmetries. It is
an element of the B-model symmetry group $PSL(4|4).$ The resulting
string theory is the B-model with target space $T\CP^1$ and four
additional local fermionic coordinates that transform as sections
of the pullback of the $\Ol(1)$ line bundle over $\CP^1.$

One aspect that is not so obvious in this construction
is the enhanced R-symmetry.
The R-symmetry group of $D=3$ super Yang-Mills with $N=8$
supersymmetry is $\Spin(7)$, but we will find
that only an $SU(4)$ subgroup
is manifest in the B-model string theory.
There is, however, another version of
twistor string theory due to Berkovits
\cite{Berkovits:2004hg}. This is an open string theory.
There is a prescription due to Berkovits and Witten
\cite{Berkovits:2004jj} which allows one to go from one picture to
the other, and we will use this somewhat extensively.
We will implement dimensional reduction from $D=4$ to $D=3$
in Berkovits's model as well, and we will derive the $D=3$
version of this string theory.
In this string theory we will be able to construct
the full R-symmetry current.

We will also describe a twistor string theory dual of a certain
massive $D=3$ super Yang-Mills theory, and this is one of our main
new results. The target space of this string theory is also
$T\CP^1$ with four additional local fermionic coordinates, but the
way in which they are fibered over $T\CP^1$ is modified from the
massless case. It corresponds to a Yang-Mills theory with massive
scalars and fermions. We will systematically study what adding
small mass terms to the fermions means for the string theory.
We will
study this question both in the
context of the B-model and Berkovits's open string theory.
In the dimensionally reduced
  $D=3$ gauge theory, mass terms have two different origins.
They either
come from mass terms in the original $D=4$ theory,
or they come by coupling the R-symmetry current to a constant
background gauge field (i.e., an R-symmetry twist).

In the B-model, the effect of a mass term for the space-time
fermions can be achieved by deforming the (super-)complex
structure of the target space. Such a deformation corresponds to a
closed string vertex operator. We will identify the vertex
operators which give rise to the mass terms. Then, using the
prescription of Berkovits and Witten, we convert these operators
to open string operators which deform the boundary of the
worldsheet. In making this transformation, we encounter a
surprise: The operators that one gets using this
  dictionary do not lie in an irreducible representation of the
R-symmetry group $\Spin(7)$, as they should. Some of the operators
have to be modified to fit into the required irreducible
representation.

We also propose that an infinitesimal mass term in
$D=4$ can be achieved by a certain
marginal deformation of the worldsheet theory.
In physical space, this corresponds to a small VEV for
a B-model closed string field, which
according to \cite{Witten:2003nn}\cite{Berkovits:2004jj}
is part of a conformal supergravity multiplet.
We identify this field.

The paper is organized as follows. In \secref{sec:R3} we review
the construction of mini-twistor space and its geometrical
interpretation as the space of oriented lines, following
\cite{Hitchin:1982gh}\cite{Baird}. We review the relation between
harmonic functions on $\R^3$ and meromorphic forms on mini-twistor
space, and we apply it to the scalar propagator. In
\secref{sec:dimred} we review the connection \cite{Hitchin:1982gh}
between the $D=3$ and $D=4$ twistor spaces, and we discuss the
supersymmetric extensions. We derive the tree-level amplitudes of
$D=3$ super Yang-Mills theory by dimensional reduction, and we
find that they have support on holomorphic curves, like their
$D=4$ counterparts. We comment on a possible physical
interpretation of this result. In \secref{sec:mass} we augment the
theory with mass terms, and we relate the infinitesimal mass terms
to worldsheet operators in the B-model and in Berkovits's model.
We conclude in \secref{sec:discussion}.


\section{The (mini-)twistor space of $\R^3$}\label{sec:R3}
The twistor space of $\R^3$ is $T\CP^1$---the tangent bundle of
$\CP^1,$ and harmonic functions on $\R^3$ can be converted into
meromorphic functions on $T\CP^1.$ The space $T\CP^1$ is called
``mini-twistor'' space \cite{Hitchin:1982gh}. We will
now explain these statements in detail and apply them to convert
the propagator of massless fields on $\R^3$ to a meromorphic
function of two points in mini-twistor space. Our initial
discussion is based mostly on the nice review paper by Paul Baird
\cite{Baird}.

Our discussion in this section is limited to scalar fields.
This is not too much of a restriction, since
in $D=3$ massless gauge fields can be converted by
duality to massless scalars, as we will review in
\secref{subsec:ft}.
Massless spinors in $D=3$ also have just one helicity state,
and solutions to the massless Dirac equation can be readily
converted to mini-twistor space. We refer the reader to
\cite{Baird} for further details.

\subsection{Harmonic functions on $\R^3$}
\label{subsec:harmonic}
Pick coordinates $x_1, x_2, x_3$ on $\R^3.$ For any fixed
$0\le\th\le 2\pi,$ the linear expression $x_1 + i x_2 \sin\th + i
x_3\cos\th$ is a harmonic function, and so is any analytic
function of this expression. We can construct a more complicated
harmonic function on $\R^3$ by taking linear combinations of
analytic functions of $x_1 + i x_3 \cos\th + i x_2\sin\th$ for
various values of $\th$. Whittaker's formula states that a
complex-valued harmonic function $\phi$ on $\R^3$ can be given by
an integral \be\label{eqn:Whit} \phi(x_1, x_2, x_3) =
\int_0^{2\pi} d\th f(x_1 + i x_2\sin\th + i x_3 \cos\th, \th), \ee
where $f(z,\theta)$ is analytic in the first variable. To prove
this formula, note that the right-hand side of \eqref{eqn:Whit} is
obviously harmonic. In order to write a Whittaker formula for an
arbitrary harmonic function $\phi$, pick polar coordinates $(r, u,
v)$ such that
$$
x_1 = r\cos u,\qquad x_2 = r\sin u\sin v,
\qquad x_3 = r\sin u\cos v.
$$
Then, for $l\ge 0,$ and $|m|\le l,$
the spherical harmonics can be written as
\bear
r^l Y_{lm}(u, v) =
\frac{\sqrt{(2l+1)(l-m)!(l+m)!}}{4\pi^{3/2} i^{3m} l!}
\int_0^{2\pi} d\th e^{i m\th}(x_1 + i x_2\sin\th + i x_3
\cos\th)^l,
\label{eqn:rplYlm}
\eear
and later we will also need the identity
\bear
\frac{1}{r^{l+1}}
Y_{l,m}(u, v) =
\pm\frac{i^{3m}l!}{4\pi^{3/2}}\sqrt{\frac{2l+1}{(l-m)!(l+m)!}}
\int_0^{2\pi} \frac{d\th e^{i m\th}}{ (x_1 + i x_2\sin\th + i
x_3\cos\th)^{l+1}},
\label{eqn:rmlYlm}
\eear
where the sign on the
right-hand side of \eqref{eqn:rmlYlm} is the same as that of
$x_1$. The formula \eqref{eqn:rplYlm} is a standard integral
representation of spherical harmonics, while \eqref{eqn:rmlYlm}
can be derived by starting from \cite{wolfram}
$$P_l^m(\cos
u)=\frac{(-1)^m}{\pi}\frac{l!}{(l-m)!}e^{m\pi
i/2}\int_0^{\pi}\frac{\cos m\theta\,d\theta}{(\cos u+i\sin
u\cos\theta)^{l+1}}\qquad(\cos u>0).
$$
\indent Any harmonic function that is regular at the origin can be
expanded as a linear combination of spherical harmonics
$$
\phi(x_1, x_2, x_3) = \sum_{l=0}^\infty \sum_{m=-l}^l C_{l m} r^l
Y_{lm}(u, v).
$$
This allows us to express $\phi$ as a Whittaker integral
\eqref{eqn:Whit}. A possible choice for the analytic function to
be used on the right-hand side of \eqref{eqn:Whit} is
$$
f(\zeta,\theta) = \sum_{l=0}^\infty \sum_{m=-l}^l C_{l m}
\frac{\sqrt{(2l+1)(l-m)!(l+m)!}}{4\pi^{3/2} i^{3m} l!}e^{im\th} \zeta^l .
$$

\subsection{Identification of mini-twistor space with $T\CP^1$}
\label{subsec:TCP1}
As explained in \cite{Hitchin:1982gh}\cite{Baird},
the twistor space of $\R^3$
can be identified with the space of oriented lines in $\R^3.$
This space
is isomorphic to the 2-complex dimensional space $T\CP^1.$
We will now review how this works.

First we rewrite Whittaker's formula \eqref{eqn:Whit} by introducing
complex coordinates
$$
w = 2 e^{i\th}(x_1 + i x_2\sin\th + i x_3\cos\th),
\qquad
z = e^{i\th}.
$$
Given the analytic function $f$ on the right-hand
side of \eqref{eqn:Whit}, it is convenient to
define a related analytic function $\vphi$ by
$$
\vphi(e^{i\th},w)\defineas
e^{-i\th}f(\tfrac{1}{2}e^{-i\th}w,\th).
$$
We assume that we can
extend $\vphi$ to an analytic function $\vphi(z,w)$ defined
in a neighborhood of the circle $\abs{z}=1.$
Formula \eqref{eqn:Whit} can now be rewritten as
\be
\label{eqn:Whitzw}
\phi(\vx) = \frac{1}{2\pi i}\oint
   \vphi(z, -[x_2 - i x_3] +2 z x_1 +z^2 [x_2 + i x_3])dz.
\ee
We take $z$ and $w$ as local coordinates on mini-twistor
space, which will be identified with $T\CP^1$ soon. Under
favorable conditions, $\vphi(z,w)$ can be analytically continued
to a meromorphic function for all $z\in\C$ and $w\in\C.$ For
simplicity of the discussion we will assume that this is the case.
(Although, in the more general case we can assume that $\vphi$ can
be analytically continued to a neighborhood around $\abs{z}=1.$ We
defer the discussion of this case till the end of this subsection,
since we will need to use sheaf cohomology.) We would actually
like to view $z$ as a coordinate on a $\CP^1$ by identifying
$\CP^1\simeq \C\cup\{\infty\}$ (say, by stereographic projection),
so that $z=\infty$ will be an allowed value. Then, for every
$\vx\in\R^3$, the equation
\be\label{eqn:increl}
w = -(x_2 - i x_3) +2 x_1 z + (x_2 + i x_3) z^2
\qquad
\text{(incidence relation)},
\ee
which is analytic in $z$, defines an algebraic
curve in $(z,w)$ space.
We get the integrand of \eqref{eqn:Whitzw} from
$\vphi(z,w)$ by setting \eqref{eqn:increl}.
The right-hand side of \eqref{eqn:increl}
has a double pole at $z=\infty.$ We define
\be\label{eqn:zpwp}
z' = \frac{1}{z},
\qquad
w' =-\frac{w}{z^2}
\ee
to be regular local coordinates on mini-twistor space near
$z=\infty$, instead of $(z,w).$
Equation \eqref{eqn:increl} then becomes
$$
w' = -(x_2 + i x_3) -2 x_1  z' +(x_2 - i x_3) {z'}^2,
$$
which has no pole at $z=\infty.$ The two coordinate systems
$(z,w)$ [$z\neq\infty$] and $(z',w')$ [$z'\neq\infty$], with the
transition rules \eqref{eqn:zpwp}, describe the holomorphic line
bundle $\cO(2)$ over $\CP^1$, $w$ being the local coordinate on
the fiber and $z$ being the coordinate on the base. $\cO(2)$ can
be identified with the tangent bundle $T\CP^1$, since from
\eqref{eqn:zpwp} it is obvious that $w$ transforms like a vector
on $\CP^1.$ Thus we identify the mini-twistor space $\Tw(\R^3)$
with $T\CP^1.$

The relation \eqref{eqn:increl} describes a holomorphic section of
the line bundle $T\CP^1$, and it varies with the point
$\vx\in\R^3.$ It is called the {\it incidence relation}
\cite{Baird}. In \secref{subsec:geom}, following \cite{Baird}, we
will give the mini-twistor space and the incidence relation a more
geometric interpretation.

{}From \eqref{eqn:Whitzw} we see that it is natural to think of
$\vphi$ as a holomorphic 1-form $\vphi(z,w)dz.$ This 1-form is
defined in the neighborhood of $|z|=1.$ A 1-form on $\CP^1$, by
definition, takes values in the holomorphic sheaf $\Omega^1\simeq
\cO(-2)$ over $\CP^1$. We can think of $\vphi dz$ as taking values
in the pullback $\pbOmega^1$ of $\Omega^1$ to the tangent bundle
$T\CP^1.$ Also, the integral \eqref{eqn:Whitzw} is unchanged if we
replace $\vphi$ with $\vphi(z,w)dz + g_0(z,w)dz -
g_\infty(z,w)dz$, for any pair of local holomorphic 1-forms $g_0$
and $g_{\infty}$ (taking values in the sheaf $\pbOmega^1$, by
definition) such that $g_0$ has no poles for all $|z|\le 1$ and
$g_\infty$ has no poles for all $|z|\ge 1$ (including $z=\infty$).
This defines an equivalence class of 1-forms $\vphi\sim\vphi+g_0
-g_\infty$, which defines the {\it sheaf cohomology} $H^1(T\CP^1,
\pbOmega^1).$ This notion will be useful in
\secref{subsec:redtree} when we derive mini-twistor tree-level
amplitudes of $D=3$ super Yang-Mills theory by dimensional
reduction of $D=4$ amplitudes.

\subsection{Geometric picture of mini-twistor space}
\label{subsec:geom}
As explained in \cite{Hitchin:1982gh}\cite{Baird},
the mini-twistor space $T\CP^1$,
whose construction was reviewed above,
can be identified with the space of oriented lines
in $\R^3$, and the incidence relation \eqref{eqn:increl} is
the condition that the line that corresponds to the
mini-twistor $(z,w)$
passes through $\vx\in\R^3.$
We will now review how this works.

\begin{figure}[t]
\begin{picture}(400,100)
\put(100,10){\circle*{4}} 
\put(102,10){$O$}
\put(100,10){\vector(-1,2){20}} 
\put(90,40){$\vA$}
\thicklines
\put(40,30){\vector(2,1){80}} 
\put(126,70){$\vn$}
\put(50,38){$\twline$}

\put(5,80){$\R^3$}
\thinlines
\put(20,75){\line(-1,0){15}}
\put(20,75){\line(0,1){15}}
\end{picture}

\caption{Mini-twistor space can be identified with the
space of oriented lines $\twline\in\R^3.$
The direction of $\twline$ is $\vn,$ and the displacement
vector is $\vA.$}

\label{fig:nA}
\end{figure}
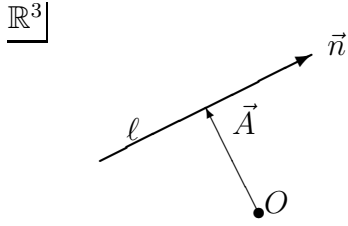

To describe an oriented line $\twline$ in $\R^3$, we need its
direction, which is a unit vector $\vn$, and its displacement
vector $\vA$, which is the vector from the origin of $\R^3$ to a
point on $\twline$ closest to the origin. Therefore, $\vA$ is
perpendicular to $\vn.$ The space of unit vectors $\vn$ in $\R^3$
is the sphere $S^2$, which we identify with $\CP^1$ by the
stereographic projection:
\be\label{eqn:stereog}
\norm{\vn}=1\Longrightarrow
z = \frac{n^2 - i n^3}{1 +n^1}\in\CP^1\simeq\C\cup\{\infty\}.
\ee
Given $\vn$, the space of vectors $\vA$ that satisfy
$\vn\cdot\vA=0$ is the tangent plane to
$\vn$ on $S^2.$ With the identification $S^2\simeq\CP^1$ we find
that the space of $(\vn,\vA)$ is $T\CP^1.$ The holomorphic
coordinate $w$ on the fiber of $T\CP^1$ can be defined as
$$
w = - A^k\frac{\partial z}{\partial n^k}
 = \frac{-(1+n^1)(A^2 - i A^3) + (n^2 - i n^3)A^1}{(1+n^1)^2},
$$
where $A^i$ and $n^i$ ($i=1,2,3$) are the components of $\vA$ and
$\vn$ respectively. Given $z$, we can recover $\vn$ by
\be\label{eqn:nzw}
\vn = \Bigl(
\frac{1-\abs{z}^2}{1+\abs{z}^2},\,
\frac{z+\bz}{1+\abs{z}^2},\,
\frac{i(z-\bz)}{1+\abs{z}^2} \Bigr),
\ee
and, noting that $\vn\cdot\vA=0$, we can recover $\vA$
from $(z,w)$:
\be\label{eqn:Azw}
\vA =
\Bigl(
2\frac{w\bz + \bw z}{(1+\abs{z}^2)^2},\,
-\frac{w(1-\bz^2)+\bw(1-z^2)}{(1+\abs{z}^2)^2},\,
-i\frac{w(1+\bz^2)-\bw(1+z^2)}{(1+\abs{z}^2)^2}
\Bigr).
\ee
One can check \cite{Baird} that the oriented line given by
$\vec{x}=(x_1,x_2,x_3)=\vA+c\vn$ is the solution set to the
incidence relation \eqref{eqn:increl}.

\subsection{The Poincar\'e group}\label{subsec:Poincare}
It will be useful for us to express the generators of the symmetry
group of $T\CP^1$ in mini-twistor variables. This will allow us to
easily check symmetry properties of various expressions that are
given in terms of $z$ and $w$. The Poincar\'e group in three
dimensions has generators $\vec{P}$ (translations) and $\vec{J}$
(rotations). It acts on $\vn$ and $\vA$ as follows:
$$
[P_i, n_j] = 0,\qquad [P_i, A_j] =i \delta_{ij} -i n_i n_j,
\qquad [J_i, n_j] =i \epsilon_{ijk} n_k,\qquad [J_i, A_j]=i\epsilon_{ijk} A_k.
$$
{}From this and \eqref{eqn:nzw}-\eqref{eqn:Azw} it is easy to find
the expressions in terms of $z$ and $w.$
We get
\bear
P_1 &=& iz\ppx{w}+i\bz\ppx{\bw},
\nn\\
P_{+} &\defineas& P_2 + i P_3 = -i\ppx{w}+i\bz^2\ppx{\bw},
\nn\\
P_{-} &\defineas& P_2 - i P_3 = iz^2\ppx{w}-i\ppx{\bw},
\label{eqn:Pgen}
\eear
and
\bear
J_1 &=& -z\ppx{z}
+\bz\ppx{\bz}-w\ppx{w}+\bw\ppx{\bw},
\nn\\
J_{+} &\defineas& J_2 + i J_3
=\ppx{z}+\bz^2\ppx{\bz}+2\bz\bw\ppx{\bw},
\nn\\
J_{-} &\defineas& J_2 - i J_3 =-z^2\ppx{z}-\ppx{\bz}-2zw\ppx{w}.
\label{eqn:Jgen}
\eear

\subsection{Extension to Superspace}\label{subsec:supersp}
We can easily extend the discussion to accommodate supersymmetry.
For $D=3$, $N=2$ supersymmetry, the generators are $Q_\pm$ and
their hermitian conjugates are $Q_\pm^\dagger\equiv\bQ_\mp.$ The
SUSY algebra is
$$
[P_i, Q_\pm] = 0,
\quad
[P_i, \bQ_\pm] = 0,
\quad
[J_1, Q_\pm] = \pm \tfrac{1}{2}Q_\pm,
\quad
[J_1, \bQ_\pm] = \pm \tfrac{1}{2}Q_\pm,
$$
$$
[J_\pm, Q_\mp] = Q_\pm,
\quad
[J_\pm, \bQ_\mp] = -\bQ_\pm,
\quad
[J_\pm, Q_\pm] = 0,
\quad
[J_\pm, \bQ_\pm] = 0,
$$
and
$$
\{Q_\pm, Q_\pm\} =
\{Q_\pm, Q_\mp\} =
\{\bQ_\pm, \bQ_\mp\} =
\{\bQ_\pm, \bQ_\pm\} = 0,
$$
and
$$
\{Q_\pm, \bQ_\pm\} = P_\pm,
\qquad
\{Q_\pm, \bQ_\mp\} = \pm P_1.
$$
We add a superspace coordinate $\th$ and its complex conjugate
$\bth.$ We can now express $Q_\pm$ and $\bQ_\pm$ in terms of
$(z,w,\th)$ and their conjugates $(\bz,\bw,\bth)$:
\be\label{eqn:Qgen}
\begin{split}
Q_{+} =
\ppx{\th} + i\bth\bz\ppx{\bw},
&\qquad
\bQ_{+} = \bz\ppx{\bth} -i
\th\ppx{w},
\\
Q_{-} = z\ppx{\th} -i\bth\ppx{\bw},
&\qquad
\bQ_{-} = \ppx{\bth} +i\th z\ppx{w}.
\end{split}
\ee
The angular momentum operators change
slightly with the fermionic contributions:
\be
\begin{split}
J_1 &= -z\ppx{z}
+\bz\ppx{\bz}-w\ppx{w}+\bw\ppx{\bw}
 - \frac{1}{2}\th\frac{\partial}{\partial \theta}
+ \frac{1}{2}\bth\frac{\partial}{\partial \bar{\theta}},
\\
J_{+} &\equiv J_2 + i J_3 =\ppx{z}+\bz^2\ppx{\bz}+2\bz\bw\ppx{\bw}
 + \bz\bth\ppx{\bth},
\\
J_{-} &\equiv J_2 - i J_3 = -z^2\ppx{z}-\ppx{\bz}-2zw\ppx{w}
 -z\th \ppx{\th}.
\end{split}
\label{eqn:JgenSUSY}
\ee
Note that $\bz$ and $\bw$ have $J_1$-eigenvalue of $+1,$
$\bth$ and $Q_{+}, \bQ_{+}$ have $J_1$-eigenvalue
$+\tfrac{1}{2}$,
$\th$ and $Q_{-}, \bQ_{-}$ have $J_1$-eigenvalue
$-\tfrac{1}{2}$, and $z, w$ have $J_1$-eigenvalue $-1.$
For $N=8$ supersymmetry, we take four copies of the
fermionic coordinates
$\th^A, \bth^A$ ($A=1,\dots,4$).
We will denote this super mini-twistor space by $\Tw_3.$
It is $T\CP^1$ with the four fermionic coordinates $\th^A$
fibered over it.

\subsection{The scalar propagator}
\label{subsec:propagator}
For a fixed point $\vx'\in\R^3$
the Green's function for Laplace's equation on $\R^3,$
$$
G(\vx, \vx') = \frac{1}{\norm{\vx-\vx'}},
$$
is harmonic away from $\vx=\vx'$, and therefore
it should be possible to express it as in
Whittaker's formula \eqref{eqn:Whitzw}.
In fact, it is not hard to check that
\be\label{eqn:Gvphi}
\frac{1}{\norm{\vx-\vx'}}\sgn(x_1'-x_1) =
\frac{1}{2\pi i}\oint
  \vphi(z, -[x_2 - i x_3] +2 z x_1 +z^2 [x_2 + i x_3]; \vx')dz,
\ee where \be\label{eqn:Gtxv} \vphi(z,w;\vx')= \frac{2}{(x'_2 + i
x'_3) z^2 +2 x'_1 z-(x'_2 - i x'_3)-w}. \ee Note that $\vphi$ has
a simple pole whenever the mini-twistor $(z,w)$ and the point
$\vx'$ satisfy the incidence relation \eqref{eqn:increl}. Thus,
the integral \eqref{eqn:Whitzw} diverges when $\vx=\vx'$, as it
should. The extra sign $\sgn(x_1'-x_1)$ on the left-hand side is
required if we take the contour of integration in
\eqref{eqn:Whitzw} to be the unit circle $\abs{z}=1.$

The left-hand side of \eqref{eqn:Gvphi} is translationally
invariant. Likewise, the mini-twistor transform $\vphi(z,w;\vx')$
is also translationally invariant in the sense that it satisfies
\bear 0 &=& \Bigl(\ppx{w}-\frac{1}{2}\ppx{x'_2}
-\frac{i}{2}\ppx{x'_3}\Bigr)\vphi(z,w;\vx'),
\nn\\
0  &=& \Bigl(z\ppx{w}+\frac{1}{2}\ppx{x'_1}\Bigr)\vphi(z,w;\vx'),
\nn\\
0  &=&
\Bigl(-z^2\ppx{w}-\frac{1}{2}\ppx{x'_2}+\frac{i}{2}\ppx{x'_3}\Bigr)\vphi(z,w;\vx').
\label{eqn:tinvphi}
\eear
Here we have used \eqref{eqn:Pgen} to
write the translation generators in terms of $z,w.$ The
translational invariance of $\vphi(z,w;\vx')$ is not a completely
trivial statement, because the integral on the right-hand side of
\eqref{eqn:Gvphi} would have been translationally invariant even
if, say, the left-hand sides of \eqref{eqn:tinvphi} were not zero
but were holomorphic functions of $z,w.$

Can we go one step further and mini-twistor transform $\vphi(z,w;
\vx')$ with respect to $\vx'$ to get a meromorphic function of two
mini-twistors $(z,w)$ and $(z',w')$? Using the familiar expansion
\be\label{eqn:xxpYY}
\frac{1}{\norm{\vx-\vx'}} =
4\pi\sum_{l=0}^\infty\sum_{m=-l}^l \frac{1}{2l+1}
\frac{\norm{\vx}^l}{\norm{\vx'}^{l+1}}
 Y_{l m}(\tfrac{\vx}{\norm{\vx}})
 Y^*_{l m}(\tfrac{\vx'}{\norm{\vx'}})
\qquad
\text{(for $\norm{\vx}<\norm{\vx'}$)}
\ee
and the
mini-twistor transforms \eqref{eqn:rplYlm}-\eqref{eqn:rmlYlm},
we arrive at the mini-twistor transform of the scalar propagator
\be\label{eqn:twprop}
\wG(\twv,\twv')\equiv\wG(z,w;z',w')=
\frac{2(w' z + w z')}{(w - w') (w z'^2 - w' z^2)},
\ee
where
$\twv\equiv (z,w)$ and $\twv'\equiv (z',w')$ are shorthand for our
twistor variables.

It can be explicitly checked that if $\norm{\vx}<\norm{\vx'}$,
then
\bear
\lefteqn{ \frac{1}{\norm{\vx-\vx'}}
=\sgn(x'_1)\oint_{\abs{z}=1} \frac{dz}{2\pi i}
\oint_{\abs{z'}=1}\frac{dz'}{2\pi i} }
\nn\\ &&
\wG( z, -[x_2 - i x_3] +2 x_1 z +[x_2 + i x_3] z^2;
z', -[x'_2 - i x'_3] +2 x'_1 z'
+[x'_2 + i x'_3] {z'}^2)
\label{eqn:IntIntGtt}
\eear
holds, unless all of the inequalities
\be\label{eqn:condxxp}
\text{$\abs{x_1'}\le\norm{\vx}\le\norm{\vx'}$
 and
$4x_1^2\le x_2^2 + x_3^2$
and
$4x_1'^2 \le x_2'^2 + x_3'^2$}
\ee
are satisfied.
If condition \eqref{eqn:condxxp} holds,
there are poles along the
integration path which need a special treatment.
Identity \eqref{eqn:IntIntGtt} does not necessarily hold for
values of $\vx$ and $\vx'$ that do not
satisfy \eqref{eqn:condxxp}. This is because when we derive
\eqref{eqn:twprop} from \eqref{eqn:rplYlm}-\eqref{eqn:rmlYlm} and
\eqref{eqn:xxpYY}, we have to change the order of integration and
summation, and for an infinite series that does not necessarily
converge.

If, on the other hand, $\norm{\vx}>\norm{\vx'}$, then
\eqref{eqn:IntIntGtt} still holds, except that $\sgn(x'_1)$ needs
to be replaced by $\sgn(x_1)$. The analog of the condition
\eqref{eqn:condxxp} is now
\be\label{eqn:condxxp2}
\text{$\abs{x_1}\le\norm{\vx'}\le\norm{\vx}$
and $4x_1^2\le x_2^2 + x_3^2$
and $4x_1'^2 \le x_2'^2 + x_3'^2$}.
\ee

The mini-twistor transform $G(\twv,\twv')$ is not uniquely
defined, because we can, for example, add an arbitrary meromorphic
function that has no poles in the region $\abs{z}\le 1,$ and we
can also add an arbitrary meromorphic function with no poles in
the region $\abs{z}\ge 1$ (including $z=\infty$). We can also add
functions with similar properties for $z'.$ Note also that the
propagator \eqref{eqn:twprop} is not invariant under translations.
The total translation generators can be read off from
\eqref{eqn:Pgen}. When acting on holomorphic functions, they
reduce to
$$
P_{+} = -i\ppx{w}-i\ppx{w'}, \qquad P_{1}=iz\ppx{w}+iz'\ppx{w'},
\qquad P_{-}=iz^2\ppx{w}+i{z'}^2\ppx{w'}.
$$
It can be checked that $P_{\pm}\wG$ and $P_1\wG$ do not vanish.

It is interesting that the off-shell propagator can be
recast in terms of mini-twistors,
in some region of parameter space.
This seems to be the $D=3$ analog of the off-shell
twistor propagator of \cite{Siegel:2004dj}.
It is possible that these formulas could be used to
convert Feynman diagram rules, which are usually expressed
in terms of momenta or coordinates, to diagrams in terms
of mini-twistor variables.
We were unsuccessful in putting such rules
to practical use. This is partly because
\eqref{eqn:IntIntGtt} only holds under the assumption \eqref{eqn:condxxp}, and partly because individual Feynman
diagrams of gauge theories are not gauge invariant.

\subsection{Minkowski space $\R^{2,1}$}\label{subsec:Minkowski}
We can also define mini-twistor space for Minkowski signature.
Pick a Majorana representation of the Clifford algebra, say,
\be\label{eqn:gamma matrices}
\Gamma^0 = \begin{pmatrix} 0 & 1 \\ -1 & 0 \\ \end{pmatrix},
\qquad
\Gamma^1 = \begin{pmatrix} -1 & 0 \\ 0 & 1 \\ \end{pmatrix},
\qquad
\Gamma^2 = \begin{pmatrix} 0 & 1 \\ 1 & 0 \\ \end{pmatrix}.
\ee
For a 3-momentum $\vec{k}=(k^0, k^1, k^2),$
set
$$
k_\u\Gamma^\u =
\begin{pmatrix} -k_1 & k_2+k_0  \\ k_2-k_0 & k_1\\ \end{pmatrix}
=
\begin{pmatrix} -k^1 & k^2-k^0  \\ k^2+k^0 & k^1\\ \end{pmatrix}
\Longrightarrow {k_\a}^\b = k_\u{(\Gamma^\u)_\a}^\b.
$$
For a lightlike $\vec{k}$ we get
$$
0 = \det (k_\u\Gamma^\u)\Longrightarrow {k_\a}^\b =
\lam_\a\tlam^\b,
$$
and if $\vec{k}$ is real, $\lam$ and $\tlam$ are also real.

We get mini-twistor space by Fourier transforming with respect to
$\tlam$:
$$
e^{i\vec{k}\cdot\vec{x}}\longrightarrow \frac{1}{(2\pi)^2}\int
d^2\tlam \ e^{i {x^\a}_\b\lam_\a\tlam^\b} e^{i\tlam^\b\mu_\b}
=\delta^{(2)}({x^\a}_\b\lam_\a+\mu_\b),
$$
where
$$
{x^\a}_\b =
\begin{pmatrix} -x^1 & x^2+x^0  \\ x^2-x^0 & x^1\\ \end{pmatrix}.
$$
The condition
$$
\mu_\b + {x^\a}_\b\lam_\a = 0
$$
is related to the incidence relation \eqref{eqn:increl} as
follows. Expanding the components we get
\be\label{eqn:mumu}
-\mu_1 = {x^1}_1\lam_1 + {x^2}_1\lam_2
 = -x^1\lam_1 + (x^2 - x^0)\lam_2,
\qquad
-\mu_2 = {x^1}_2\lam_1 + {x^2}_2\lam_2
 =(x^2+x^0)\lam_1 +x^1\lam_2.
\ee
Now set
\be\label{eqn:minzw}
z=\frac{\lam_1}{\lam_2},
\qquad
w=\frac{\mu_1\lam_2-\mu_2\lam_1}{(\lam_2)^2}.
\ee
Then we find from \eqref{eqn:mumu} that
\be\label{eqn:minkrel}
w = 2 z x^1 -(x^2-x^0) +(x^2+x^0)z^2.
\ee
Renaming $x^0\rightarrow i x^3$, the
Minkowski incidence relation \eqref{eqn:minkrel} becomes the
Euclidean version \eqref{eqn:increl}. For Minkowski space, $z$ and
$w$ are real. $z$ takes values in $\RP^1\simeq S^1$ and $w$ takes
values in its tangent space, so together they parameterize the
tangent bundle $T S^1\simeq \R\times S^1.$

\section{Dimensional reduction}\label{sec:dimred}
We can calculate amplitudes of $D=3$ super Yang-Mills theory by
dimensional reduction of $D=4$ amplitudes. For this purpose, we
need to understand the connection \cite{Hitchin:1982gh} between
the twistor space $\CP^3\setminus\CP^1$ of $\R^4$ and the
mini-twistor space $T\CP^1$ of $\R^3.$ We will review this
connection in \secref{subsec:dimred}, after a brief review of
dimensional reduction for super Yang-Mills theory. For a recent
comprehensive review of various aspects of $D=3$ Yang-Mills theory
see \cite{Nair:2003em}.

\subsection{Field theory}\label{subsec:ft}
Euclidean
$D=4$ Yang-Mills theory with $N=4$ supersymmetry has an $SU(4)$
R-symmetry group and $SU(2)_L\times SU(2)_R$
is the (double cover of) the rotation group.
We will now review the $SU(2)_L\times SU(2)_R\times SU(4)$
representations of the fields, and at the same time introduce our
notation. The field content is given by: a bosonic gauge field
$A_i$ ($i=1,\dots,4$) in $(\rep{2},\rep{2},\rep{1})$, bosonic
scalars $\Phi^I$ ($I=1,\dots, 6$) in $(\rep{1},\rep{1},\rep{6})$,
fermionic left-spinors $\psi^A_\a$ ($A=1,\dots,4$ and $\a=1,2$) in
$(\rep{2},\rep{1},\rep{4})$, and fermionic right-spinors
$\psi_{\dta A}$ ($\dta=\dot{1},\dot{2}$) in
$(\rep{1},\rep{2},\overline{\rep{4}})$.

Dimensional reduction to $D=3$ proceeds by taking
all the fields to be independent
of $x_4$ and defining $\Phi^7\defineas A_4.$
The (double cover of the)
rotation group in (Euclidean)
$D=3$ is $SU(2)$ and the distinction between
dotted and undotted spinors disappears.
The $D=3$ R-symmetry group is $\Spin(7).$
We will now review the $SU(2)\times\Spin(7)$
representations of the $D=3$ fields.
The fermions $\psi^A_\a$ and $\psi_{\dta A}$ combine to
form fields in the $(\rep{2},\rep{8})$,
the gauge field components $A_1,\dots, A_3$
form a gauge field in the
$(\rep{3},\rep{1})$ [that is dual to a scalar in $D=3$],
and the scalars $\Phi^I$ ($I=1,\dots,7$) are in the
$(\rep{1},\rep{7})$ representation.

We will denote the $D=3$ fermions by $\chi^a_\a$ with
$a=1,\dots,8$ being the index of the spinor representation of
$so(7)$ and $\a=1,2$ the $D=3$ spacetime spinor index. The $D=3$
Lagrangian is given by
\bear g_3^2\Lag &=& \tr\Bigl(
\frac{1}{4}F_{ij}F^{ij} +\frac{1}{2}\sum_{i=1}^7 D_i\Phi^I
D^i\Phi^I
-\frac{1}{4}\sum_{I,J} [\Phi^I,\Phi^J]^2 \nn\\
&&\qquad\qquad\qquad
+\sum_{a=1}^8\chi^a_\a\sigma^{i\,\a\b}\px{i}\chi^a_\b
+\sum_{a,b,I}\epsilon^{\a\b}\Gamma^I_{ab}\Phi^I\chi^a_\a\chi^b_\b
\Bigr), \label{eqn:Lagrangian}
\eear
where $g_3$ is the $D=3$
coupling constant, $\epsilon^{\a\b}$ is the standard antisymmetric
lowering and raising matrix for 2-component spinors,
$\sigma^{i\a\b}$ are Pauli matrices,
and $\Gamma^I_{ab}$ are $so(7)$ Dirac matrices.

Helicity in $D=3$ is defined as follows.
We Wick rotate to Minkowski metric $\R^{2,1}$ and let $k^\u$ ($\u=0,\dots,2$)
be a 3-momentum of a massless particle.
Choose a reference frame and let $\vk \equiv k \vn\in \R^2$
be the spatial component of $k^\u$, with $\vn$ a unit vector.
Then, in the temporal gauge $A_0=0$,
a $(\pm)$ helicity  photon has a wave-function $\vA$ (the two-component
spatial part of $A_i$) satisfying $\vn\times\vA = \pm i\Phi^7.$
This can be described more conveniently as follows.
In $D=3$ a photon is equivalent to a scalar $\Phi^8$, by duality.
The field-strength is then given by
$$
F_{ij} = \epsilon_{ijl}\px{l}\Phi^8.
$$
A $(\pm)$ helicity photon then
satisfies $\Phi^7 =\pm i \Phi^8.$ Note
that the condition for a particular helicity breaks the $\Spin(7)$
R-symmetry to $\Spin(6)\simeq SU(4).$ We will see later in
\secref{subsec:dimsup} that, indeed, our super-mini-twistor space
only has a manifest $SU(4)$ symmetry and not $\Spin(7).$

\subsection{Dimensional reduction of $D=4$ twistor space}
\label{subsec:dimred}
What is the relation between the twistor space
$\CP^3\setminus\CP^1$ of $\C^4$
(regarded here as complexified $\R^{2,2}$)
and the mini-twistor space $T\CP^1$ of $\R^3$?
The answer is that $\CP^3\setminus\CP^1$
is a fibration over $T\CP^1$ \cite{Hitchin:1982gh}.
To see this, consider an arbitrary
lightlike 4-momentum $k^\u$ in $\C^4.$ It can be written as
$$
k_{\a\dta} = \lam_\a\tlam_\dta.
$$
In accordance with the Clifford algebra in \eqref{eqn:gamma
matrices}, we choose the Majorana-Weyl spinor
representation of $\R^{2,2}$ as
\be\label{eqn:Weylspinor}
\sigma^0_{\a\dot{\a}} =
\begin{pmatrix} 0 & -1 \\ 1 & 0 \\ \end{pmatrix},
\quad
\sigma^1_{\a\dot{\a}} =
\begin{pmatrix} 1 & 0 \\ 0 & -1 \\ \end{pmatrix},
\quad
\sigma^2_{\a\dot{\a}} =
\begin{pmatrix} 0 & -1 \\ -1 & 0 \\ \end{pmatrix},
\quad
\sigma^{3}_{\a\dot{\a}} =
\begin{pmatrix} -1 & 0 \\ 0 & -1 \\ \end{pmatrix},
\ee
where we are working in signature $(+--+)$. Now suppose that we set
$k^{3}=0$, so that $k^\u$ will lie in $\C^3$
(complexified $\R^{2,1}$).
This implies the constraint
\be\label{eqn:k3eq0}
k^{\a\dta}\sigma^{3}_{\a\dta} =
0\Longrightarrow \lam^1\tlam^{\dot{1}}+\lam^2\tlam^{\dot{2}}=0.
\ee
In general, we will refer to the direction on which we
dimensionally reduce as the {\it $4^{th}$ direction}.
We can now see how
a choice of direction for dimensional reduction naturally defines
a fibration structure of $\CP^3\setminus\CP^1$.
For example, the choice of $4^{th}$ direction
in \eqref{eqn:k3eq0} defines the fibration by the condition
that two twistors $(\lam,\mu)$ and $(\lam',\mu')$ in
$\CP^3\setminus\CP^1$ belong to the same fiber if
\be\label{eqn:fibeq}
\lam'_\a=\lam_\a,
\qquad
\mu'_{\dot{1}} =
\mu_{\dot{1}} - t\lam^1,
\quad
\mu'_{\dot{2}} = \mu_{\dot{2}}-t\lam^2,
\ee
for some $t\in\C.$ The equivalence relation
\eqref{eqn:fibeq} arises naturally from \eqref{eqn:k3eq0} if we
recall that the twistor transform of $\R^{2,2}$ is the Fourier
transform from $\tilde{\lam}$  to $\mu$. In general, had we chosen
another direction $n^\mu$ on which to dimensionally
reduce [instead of a unit vector $(0,0,0,1)$], we would have
gotten the condition
$$
\lam'_\a=\lam_\a,\qquad
\mu'_{\dta} = \mu_{\dta}- t n_\u\sigma^\u_{\a\dta}\lam^\a.
$$

To see that \eqref{eqn:fibeq} gives $T\CP^1$ as the base of the
fibration, consider the two patches of $\CP^3\setminus\CP^1$,
defined by the conditions $\lam_1\neq 0$ and $\lam_2\neq 0$,
respectively. If $\lam_2\neq 0$ we can set $z=\lam_1/\lam_2,$ and
rescale by $\lam_2$ to get
$$
(\lam_1, \lam_2, \mu^{\dot{1}}, \mu^{\dot{2}})
\rightarrow
(z, 1,
\mu^{\dot{1}}/\lam_2, \mu^{\dot{2}}/\lam_2).
$$
After raising and lowering indices, the equivalence relation
\eqref{eqn:fibeq} can be written as
\be\label{eqn:fibeq2}\lam'_\a=\lam_\a,\qquad\mu'^{\,\dot{1}} =
\mu^{\dot{1}} + t\lam_1, \quad \mu'^{\,\dot{2}} = \mu^{\dot{2}}
+t\lam_2.\ee Therefore, it has a unique representative given by
$\mu^{\dot{2}}=0.$ We get to that representative by picking
$t = -\mu^{\dot{2}}/\lam_2$ in \eqref{eqn:fibeq2}, and
using $\mu'$ instead of $\mu.$ Thus, the twistor
$$
\Bigl(z,\, 1,\, \frac{\mu^{\dot{1}}}{\lam_2} + t z,\,
\frac{\mu^{\dot{2}}}{\lam_2} + t\Bigr) = \Bigl(z,\, 1,\,
\frac{\mu^{\dot{1}}\lam_2 - \lam_1\mu^{\dot{2}}}{(\lam_2)^2},\,
0\Bigr)
$$
represents the equivalence class \eqref{eqn:fibeq}. We set
\be\label{eqn:wlammu}
w \defineas
\frac{\mu^{\dot{1}}\lam_2 -\lam_1\mu^{\dot{2}}}{(\lam_2)^2}.
\ee
Now consider the other patch
$\lam_1\neq 0.$ By similar arguments
$$
\Bigl(1,\, \frac{1}{z},\, 0,\, \frac{\mu^{\dot{2}}\lam_1 -
\lam_2\mu^{\dot{1}}}{(\lam_1)^2}\Bigr)
$$
represents \eqref{eqn:fibeq}. In this patch we choose the
coordinates
$$
z' = \frac{\lam_2}{\lam_1}=\frac{1}{z}, \qquad w' =
\frac{\mu^{\dot{2}}\lam_1 - \lam_2\mu^{\dot{1}}}{(\lam_1)^2}
 = -\frac{w}{z^2}.
$$

A given spacetime point $x\in\R^{2,2}$ corresponds to a
holomorphic section on $\Tw(\R^{2,2})$
through the \emph{incidence relation} \cite{Witten:2003nn}:
$$
\mu_{\dot{\a}}+x_{\a\dot{\a}}\lam^\a=0.
$$
In particular, if $x\in\R^{2,1}$ ($x^{0'}=0$),
the incidence relation gives
$$
\mu^{\dot{1}}
= x^{1\dot{1}}\lam_1 + x^{2\dot{1}}\lam_2
= x^1\lam_1 - (x^2 - x^0)\lam_2,
\qquad
\mu^{\dot{2}}
= x^{1\dot{2}}\lam_1 + x^{2\dot{2}}\lam_2
= -(x^2+x^0)\lam_1 - x^1\lam_2,
$$
where we used the conventions \eqref{eqn:Weylspinor}.
This together with
\eqref{eqn:wlammu} leads to the same
three-dimensional incidence relation as \eqref{eqn:minkrel},
except that $z$ and $w$ are now complex
numbers. Again, \eqref{eqn:minkrel} becomes \eqref{eqn:increl} by
taking $x^0 \rightarrow ix^3$. Therefore, $(z,w)$ and $(z',w')$
parameterize the mini-twistor space $T\CP^1$ of $\R^3$
exactly as described in \secref{subsec:TCP1}.

Thus, the twistor space $\CP^3\setminus\CP^1$ of $\C^4$ is a
fibration over the mini-twistor space $T\CP^1$ of $\R^3.$ The
fiber is $F\simeq\C$ and the structure group is the group $\C$ of
translations of $\C.$ To see this, note that on the patch of
$T\CP^1$ where $z$ is a good coordinate,
$u\defineas\mu^{\dot{2}}/\lam_2$ is a good coordinate on $F$, and
on the other patch where $z'$ is a good coordinate,
$u'\defineas\mu^{\dot{1}}/\lam_1$ is a good coordinate on $F$. On
the intersection of the two patches, where both $z$ and $z'$ are
good coordinates,
$$
u' = \frac{\mu^{\dot{1}}}{\lam_1} = u + \frac{w}{z}.
$$
We will see in \secref{subsec:redtree} that tree-level $D=3$
amplitudes in
mini-twistor space $T\CP^1$ can be obtained from tree-level
$D=4$ amplitudes in twistor space $\CP^3\setminus\CP^1$ by integration
over the fiber $F.$

To understand the fibration structure more geometrically, note
that $\mu=0$ defines a rational curve (which is homeomorphic to
$\CP^1$) in $\CP^3\setminus\CP^1$. The normal bundle to the curve
$\mu=0$ is isomorphic to the direct sum $\cO(1)\oplus\cO(1)$ of
line-bundles. [Embedded in $\CP^3\setminus\CP^1$, the normal
bundle to $\mu=0$ can be parameterized as
$(\lam_1,\lam_2,d\mu^{\dot 1},d\mu^{\dot 2})\sim(z,1,\xi_1,\xi_2)$
for $z= \lam_1/\lam_2\neq\infty$ and
$(\lam_1,\lam_2,d\mu^{\dot{1}},d\mu^{\dot{2}})
\sim(1,z',\xi_1',\xi_2')\equiv(1,1/z,\xi_1/z,\xi_2/z)$
for $z=1/z'\neq 0$.
Thus, the normal bundle is $\cO(1)\oplus\cO(1)$.
See also \cite{Popov:2004rb}-\cite{Wolf:2004hp}
for a related discussion and further details.]
This holomorphic vector bundle has nowhere-vanishing holomorphic
sections. For example, we define a section $\OOs$ of
$\cO(1)\oplus\cO(1)$ as follows. Set
 $\OOs(z,1)=(z,1,z,1)$ for $z\neq\infty$ and ${\mathfrak
s}(1,z')=(1,z',1,z')$ for $z'=1/z\neq\infty$. Obviously, the
section values agree on the intersection of the two patches
because
$(z,1,z,1)\sim(\lam_1,\lam_2,\lam_1,\lam_2)\sim(1,z',1,z'),$ and
$\OOs$ is nowhere zero.

Based on the section $\OOs$, we can define the sub-bundle
of $\cO(1)\oplus\cO(1)$ as $c\OOs$ ($c\in\C$), which is a
trivial line-bundle over $\mu=0$,
since $c$ is a good global coordinate for the line fibers.
Modding out $\cO(1)\oplus\cO(1)$ by this
trivial line bundle, we get a quotient space.
This means that we impose the equivalence
relation on $\cO(1)\oplus\cO(1)$ by $c\OOs$:
\be
(\xi_1,\xi_2)+c(z,1)\sim(\xi_1,\xi_2),\; z\neq\infty;\qquad
(\xi'_1,\xi'_2)+c'(1,z')\sim(\xi'_1,\xi'_2),\; z'\neq\infty.
\ee
We get the representatives $(\eta,0)=(\xi_1-\xi_2z,0)$ and $(0,\eta')=(0,\xi_2'-\xi'_1z')$
by choosing $c=-\xi_2$ and $c'=-\xi'_1$. It can be easily shown that $\eta'=-\eta/z^2$ and
therefore the quotient space is a line bundle $\cO(2)$.

The resulting $\cO(2)$ line bundle can be identified with the
mini-twistor space $T\CP^1$. The $\mu=0$ rational curve is then
identified with the base of $T\CP^1.$ We denote the projection by
\be\label{eqn:projTCP}
\pi:\CP^3\setminus\CP^1\rightarrow T\CP^1.
\ee
These observations can be readily modified to spacetime $\R^{2,2}$
with signature $(++--)$, for which the twistor space is
$\RP^3\setminus\RP^1.$ All the equations above still apply, but
$z, w$ and $\lam, \mu$ should be real. In particular, if we choose
a timelike $4^{th}$ direction ($0'$) for dimensional reduction we
get $D=3$ Minkowski space $\R^{2,1}.$ The choice of $4^{th}$
direction defines a fibration structure on twistor space
$\RP^3\setminus\RP^1$, with the projection 
\be\label{eqn:projTRP}
\pi':\RP^3\setminus\RP^1\rightarrow T\RP^1. 
\ee 
The base of this
fibration is the mini-twistor space of $D=3$ Minkowski space
$T\RP^1$. The incidence relation is \eqref{eqn:minkrel}, as
discussed in \secref{subsec:Minkowski}.

\subsection{Dimensional reduction of $D=4$ super-twistor space}
\label{subsec:dimsup}
The dimensional reduction of supertwistor space $\CP^{3\lvert 4}$
proceeds in a similar fashion.
We take homogeneous variables
$$
(\lam_1, \lam_2, \mu^{\dot{1}}, \mu^{\dot{2}},
\theta^1, \theta^2, \theta^3, \theta^4)
$$
on $\CP^{3\lvert 4}$ and begin with the patch $\lam_2\neq 0.$
At the end of \secref{subsec:dimred} we identified the
three-dimensional mini-twistor space $T\CP^1$ with a quotient of
the normal bundle to the $\mu=0$ rational curve, where we modded
out by a trivial sub-bundle. We can repeat the same procedure for
$\CP^{3\lvert 4}.$ Pick the rational curve given by $\mu=0$ and
$\theta=0.$ The normal superspace is a sum of $\cO(1)\oplus\cO(1)$
corresponding to the $\mu$ directions, and four copies of
anti-commuting $\cO(1)$ spaces, corresponding to the $\theta$
directions. As before, a choice of $4^{th}$ direction on which to
dimensionally reduce defines, as in \eqref{eqn:fibeq}, a trivial
sub-bundle of the commuting $\cO(1)\oplus\cO(1)$ vector bundle.
Modding out by this subspace leaves
$$
\cO(2|0)\oplus \cO(0|1)^4,
$$
where the first factor is commuting and the last four are
anticommuting. This is the $D=3$ super-mini-twistor space. It can
be covered with two patches $U_1$ and $U_2.$
On the first patch $U_1$
the local coordinates are $(z,w,\theta_1,\dots,\theta_4)$
(where $z,w$ are commuting),
and on the second patch $U_2$ the local
coordinates are $(z',w',\theta'_1,\dots,\theta'_4)$,
with transition functions
\be\label{eqn:strzw}
z' = \frac{1}{z}, \qquad
w' = -\frac{w}{z^2},\qquad
\theta_i' = \frac{1}{z}\theta_i\quad
(i=1\dots 4),
\ee
generalizing \eqref{eqn:zpwp}.
Note that this super-mini-twistor space is a noncompact Calabi-Yau
supermanifold \cite{Sethi:1994ch}. Defining $x$ to be the
generator of the cohomology group $H^2(\CP^1,\Z),$ we find that
the first Chern class of $\CP^1$ is $2x$, the $\cO(2|0)$ factor
contributes another $2x$, and the four $\cO(0|1)$ anti-commuting
factors contribute $-x$ each, to a total of $0.$

\subsection{Tree-level amplitudes}\label{subsec:redtree}
For simplicity, we will start with $D=4$ with signature $\R^{2,2}$
and dimensionally reduce by picking a timelike $4^{th}$ direction
to obtain $D=3$ Minkowski space $\R^{2,1}.$
We have seen in \secref{subsec:dimred}
that the twistor space $\RP^3\setminus\RP^1$ is a fibration over $T\RP^1$.
We will now argue that
the tree-level $D=3$ amplitudes are obtained by integrating
the $D=4$ tree-level amplitudes over the fiber.

In \cite{Witten:2003nn}, the twistor space $\RP^3\setminus\RP^1$
is obtained as the parameter space of the Fourier transforms of
functions of $k^\u_{\a\dot{\a}}=\lam_\a\tilde{\lam}_{\dot{\a}}$
with respect to $\tlam.$ Alternatively, if we have a function
$\tF(\lam,\mu)$ on the twistor space, we can get back a function
$F(\lam,\tlam)$ by Fourier transforming with respect to $\mu.$

To dimensionally reduce a given amplitude of the $D=4$ theory in its
twistor space $\RP^3\setminus\RP^1$, we need to enforce the
condition \eqref{eqn:k3eq0}.
Let us integrate $\tF(\lam,\mu)$ over the fiber 
of the fibration \eqref{eqn:projTRP}.
We get
\bear
\frac{1}{2\pi} \int dt
\tF(\lam_1, \lam_2, \mu_{\dot{1}}+t\lam^1, \mu_{\dot{2}}+ t\lam^2)
&=& \frac{1}{2\pi} \int dt
d^2\tilde{\lambda}\,
e^{i\tlam^{\dot{1}}(\mu_{\dot{1}} +t\lam^1) +
   i\tlam^{\dot{2}}(\mu_{\dot{2}}+t\lam^2)}
F(\lam_1, \lam_2, \tlam^{\dot{1}}, \tlam^{\dot{2}})
\nn\\
&=& \int d^2 \tilde{\lambda}\, e^{i\tlam^\dta\mu_\dta}
\delta(\lam^1\tlam^{\dot{1}}+\lam^2\tlam^{\dot{2}})
F(\lam_1, \lam_2, \tlam^{\dot{1}}, \tlam^{\dot{2}})
\label{eqn:FT}
\eear
Thus, integrating over the fiber is equivalent to inserting a
delta function
$\delta(\lam^1\tlam^{\dot{1}}+\lam^2\tlam^{\dot{2}})=\delta(k^3).$
To transform an $n$-particle $D=4$ tree-level amplitude
$A_4(\lam^{(1)},\mu^{(1)}; \cdots; \lam^{(n)},\mu^{(n)})$ on
$\RP^3\setminus\RP^1$ to $T\RP^1$, all we need then is to
integrate $A_4$  over the fibers of $n$ twistors:
\be
A_3(\lam^{(1)},\mu^{(1)}; \cdots; \lam^{(n)},\mu^{(n)})=
\int\prod_{j=1}^n dt_j A_4(\lam^{(1)},\mu^{(1)}+t_1\lam^{(1)};
\cdots; \lam^{(n)},\mu^{(n)}+t_n\lam^{(n)}).
\ee
In momentum space,
each $dt_j$ integration inserts a $\delta(k^3_j)$, as in
\eqref{eqn:FT}. Because of total 4-momentum conservation, this is
one $\delta(k_j^3)$ too many,
$$
\delta(\sum_{j=1}^n k_j^3)
\prod_{j=1}^n \delta(k_j^3)
=\delta(0)\prod_{j=1}^n \delta(k_j^3).
$$
The infinite factor $\delta(0)$ can be regularized if we take a
compact $4^{th}$ dimension. The singular $\delta(0)$ can then be
interpreted as the spatial size ($\sim 2\pi R$) in the $4^{th}$
direction, which is absorbed in the $D=3$ coupling constant
$g_3=g_4/2\pi R.$ (Here $g_4$ is the $D=4$ coupling constant.) The
resulting amplitude $A_3$ is invariant under the equivalence
relation \eqref{eqn:fibeq} for each particle separately. It can
therefore be written as a function of only
$\twv_j\defineas\pi'(\lam^{(j)},\mu^{(j)})$ (for $j=1\dots n$)
[where the projection $\pi'$ was defined in \eqref{eqn:projTRP}].

We are now ready to show that tree-level amplitudes have support
on algebraic curves, as in $D=4$ \cite{Witten:2003nn}. Let
$A_3(\twv_1, \dots, \twv_n)$ be a tree-level $n$-point $D=3$
amplitude as a function of $n$ mini-twistors in $T\RP^1.$ As
argued above, this amplitude can be obtained from a $D=4$
amplitude $A_4(\wtfd_1,\dots, \wtfd_n)$ where $\wtfd_i$ is a
twistor in $\RP^3\setminus\RP^1$ that projects to $\twv_i$ in the
fibration above. That is, $A_3(\twv_1, \dots, \twv_n)$ is obtained
from $A_4(\wtfd_1,\dots, \wtfd_n)$ by integrating with respect to
$\wtfd_1, \dots, \wtfd_n$ over the fibers above $\twv_1, \dots,
\twv_n.$ In the notation of \eqref{eqn:projTRP} we have
$\pi'(\wtfd_i)=\twv_i.$ According to \cite{Witten:2003nn}, $A_4$
is nonzero only if its arguments $\wtfd_1, \dots, \wtfd_n$ lie on
an algebraic curve of degree $d=q-1$, where $q$ is the number of
negative helicity gluons. Since this curve is given by algebraic
equations in the homogeneous coordinates of $\RP^3$, it can be
analytically continued to $\CP^3.$ Let this analytically continued
curve be $\wCurve\subset\CP^3.$ According to \cite{Witten:2003nn},
this curve must be of genus $0$, otherwise the amplitude vanishes.
We denote its projection by $\Curve=\pi(\wCurve)\subset T\CP^1$,
where we used  $\pi$ from \eqref{eqn:projTCP}. The genus of a
projection of a sphere cannot be $> 0$, so $\Curve$ is also of
genus $0$. In the local coordinates $(z,w)$ on $T\CP^1$, $\Curve$
can be expressed as a polynomial equation
\be\label{eqn:Czw}
0 = \sum_{r,s} C_{r,s} z^r w^s,
\ee
with some coefficients
$C_{r,s}\in\C.$ In order to reduce Witten's conjectures
\cite{Witten:2003nn} to $D=3$, we need to define the degree of
$\Curve.$ We can do that by identifying $T\CP^1$ with an open
subset of weighted projective space $W\CP^{1,1,2}$ as follows. Let
$\xi_1, \xi_2, \xi_3$ be projective coordinates on $W\CP^{1,1,2}$
defined with the equivalence relation
\be\label{eqn:WCP}
(\xi_1, \xi_2, \xi_3)\sim
(\zeta\xi_1, \zeta\xi_2, \zeta^2\xi_3), \qquad
0\neq \zeta\in\C.
\ee
If we take the singular point
$\xi_1=\xi_2=0$ out of $W\CP^{1,1,2}$, we get the mini-twistor
space $T\CP^1$ as follows,
$$
z = \frac{\xi_2}{\xi_1},\qquad
w = \frac{\xi_3}{\xi_1^2}.
$$
The equation \eqref{eqn:Czw} for $\Curve$
becomes
\be\label{eqn:CzwWCP}
0 =
\sum_{r,s} C_{r,s} \xi_1^{-r-2s}\xi_2^r\xi_3^s.
\ee
We define the degree of $\Sigma$ by
\be\label{eqn:defdeg}
\tilde{d}(\Curve) \defineas
\max_{r,s} (r + 2 s).
\ee
Multiplying \eqref{eqn:CzwWCP} by $\xi_1^{\tilde{d}(\Curve)}$
we see that $\Curve$ can be represented by
a homogeneous polynomial of degree $\tilde{d}(\Curve)$
in weighted projective space $W\CP^{1,1,2}.$

We will see below that the dimensional reduction of a $D=4$
twistor amplitude that corresponds to twistors lying on a
holomorphic curve of degree $d$ in $\CP^3$ reduces to a
mini-twistor amplitude with the mini-twistors restricted to lie on
a curve of the form \eqref{eqn:Czw} and of degree
$\tilde{d}(\Curve) = 2d.$ Thus, it follows immediately from the
observations in \cite{Witten:2003nn} that the $D=3$ tree-level
mini-twistor amplitude with $q$ negative helicity gluons and $n-q$
positive helicity gluons
 is nonzero only if
the $n$ mini-twistors lie on a genus $0$ curve $\Curve$ of degree
$\tilde{d}(\Curve)=2 (q-1)$ in $T\CP^1.$
The Minkowski tree-level amplitude is nonzero if the $n$
mini-twistors
lie on an algebraic curve in
$T\RP^1\simeq \R^1\times S^1$
with the same degree and genus as above.

Suppose that the $D=4$ tree-level amplitude is supported on a
``complete intersection,'' which is defined as the zero set of two
homogeneous polynomials $f_1(\lambda_\a,\mu^\dta)$ and
$f_2(\lambda_\a,\mu^\dta)$, of degrees $d_1$ and $d_2$,
respectively. Then the degree of such a curve is $d=d_1d_2$. The
$D=4$ amplitude can be schematically written as
$$
A_4(\wtfd_1,\ldots,\wtfd_n)=\int_{\mathcal{M}}[df_1][df_2]
\prod_{i=1}^n
\delta\bigl(f_1(\lambda^{(i)}_\a,\mu^{(i)\dta})\bigr)
\delta\bigl(f_2(\lambda^{(i)}_\a,\mu^{(i)\dta})\bigr)
\mathcal{A}(\lambda^{(i)}_\a,\mu^{(i)\dta}),
$$
where the integration is performed over the moduli space
$\mathcal{M}$ of genus 0, degree $d$ curves in $\CP^3$ of this
particular ``complete intersection'' form (or equivalently, over
the space of polynomials $f_1(\lambda_\a,\mu^\dta)$ and
$f_2(\lambda_\a,\mu^\dta)$ with specified degrees, after some
appropriate identifications). According to our previous
discussion, the $D=3$ amplitude can then be obtained by
integrating the $D=4$ amplitude over the fibers of the twistors
$\wtfd_i$:
$$
A_3(\twv_1,\ldots,\twv_n)= \int_{\mathcal{M}}[df_1][df_2]
\int\prod_{i=1}^{n} dt_i
\delta\bigl(f_1(\lambda^{(i)}_\a,\mu^{(i)\dta}+t_i\lambda^{(i)}_\a)\bigr)
\delta\bigl(f_2(\lambda^{(i)}_\a,\mu^{(i)\dta}+t_i\lambda^{(i)}_\a)\bigr)
\mathcal{A}(\lambda^{(i)}_\a,\mu^{(i)\dta}+t_i\lambda^{(i)}_\a).
$$
Therefore, our $D=3$ amplitude will not vanish provided there
exist two polynomials $f_1(\lambda_\a,\mu^\dta)$ and
$f_2(\lambda_\a,\mu^\dta)$ and some value $t_i$ for each $i$ so
that both
$f_1(\lambda^{(i)}_\a,\mu^{(i)\dta}+t_i\lambda^{(i)}_\a)$ and
$f_2(\lambda^{(i)}_\a,\mu^{(i)\dta}+t_i\lambda^{(i)}_\a)$ are
non-zero for each $i$.\par Now, for two fixed polynomials
$f_1(\lambda_\a,\mu^\dta)$ and $f_2(\lambda_\a,\mu^\dta)$, what is
the condition for such values $t_i$ to exist? To answer this
question, let us expand the two polynomials
$f_1(\lambda^{(i)}_\a,\mu^{(i)\dta}+t_i\lambda^{(i)}_\a)$ and
$f_2(\lambda^{(i)}_\a,\mu^{(i)\dta}+t_i\lambda^{(i)}_\a)$ and
group the terms in decreasing order in $t_i$. As these polynomials
are homogeneous of degrees $d_1$ and $d_2$, we will get two
polynomial equations, \be
\begin{split}
a_0 t_i^{d_1}+a_1 t_i^{d_1-1}+\cdots +a_{d_1}=&0,\\
b_0 t_i^{d_2}+b_1 t_i^{d_2-1}+\cdots +b_{d_2}=&0,
\end{split}
\ee where the coefficients $a_0,\ldots,a_{d_1}$ and
$b_0,\ldots,b_{d_2}$ are of degrees $d_1$ and $d_2$, respectively,
in $\lambda^{(i)}_\a$ and $\mu^{(i)\dta}$. But the condition for
the existence of a simultaneous solution $t_i$ for the above two
polynomial equations is that the resultant of the polynomials be
zero. By explicitly writing down the resultant, one can easily
check that each term of the resultant is of degree $d_2$ in
$a_0,\ldots,a_{d_1}$ and of degree $d_1$ in $b_0,\ldots,b_{d_2}.$
Hence the condition for the $D=3$ amplitude to be non-zero is that
the twistors $\wtfd_i=(\lambda^{(i)},\mu^{(i)})$ satisfy a certain
polynomial equation of degree
$$\tilde{d}=d_1d_2+d_2d_1=2d_1d_2=2d.$$
As $z$ is linear and $w$ is quadratic in $\lambda_\a$ and
$\mu^\dta$ after setting $\lam_2=1$ in \eqref{eqn:wlammu}, the
degree of the resulting polynomial when expressed in terms of $z$
and $w$ becomes $\tilde{d}=2d$, according to our definition in
\eqref{eqn:defdeg}.

We will end this subsection with an analysis of dimensional
reduction for Euclidean signature. We have seen at the end of
\secref{subsec:TCP1} that a harmonic scalar function on $\R^3$ is
mini-twistor transformed to an element of the sheaf cohomology
$H^1(T\CP^1, \pbOmega^1)$, where $\pbOmega$ was the pullback of
the sheaf $\cO(-2)$ over $\CP^1.$ By the arguments of
\cite{Witten:2003nn}, for each external particle the scattering
amplitude must be an element of the dual space, which in our case
also happens to be $H^1(T\CP^1, \pbOmega^1)$. As explained in
\S2.5 of \cite{Witten:2003nn}, for each external particle of
helicity $h$ the $D=4$ amplitude is an element of the sheaf
cohomology $H^1(\CP^3\setminus\CP^1,\cO(h-2)).$ We have seen in
\secref{subsec:dimred} that $\CP^3\setminus\CP^1$ is a fibration
over $T\CP^1$ with $\C$ fibers.

Given a $D=3$ harmonic function, we can lift it
to a $D=4$ harmonic function that is invariant under
translations in the $4^{th}$ direction.
What is the corresponding statement for the
twistor transforms?
The pull-back of an element of $H^1(T\CP^1,\pbOmega^1)$,
with respect to the projection \eqref{eqn:projTCP},
is an element of $H^1(\CP^3\setminus\CP^1,\cO(-2))$
that is invariant under translations
along the fiber.

\subsection{Explicit examples of amplitudes}
\label{subsec:Explicit}
We will now give a few examples of dimensionally reduced tree-level amplitudes.

The tree-level maximally helicity-violating (MHV) twistor
amplitude with 2 gluons of negative helicity and $(n-2)$ gluons of
positive helicity is given by formula (3.3) of
\cite{Witten:2003nn}, \be\label{eqn:4dmhv} A(\lam^{(i)},
\mu^{(i)}) = i g_4^{n-2}\int d^4 x\prod_{i=1}^n
\delta^2(\mu^{(i)}_{\dta}+ x_{\a\dta}\lam^{(i)\a})
\frac{\langle\lam^{(r)}, \lam^{(s)}\rangle^4}{
\prod_{i=1}^n\langle\lam^{(i)}, \lam^{(i+1)}\rangle}, \ee where we
use the standard notation
$$
\langle\lam^{(i)}, \lam^{(j)}\rangle\defineas
\lam^{(i)}_{\a}\lam^{(j)\a}, \qquad \lam^{(n+1)}\defineas
\lam^{(1)}.
$$
Here, the $r$-th and $s$-th gluons are of negative helicity. To
get the $D=3$ twistor amplitude, we have to replace the $D=4$
coupling constant $g_4$ by the $D=3$ coupling constant $g_3$,
replace $d^4 x$ by $d^3 x$, and integrate over the fibers:
$$
i g_3^{n-2}\int d^3 x\prod_{i=1}^n dt_i \delta(\mu^{(i)}_{\dot{1}}
+ t\lam^{(i)1} + x_{\a\dot{1}}\lam^{(i)\a})
\delta(\mu^{(i)}_{\dot{2}} + t\lam^{(i)2} +
x_{\a\dot{2}}\lam^{(i)\a})\frac{\langle\lam^{(r)},
\lam^{(s)}\rangle^4}{\prod_{i=1}^n\langle\lam^{(i)},
\lam^{(i+1)}\rangle}.
$$
Note that if we first perform the fiber integrations over $dt_i$
and leave the $d^3 x$ integral for last, the integrand will be
independent of the $4^{th}$ component of $x.$ This is because
momentum conservation in the $4^{th}$ direction is already
enforced, since the $dt_i$ integrations make sure that the
$4^{th}$ component of momentum is zero.

We calculate
$$
\int dt
\delta(\mu_{\dot{1}} + t\lam^{1} + x_{\a\dot{1}}\lam^\a)
\delta(\mu_{\dot{2}} + t\lam^{2} + x_{\a\dot{2}}\lam^\a)
=
\frac{1}{(\lam_2)^2}
\delta(w + [x^2 - i x^3] - 2 x^1 z - [x^2 + i x^3] z^2),
$$
where we used \eqref{eqn:wlammu} and $z=\lam_1/\lam_2.$
The argument of the $\delta$-function enforces the incidence relation
\eqref{eqn:increl}, which is a polynomial of degree 2.
Note also that
$$
\langle\lam^{(i)}, \lam^{(i+1)}\rangle = (z_{i+1} -
z_{i})(\lam^{(i)}_2\lam^{(i+1)}_2).
$$
The $D=3$ MHV amplitude is therefore
$$
i g_3^{n-2}\int d^3 x
\prod_{i=1}^n\delta(w_i + [x^2 - i x^3] - 2 x^1 z_i - [x^2 + i x^3] z_i^2)
\frac{(z_r - z_s)^4}{\prod_{i=1}^n(z_{i+1} - z_{i})}.
$$
Here, we used the fact that on the patch where $\lambda_2\neq 0$
we can scale the factor $\lambda_{2}$ to $1$. If, instead, we had
included all the $\lambda_2$ factors, we would have needed an
extra factor of ${(\lam^{(r)}_{2})^4
(\lam^{(s)}_2)^4}/{\prod_{i=1}^n (\lam^{(i)}_2)^4}.$ This factor
indicates that the MHV amplitude in $D=3$ is homogeneous of degree
$-4$ and $4$ for each negative and positive helicity particle,
respectively. On the patch $\lambda_1\neq 0$ we define the twistor
amplitude as \be\label{eqn:MHVa} A_3(w_i', z_i') = i g_3^{n-2}\int
d^3 x \prod_{i=1}^n \delta(w_i' + [x_2 - i x_3] - 2 x_1 z_i' -
[x_2 + i x_3] z_i'^2) \frac{(z_r' -
z_s')^4}{\prod_{i=1}^n(z'_{i+1} - z'_{i})}. \ee The
$\delta$-functions enforce the incidence relations. In the
geometrical language of \secref{subsec:geom}, the amplitude is
nonzero only if all $n$ mini-twistors, which are oriented lines in
$\R^3$, intersect at a common point.

Our next example is the ``googly'' description of the tree-level
amplitudes with helicities $---++$. It was shown in
\cite{Witten:2003nn} that these are supported on genus zero,
degree two curves in the $D=4$ twistor space $\RP^3$. It was also
shown there that this condition is equivalent to saying that the
amplitude is nonzero only if (i) the five points
$\wtfd_i=(\lambda^{(i)},\mu^{(i)})$ lie on a common
$\mathbb{RP}^2\subset\mathbb{RP}^3$, and (ii) the five points lie
on a common conic section contained in that $\mathbb{RP}^2$. But
once the first condition is satisfied, the second one is
automatic, because a generic set of five points in $\mathbb{RP}^2$
is contained in a unique conic section. Therefore we can
schematically write the $D=4$ amplitude as\footnote{A precise
formula for the googly amplitude was derived from B-model
calculation in \cite{Roiban:2004vt}. For our purposes, this
schematic form is enough.} \be\label{eqn:4dgoogly}
A_4(\wtfd_1,\ldots,\wtfd_5)=
\int_{\mathcal{M}}[da][db]\prod_{i=1}^5
\delta\bigl(\sum_{I=1}^4a_I Z_i^I\bigr)
\delta\bigl(\sum_{I,J=1}^4b_{IJ}Z_i^I Z_i^J\bigr)
\mathcal{A}(Z_1^I,\dots,Z_5^I), \ee where
$$
(Z_i^1,Z_i^2,Z_i^3,Z_i^4)= (\lambda^{(i)}_\a, \mu^{(i)\dta})
$$
are the coordinates of the five twistors $\wtfd_i$, and the
integration is to be performed over the moduli space of genus
zero, degree two algebraic curves in $\RP^3$. (This space
is parameterized by the coefficients $a_I$ and $b_{IJ}$, with
appropriate identification.)

As before, dimensional reduction in twistor space is achieved by
replacing $\mu^{(i)\dta}$ with $\mu^{(i)\dta}+t_i\lambda^{(i)}_\a$
and integrating over $t_i$ for each $i$. So in the $D=3$
amplitude, the first delta function for each $i$ in the above
expression becomes
$$
\delta\bigl(
 a_1\lambda^{(i)}_1
+a_2\lambda^{(i)}_2 +a_3(\mu^{(i)\dot{1}} +t_i\lambda^{(i)}_1)
+a_4(\mu^{(i)\dot{2}} +t_i\lambda^{(i)}_2)\bigr).
$$
Integration over $t_i$ then amounts to solving the equation
$$
a_1\lambda^{(i)}_1 +a_2\lambda^{(i)}_2 +a_3(\mu^{(i)\dot{1}}
+t_i\lambda^{(i)}_1) +a_4(\mu^{(i)\dot{2}} +t_i\lambda^{(i)}_2)=0
$$
for $t_i$ and plugging it into the $t_i$ in the argument of the
second delta function. Therefore, we end up with a product of
delta functions whose arguments look like
$$
 b_{11}(\lambda^{(i)}_1)^2
+b_{12}\lambda^{(i)}_1\lambda^{(i)}_2
+b_{13}\lambda^{(i)}_1(\mu^{(i)\dot{1}}+t_i\lambda^{(i)}_1)
+b_{14}\lambda^{(i)}_1(\mu^{(i)\dot{2}}+t_i\lambda^{(i)}_2)
+\cdots +b_{44}(\mu^{(i)\dot{2}} +t_i\lambda^{(i)}_2)^2,
$$
where $t_i$ are to be replaced with the solution of the above
equation. After clearing the denominators, these arguments become
homogeneous polynomials of degree four in $\lambda^{(i)}_\a$ and
$\mu^{(i)\dta}$. Then from \eqref{eqn:wlammu}, we see that when
expressed in terms of $z_i$ and $w_i$, they will become a
polynomial of degree four in $T\RP^1$ according to our definition
of degree in \eqref{eqn:defdeg}. We conclude that dimensionally
reduced tree-level $---++$ amplitudes are nonvanishing only if the
five twistors $\twv_i$ lie on a common curve of degree four in
$D=3$ twistor space. This result agrees with our claim in
\secref{subsec:redtree}.

But this condition is actually trivial, because a generic set of
five points in $T\RP^1$ always lies on a common curve of degree
four. To see this, it suffices to simply write down the most
general form of degree four curves in $T\RP^1$:
$$f(z,w)=w^2+c_1z^2w+c_2z^4+c_3z^3+c_4zw+c_5z^2+c_6w+c_7z+c_8=0.$$
We have eight parameters $c_1,\ldots,c_8$ at our disposal but have
to satisfy only five constraints $f(z_i,w_i)=0$, so generically
such curves exist.

In summary, we have seen that (a) for MHV amplitudes, the $D=3$
twistor amplitudes after dimensional reduction are nonvanishing
only if the twistors lie on a common algebraic curve with genus 0
and degree 2 in $T\RP^1$, while (b) for the googly description of
$---++$ amplitudes, there does not exist such nontrivial
criterion. The difference between these two cases is easy to
understand in geometrical terms. In the MHV case, the $D=4$
amplitude in \eqref{eqn:4dmhv} contains two $\delta$-functions for
each $i$. We can think of the first set of $\delta$-functions as
enforcing the $n$ points
$(\lambda^{(i)},\mu^{(i)}+t_i\lambda^{(i)})$ to lie on a common
$\RP^2$. Then the second set of $\delta$-functions further demands
that the $n$ points lie on a common line contained in that
$\RP^2$. We can always pick $t_i$ for each $i$ so that the $n$
points lie on a common $\RP^2$, but after fixing $t_i$, these
points will in general not lie on a common line if $n>2.$
Therefore, we still have a nontrivial criterion for nonvanishing
amplitudes after dimensional reduction.

In contrast, for $---++$ amplitudes, the first set of
$\delta$-functions in \eqref{eqn:4dgoogly} requires that the five
twistor points lie on a common $\RP^2$. Again, this can always be
achieved by a judicious choice of $t_i$. But then the second set
of $\delta$-functions demands that the five points lie on a common
conic section, which is satisfied automatically. So there is no
nontrivial criterion for nonvanishing amplitudes.

\subsection{Physical interpretation of the holomorphic curves}
\label{subsec:evolute}
In this subsection we will present a geometrical and physical
interpretation of the holomorphic curves in mini-twistor space. 
There is an interesting connection
between holomorphic curves in mini-twistor space $T\CP^1$
and (real) surfaces of minimal area in the physical space $\R^3.$
A minimal area surface is constructed from 
a holomorphic curve $\Curve\subset T\CP^1$
as follows \cite{Hitchin:1982gh}.
Given a point $(z_0,w_0)$ on $\Curve$, we can write the equation
for $\Curve$ near $(z_0,w_0)$ as \be\label{eqn:oscul} w = w_0 +
a_1 (z-z_0) + a_2 (z-z_0)^2 + O(z-z_0)^3 = (w_0 -a_1 z_0 + a_2
z_0^2) + (a_1-2a_2 z_0)z + a_2 z^2 +O(z-z_0)^3. \ee Dropping the
$O(z-z_0)^3$ terms, we can approximate the curve by a quadratic
equation and find a vector $\vx\in\C^3$ such that this quadratic
equation will look like the incidence relation \eqref{eqn:increl}.
Given the coefficients of $1, z, z^2$ in \eqref{eqn:oscul}, we
therefore define $\vx\equiv (x_1, x_2, x_3)\in\C^3$ as the unique
solution to the linear equations \be\label{eqn:vxoscE} -(x_2-i
x_3) = (w_0 -a_1 z_0 + a_2 z_0^2), \quad 2x_1 = (a_1-2a_2 z_0),
\quad x_2 + i x_3 = a_2. \ee Thus, each point on $\Curve$ defines
a point $\vx\in\C^3$ and the collection of these points defines a
holomorphic curve in $\C^3.$ The projection of this curve to
$\R^3,$ i.e., the collection of points $\Rex\vx\equiv (\Rex x_1,
\Rex x_2, \Rex x_3),$ is a minimal area (real) surface in $\R^3.$
Furthermore, there is a one-to-one map from minimal area surfaces
in $\R^3$ to holomorphic curves in $T\CP^1.$ For more details, we
refer the reader to the appendix of \cite{Hitchin:1982gh}.

We saw in \secref{subsec:redtree} that a physical
amplitude defines a holomorphic curve in $T\CP^1.$
{}From the above discussion it follows that a physical
amplitude defines a minimal area surface in $\R^3.$
What is its significance?

We will address the Minkowski variant of this question. We have
seen in \secref{subsec:Minkowski} that the mini-twistor space of
$\R^{2,1}$ is $T\RP^1.$ For $\R^{2,1}$ the coordinates $z$ and $w$
are real, and so are the coefficients $a_1, a_2,$ in
\eqref{eqn:oscul}. We define $\vx\in\R^3$ by comparison with the
Minkowski incidence relation \eqref{eqn:minkrel},
\be\label{eqn:vxoscM} -(x_2-x_0) = w_0 -a_1 z_0 + a_2 z_0^2, \quad
2x_1 = a_1-2a_2 z_0, \quad x_2 + x_0 = a_2. \ee The solution to
\eqref{eqn:vxoscM} defines a unique point in $\R^{2,1}$ for every
point on the real curve $\Curve\in T\RP^1.$ The collection of
these points form a curve $\Focal(\Curve)$ in $\R^{2,1}.$ What is
the physical significance of $\Focal(\Curve)$?

Take a particular amplitude with $n$ mini-twistors
$\twv_1,\dots,\twv_n,$ and let us consider a particular scattering
experiment to which this amplitude would contribute. Thus, we pick
$m<n$ twistors $\twv_1,\dots,\twv_m$ to describe incoming
particles, and assume that $\twv_{m+1},\dots,\twv_n$ describe
outgoing particles. We also assume that the number of negative
helicities $q$ is fixed and that $m$ and $\twv_1,\dots,\twv_m$ are
chosen so that there is a unique holomorphic curve, of the
corresponding degree $2(q-1),$ that passes through all the
mini-twistors $\twv_1,\dots,\twv_m.$ The amplitude will then be
nonzero only if $\twv_{m+1},\dots,\twv_n$ lie on that curve.

A mini-twistor $\twv=(z,w)$ describes an incoming planar
shockwave of the form \be\label{eqn:shockw} \phi_\twv(\vx)\sim
\delta(w - 2 z x^1 +[x^2-x^0] -[x^2+x^0]z^2), \ee that travels at
the speed of light. The scattering process is therefore a
collision of $m$ incoming shockwaves. What comes out?

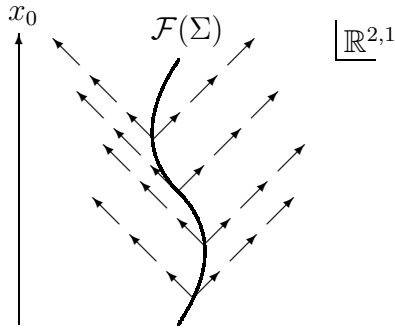
\begin{figure}[t]
\begin{picture}(400,120)
\thinlines
\put(133,113){$\R^{2,1}$}
\put(130,110){\line(0,1){15}}
\put(130,110){\line(1,0){15}}

\put(10,10){\vector(0,1){110}} 
\put(6,125){$x_0$}
\thicklines
\qbezier(70,10)(90,40)(70,60) %
\qbezier(70,60)(50,80)(70,110) %
\put(60,118){$\Focal(\Curve)$}
\thinlines
\multiput(76,20)(-14,14){3}{\vector(-1,1){10}} %
\multiput(76,20)(14,14){3}{\vector(1,1){10}} %

\multiput(80,40)(-14,14){3}{\vector(-1,1){10}} %
\multiput(80,40)(14,14){3}{\vector(1,1){10}} %

\multiput(70,60)(-14,14){3}{\vector(-1,1){10}} %
\multiput(70,60)(14,14){3}{\vector(1,1){10}} %

\multiput(61,80)(-14,14){3}{\vector(-1,1){10}} %
\multiput(61,80)(14,14){3}{\vector(1,1){10}} %

\end{picture}

\caption{The outgoing waves of the
scattering process can be described as
a physical disturbance that is emanating from
the ``focal curve'' $\Focal(\Curve).$}
\label{fig:Focal}
\end{figure}
For fixed mini-twistors $\twv_{m+2},\dots,\twv_n,$ the outgoing
wave-function of the $(m+1)^{st}$ particle is a linear combination
of shock-waves of the form \eqref{eqn:shockw}, and in general all
mini-twistors $\twv$ that lie on $\Curve$ can contribute. Suppose
a particular $\twv\equiv\twv_{m+1}\equiv (z_0, w_0)$ contributes
to the outgoing wave-function. Then nearby mini-twistors
$\twv+\delta\twv\equiv (z_0+\delta z, w_0 + \delta w)$ will also
contribute provided that they lie on $\Curve.$ But near $(z_0,
w_0)$ the curve $\Curve$ looks like the parabola
\eqref{eqn:oscul}. This implies that up to second order in
$\delta\twv$, all shock-waves $\phi_{\twv+\delta\twv}$ are nonzero
at the point $\vx$ defined by \eqref{eqn:vxoscM}. Thus, the
outgoing wave-function of the $(m+1)^{st}$ particle is a linear
combination of shock-waves, all of which pass through the ``focal
curve'' $\Focal(\Curve).$ The outgoing wave-function is therefore
a disturbance emanating from $\Focal(\Curve).$ (See
\figref{fig:Focal}.)

As an example, take the curve $\Curve$ given by
$$
w = 2 z^3.
$$
Then from \eqref{eqn:oscul} and \eqref{eqn:vxoscM} we
get the parametric equation for $\Focal(\Curve)$ in the form
$$
\vx\equiv (x_0, x_1, x_2)
= (z^3 + 3z,-3z^2,3z-z^3).
$$
Note that the tangent to this curve is null in $\R^{2,1}.$
It is easy to see that this is true for a generic curve $\Curve.$
In the special case of MHV amplitudes, the focal curve 
$\Focal(\Curve)$ degenerates to a point.

\subsection{Dimensional reduction in B-model twistor string theory}
\label{subsec:drstr}
In \cite{Witten:2003nn}, Witten proposed a reformulation of $D=4$
super Yang-Mills theory as a B-model on supertwistor space
$\CP^{3|4}.$ An alternative approach was presented in
\cite{Berkovits:2004hg}\cite{Berkovits:2004tx}. In this section we
will make a few observations about the twistor string theory of
$D=3$ super Yang-Mills theory.

The twistor string theory that describes $D=3$ super Yang-Mills
with $N=8$ supersymmetry is the topological B-model on the $D=3$
supertwistor space from \secref{subsec:supersp}. It can be
obtained from the $D=4$ twistor space $\CP^{3|4}$ by dimensional
reduction. For the B-model, dimensional reduction is implemented
by gauging one generator of $SL(4|4)$ {\footnote{We are now
considering the 4-d space to have signature $(+--+)$ and will
reduce along one of the time-like directions.}} that corresponds to translations in the $4^{th}$ direction.
(This is somewhat
reminiscent of the construction of topological $\sigma$-models in
\cite{Montano:1988dr}.) Let 
\be\label{eqn:coCP3} 
Z_1\equiv\lam_1,
Z_2\equiv\lam_2, 
Z_3\equiv\mu_{\dot{1}}, 
Z_4\equiv\mu_{\dot{2}},
\theta^1,\dots,\theta^4 
\ee 
be projective coordinates on
$\CP^{3|4}$ with the equivalence relation
\be\label{eqn:CPscale} 
(Z_1,\dots,Z_4,\theta^1,\dots,\theta^4)\sim
(\zeta Z_1,\dots, \zeta Z_4, \zeta\theta^1,\dots,\zeta \theta^4).
\ee 
We choose the basis so
that the translation generator $P_4$ in the $4^{th}$ direction
acts as
\be\label{eqn:P3ws} 
P_4:\quad \delta Z_1 = \delta Z_2 = 0,
\quad
\delta Z_3 = \epsilon Z_1, \quad
\delta Z_4 = -\epsilon Z_2, \quad
\delta\theta^1 = \cdots = \delta\theta^4 = 0.
\ee 
The transformation \eqref{eqn:P3ws} is a symmetry of
the B-model, since it preserves the complex structure of
$\CP^{3|4}$ and the holomorphic measure.

The resulting topological B-model 
on the $D=3$ mini-twistor space can
also be viewed as a limit of a discrete orbifold of the B-model on
$\CP^{3|4}.$ To construct this orbifold, pick a constant parameter
$r > 0$ and define the group $\Gamma_r\simeq\Z$ generated by
$\gamma_r = \exp(2\pi i r P_4).$ It acts on $\CP^{3|4}$ as 
\bear
\gamma_r\equiv e^{2\pi i r P_4}&:& Z_1 \mapsto Z_1,\quad
Z_2\mapsto Z_2,\quad Z_3\mapsto Z_3 +2\pi r Z_1, 
\nn\\ &&
Z_4\mapsto Z_4 -2\pi r Z_2, \quad 
\theta^A\mapsto\theta^A\quad
(A=1,\dots,4). 
\label{eqn:Orbp} 
\eear 
This map is compatible with
the equivalence relation \eqref{eqn:CPscale}, and it also
preserves the holomorphic superform 
$\epsilon^{IJKL}Z_I dZ_J\wedge
dZ_K\wedge dZ_L\wedge d\theta^1\wedge\cdots\wedge d\theta^4.$ 
The orbifold $\CP^{3|4}/\Gamma_r$ is therefore a Calabi-Yau
superspace. The fixed-point set of $\Gamma_r$ is the subset
$Z_1=Z_2=0$ which is the $\CP^{1|4}$ that is excised. The physical
interpretation of the B-model on the $\Gamma_r$-orbifold is, of
course, the twistor string worldsheet theory for $D=4$ super
Yang-Mills theory compactified on a circle of radius $r.$ In the
limit $r\rightarrow 0$ we recover the $D=3$ mini-twistor space.
\begin{table}[t]
\begin{tabular}{|cccc|}
\hline\hline
field  & Variable  & Worldsheet &  $SU(4)$ representation \\
\hline
 &&& \\
$\tZ, \tW, \btZ, \btW$ & commuting & scalars     & $\rep{1}$ \\
$\oeta^\bz, \oeta^\bw,$ &
  anti-commuting & scalars     & $\rep{1}$ \\
$\varth_z, \varth_w$ &
  anti-commuting & scalars     & $\rep{1}$ \\
$\rho^z, \rho^w$ &
  anti-commuting & 1-forms     & $\rep{1}$ \\
$\Theta^A$ & anti-commuting & scalars & $\rep{4}$ \\
$\oTheta_A$ & anti-commuting & scalars & $\rep{\overline{4}}$ \\
$\oeta_A$ & commuting & scalars & $\rep{\overline{4}}$ \\
$\varth_A$ & commuting & scalars & $\rep{\overline{4}}$ \\
$\rho^A$ & commuting & 1-forms & $\rep{4}$ \\
\hline\hline
\end{tabular}
\caption{The fields of the worldsheet B-model.
Note that the index $z$ on some of the
fields refers
to the target space coordinate, and should not be confused
with the worldsheet coordinate that is implicit.}
\label{table:fields}
\end{table}
The resulting worldsheet fields of the B-model
on mini-twistor space are listed in Table~\ref{table:fields}.
The BRST transformation laws are
\be\label{eqn:BRST}
\begin{split}
\delta \btZ = \oeta^\bz,\quad
\delta \btW = \oeta^\bw,\quad
\delta\oTheta_A =& \oeta_A,
\\
\delta\rho^z = d\tZ,\quad
\delta\rho^w = d\tW,\quad
\delta\rho^A =& d\Theta^A.
\end{split}
\ee
The transformation laws of the remaining fields are zero.

\subsection{Dimensional reduction in Berkovits's
twistor string theory}
\label{subsec:drstrBerk}
Dimensional reduction can be performed similarly in Berkovits's
model of the twistor string theory \cite{Berkovits:2004hg}. In
this model there are separate left- and right-moving worldsheet
fields. The $SL(4|4)$ {\footnote 
{We are still working in signature $(+--+)$.}} 
charged fields are $Z^i_L, Z^i_R, Y_{i L}, Y_{i R}$ 
($i=1,\dots,4$) and 
$\Theta^A_L, \Theta^A_R, \Upsilon_{A L}, \Upsilon_{A R}$ 
($A=1,\dots,4$), where $L$ ($R$) denotes a
left-moving (right-moving) field.
There is an additional $GL(1)$
gauge field $A$ under which 
$Z^i, \Theta^A$ have $+1$ charge and
$Y_i, \Upsilon_A$ have $-1$ charge.
Also, there are left-moving
and right-moving chiral current algebras that give rise to the
spacetime $SU(N)$ quantum numbers. The action is
\cite{Berkovits:2004hg} 
\bear 
S &=& \int
d^2\zws\Bigl[ 
Y_{L i}\nabla_\bzws Z_L^i 
+\Upsilon_{L A}\nabla_\bzws\Theta^A_L 
+Y_{i R}\nabla_\zws Z^i_R +\Upsilon_{A R}
\nabla_\zws\Theta^A_R \Bigr] + S_C, 
\label{eqn:BerkA} 
\eear
where $\nabla_\zws = \px{\zws}-A_\zws$ and $\nabla_\bzws =
\px{\bzws}-A_\bzws$ are the covariant derivatives, and $S_C$ is
the action of the chiral current algebras.

It is important to recall that even though \eqref{eqn:BerkA} has
cubic gauge interactions it is a conformally invariant
theory. In fact, the equations of motion are
\be\label{eqn:Berkeoms}
\begin{split}
\nabla_\bzws Z_L^i = 0,\qquad
\nabla_\bzws Y_{L i} = 0,\qquad
\nabla_\zws Z_R^i =0,\qquad
\nabla_\zws Y_{R i} = 0,
&\qquad i=1,\dots,4,
\\
\nabla_\bzws \Theta_L^A = 0,\qquad
\nabla_\bzws \Upsilon_{L A} = 0,\qquad
\nabla_\zws \Theta_R^A =0,\qquad
\nabla_\zws \Upsilon_{R A} = 0,
&\qquad A=1,\dots,4,
\end{split}
\ee and \be\label{eqn:BerkConstr} 
0 = \sum_i Z_L^i Y_{L i}+\sum_A\Upsilon_{L A}\Theta_L^A,
\qquad
0 = \sum_i Z_R^i Y_{R i}+\sum_A\Upsilon_{R A}\Theta_R^A,
\ee
and the gauge fields can be
solved in terms of the other fields. For $Z^1_L\neq 0$ we can set
$$
A_\bzws = \px{\bzws}Z_L^1/Z_L^1.
$$
Eliminating $A_\bzws$ from all the equations of motion
\eqref{eqn:Berkeoms}, we find that all left-moving gauge invariant
combinations (for example $Z_L^2/Z_L^1$) are holomorphic, and all
right-moving gauge invariant combinations are anti-holomorphic.
This holomorphicity condition, together with the analytic
constraints \eqref{eqn:BerkConstr}, completely captures all the
equations of motion. The theory is therefore conformally
invariant, since the gauge invariant fields are holomorphic.

As was explained in \cite{Berkovits:2004tx}, the path integral
splits into sectors that are labeled by an ``instanton
number'' $d.$ This number represents the total $U(1)\subset GL(1)$
flux. Following
\cite{Berkovits:2004tx}, we gauge-fix the Weyl transformations on
the worldsheet and the $GL(1)$ gauge field by setting the
worldsheet metric in such a way that a field of $GL(1)$ charge $q$
and conformal dimension $h$ will be equivalent to a gauge neutral
holomorphic field of conformal dimension $h+(d/2)q.$
This is a particularly convenient gauge fixing, because
as explained in \cite{Berkovits:2004tx}, all the left-moving
fields are holomorphic and all the right-moving field are
anti-holomorphic. We can therefore use conformal field theory OPEs
to calculate commutators.

(To avoid clutter, unless otherwise specified
we will from now on write formulas only for the 
left-movers and suppress the $L$ subscripts.)
Dimensional reduction proceeds in a similar fashion as for the
B-model. Instead of \eqref{eqn:Orbp}, we have 
\bear 
\gamma_r\equiv
e^{2\pi i r P_4}&:& Z^1\mapsto Z^1,\quad Z^2\mapsto Z^2,
\quad
Z^3\mapsto Z^3 +2\pi r Z^1,\quad Z^4\mapsto Z^4 -2\pi r Z^2, 
\nn\\
&& Y_1\mapsto Y_1 -2\pi r Y_3,\quad Y_2\mapsto Y_2 +2\pi r
Y_4,\quad Y_3\mapsto Y_3,\quad Y_4\mapsto Y_4, 
\nn\\ &&
\Theta^A\mapsto\Theta^A,\quad 
\Upsilon_A\mapsto\Upsilon_A\quad
 (A=1,\dots,4),
\label{eqn:OrbpB}
\eear
In the limit $r\rightarrow 0$ we gauge a continuous symmetry.
We can do this by
introducing an extra auxiliary gauge field $\Br_\zws, \Br_\bzws$
and modifying the covariant derivatives of $Z^3, Z^4$ to
\be\label{eqn:NabBr}
\nabla_\bzws Z^3 =
  \px{\bzws}Z^3 -A_\bzws Z^3 -\Br_\bzws Z^1,
\qquad
\nabla_\bzws Z^4 =
  \px{\bzws}Z^4 -A_\bzws Z^4 +\Br_\bzws Z^2,
\ee 
and similarly for the right-moving fields. Inserting these
covariant derivatives into the action \eqref{eqn:BerkA} and
integrating over $\Br_\bzws$, we get the constraint
\be\label{eqn:YZYZ} 
Y_3 Z^1 - Y_4 Z^2 = 0.
\ee 
Out of $Z^1,\dots,Z^4,\Theta^1,\dots,\Theta^4$ we can make the
following $\Br$-gauge invariant combinations: $W\defineas Z^3 Z^2
+ Z^4 Z^1$ with $GL(1)$ charge $+2$, and $Z^1, Z^2, \Theta^A$ with
$GL(1)$ charge $+1.$ The constraint \eqref{eqn:YZYZ} 
allows us to define
\be\label{eqn:rU} 
\rU\defineas \frac{Y_3}{Z^2} =\frac{Y_4}{Z^1}.
\ee 
This field $\rU$ has $GL(1)$ charge $-2.$ It is well-defined
provided that either $Z^1\neq 0$ or $Z^2\neq 0$, which is indeed
always the case. We also define 
\be\label{eqn:rY} 
\rY_1 = Y_1 -\rU Z^4, \qquad \rY_2 = Y_2 -\rU Z^3. 
\ee 
The action \eqref{eqn:BerkA}
together with the constraint \eqref{eqn:YZYZ} becomes, 
\bear 
S &=&
\int d^2\zws\Bigl[ 
\rY_1\nabla_\bzws Z^1 +\rY_2\nabla_\bzws Z^2 
+\rU\nabla_\bzws W +\Upsilon_A\nabla_\bzws\Theta^A 
\Bigr]
+\text{(right-movers)}
 + S_C, 
\label{eqn:BerkDR} 
\eear 
where $S_C$ is the action of the
current algebra and $\nabla_\bzws\defineas\px{\bzws}-A_\bzws$
is the covariant $GL(1)$ derivative.

The resulting theory is easy to interpret if we recall
that mini-twistor space is equivalent to $W\CP^{1,1,2}$
[see \eqref{eqn:WCP}].
The fields $Z^1, Z^2$ and $W$ correspond to
standard projective coordinates on $W\CP^{1,1,2}$
with weights $1,1$ and $2$, respectively.
The $GL(1)$ charges of those
fields correspond to these weights.
In addition, we have four anti-commuting fields
$\Theta^A$ ($A=1,\dots,4$) with $GL(1)$ charges $1.$
Each of the fields above has a canonical conjugate field.
$\rY_1, \rY_2$ are the conjugates of $Z^1, Z^2$,
and $\rU$ is the conjugate of $W.$
The conjugate of $\Theta^A$ is $\Upsilon_A.$
In addition, there are
also left and right moving current algebras.
As Berkovits and Motl \cite{Berkovits:2004tx} explained,
the worldsheet path integral splits into discrete sectors
labeled by the flux $d$ of the $GL(1)$ gauge field.
The left-moving fields are listed
in Table~\ref{table:Berkovits}.
The right-moving fields have similar quantum numbers.

The generators of the Poincar\'e algebra
\eqref{eqn:Pgen},\eqref{eqn:JgenSUSY} can
easily be expressed in terms of the fields $Z^1, Z^2, W$
and their conjugates $\rY_1, \rY_2, U.$
The worldsheet currents that correspond to the
translation generators
$P_{+}, P_{-}$ and $P_1$ are
\be\label{eqn:PgenCu}
\Pc_{+} = -\rU (Z^1)^2,\qquad
\Pc_1 = \rU Z^1 Z^2,\qquad
\Pc_{-} = \rU (Z^2)^2.
\ee
The transformation properties can be determined from
the OPEs
\be
\begin{split}
Z^i(\zws)\rY_j(0)\sim \frac{1}{\zws},\qquad
W(\zws)\rU(0)\sim &\frac{1}{\zws},
\\
Z^i(\zws)\rU(0)\sim
Z^i(\zws)Z^j(0)\sim \rY_i(\zws)\rY_j(0)\sim
W(\zws) W(0)\sim\rU(\zws)\rU(0)
\sim &\,\text{regular}.
\end{split}
\ee
\begin{table}[t]
\begin{tabular}{|cccc|}
\hline\hline
field  & statistics & worldsheet  &  $GL(1)$ charge \\
\hline
$Z^1, Z^2$ & commuting & scalars & $1$ \\
$W$ & commuting & scalar & $2$ \\
$\rY_1, \rY_2$ & commuting & vectors & $-1$ \\
$\rU$ & commuting & vector & $-2$ \\
$\Theta^A$ & anti-commuting& scalars & $1$ \\
$\Upsilon^A$ & anti-commuting& vectors & $-1$ \\
\hline\hline
\end{tabular}
\caption{The left-moving worldsheet fields of mini-twistor
string theory \`a la Berkovits.}
\label{table:Berkovits}
\end{table}
The supersymmetry generators \eqref{eqn:Qgen}
can also be expressed as currents.
The left-moving currents corresponding to the SUSY generators
are
\be\label{eqn:QgenCu}
\Qc_{A +} = Z^1\Upsilon_A,
\quad
\bQc^A_{+} = -i Z^1\Theta^A\rU,
\quad
\Qc_{A -} = Z^2\Upsilon_A,
\quad
\bQc^A_{-} = i Z^2\Theta^A\rU.
\ee

\subsection{$\Spin(7)$ R-symmetry}
\label{subsec:Spin7}
The R-symmetry group of our $D=3$ super Yang-Mills theory is
$\Spin(7).$
 It acts linearly on the $8$ supersymmetry generators,
which transform as the spinor representation $\rep{8}$ of
$\Spin(7).$ However, only an $SU(4)$ subgroup is manifest in
mini-twistor string theory. This is the subgroup that acts
linearly on $\Theta^A$ and can be identified with the $D=4$
R-symmetry, before dimensional reduction.\footnote{We are
now working in signature $(-+++)$ again. 
Had we started instead in signature $(-++-)$
and dimensionally reduced along a timelike direction,
we would have had to work with the noncompact R-symmetry group
$\Spin(4,3)$, which extends the $D=4$ group $\Spin(3,3)\sim SL(4).$}

It is not a big surprise that full $\Spin(7)$ symmetry
is not explicit.
Some symmetries are obscure in twistor string theory.
A nice example is parity, which is not at all manifest
in the B-model \cite{Witten:2004cp},
but was cleverly identified by
Berkovits and Motl \cite{Berkovits:2004tx}
in the open twistor string model.
In this subsection we will identify the full $\Spin(7)$
R-symmetry generators in Berkovits's model.

Let us first write down the commutation relations for the
$\Spin(7)$ R-symmetry generators. We take a basis for the $so(7)$
Lie algebra that consists of $su(4)$ generators, which we denote
by ${T^A}_B$, and 6 additional generators, which we denote by
$T^{AB} = -T^{BA}$ ($A,B=1,\dots, 4$). The commutation relations
are
\be\label{eqn:CommRels}
\begin{split}
[{T^A}_B, {T^C}_D] = \delta^A_D {T^C}_B -\delta^C_B {T^A}_D,
\qquad [{T^A}_B, T^{CD}] = \delta^D_B T^{AC}-\delta^C_B T^{AD},
\\
[T^{AB}, T^{CD}] = -\epsilon^{ABCE}{T^D}_E
+\epsilon^{ABDE}{T^C}_E +\epsilon^{CDAE}{T^B}_E
-\epsilon^{CDBE}{T^A}_E.
\end{split}
\ee
To see how the missing R-symmetry generators $T^{AB}$ operate,
let us look at a particular example of a $\Spin(7)$ multiplet of
fields. According to \cite{Berkovits:2004jj}, $D=4$ twistor string
theory contains, in addition to the super Yang-Mills theory, a
sector that describes conformal supergravity. Our example of a
$\Spin(7)$ multiplet of fields will comprise of the dimensional
reduction of some of these $D=4$ conformal supergravity fields. We
will take a multiplet of fields that transform in the irreducible
representation $\rep{35}$ (anti-symmetric 3-tensors) of
$\Spin(7).$ This irreducible representation decomposes under
$SU(4)\subset\Spin(7)$ as \be\label{eqn:rep35} \rep{35} = \rep{15}
+ \rep{10} + \rep{\overline{10}}. \ee Thus, we can construct our
multiplet by combining fields that transform in the three
irreducible representations on the right hand side of
\eqref{eqn:rep35}. 

The fields of $D=4$ conformal supergravity are
listed in Table~1 in \S4.2 of \cite{Berkovits:2004jj}. We can
obtain our multiplet by combining three irreducible $SU(4)_R$
representations from that list. In the notation of
\cite{Berkovits:2004jj}, we pick the fields $E_{AB}=E_{BA},$
$\overline{E}^{AB}=\overline{E}^{BA}$ and ${{V_\u}^A}_B.$ The
first is a $D=4$ space-time scalar in the $\rep{\overline{10}}$ of
$SU(4)_R$; the second is a $D=4$ space-time scalar in the
$\rep{10}$ of $SU(4)_R$, and the last one is a $D=4$ space-time
vector in the $\rep{15}$ adjoint representation of $SU(4)_R.$
After dimensional reduction to $D=3$, the scalars $E_{AB},
\overline{E}^{AB}$ and the $4^{th}$ component ${{V_4}^A}_B$ of
$V_\u$ form a $\Spin(7)$ multiplet in the irreducible
representation $\rep{35}.$

Now let us see how $\Spin(7)$ acts on $E_{AB}, \overline{E}^{AB}$
and ${{V_4}^A}_B.$ The action of the $SU(4)\subset\Spin(7)$
generators ${T^A}_B$ is obvious. The $\Spin(7)$ generators
$T^{AB}$ must also transform these states to each other. It is,
however, difficult to identify the $T^{AB}$ generators in the
B-model version of mini-twistor string theory. Part of the problem
is that there are no perturbative B-model string vertex operators
that correspond to the fields $E_{AB}.$ The vertex
operator for the dimensionally reduced 
$\overline{E}^{BA}$ and the vertex
operator for ${{V_4}^A}_B$ were constructed in
\cite{Berkovits:2004jj}, and we will recall them below, in
\eqref{eqn:VtxMass10} and \eqref{eqn:VtxMass15}. But the vertex
operators for $E_{AB}$ are absent in the perturbative
B-model. To understand this, recall that the states of a
perturbative B-model with target space $X$ correspond to the sheaf
cohomology $H^p(X,\wedge^q TX)$ where $TX$ is the holomorphic
tangent bundle. Nonperturbatively, it is conjectured
\cite{Nekrasov:2004js} that the B-model is S-dual to the A-model
and therefore also has states that correspond to 
Dolbeault cohomology $H^p(X,\Omega^q X)$ 
[where $\Omega^q X$ is the sheaf of holomorphic
$(q,0)$-forms]. The states $E_{AB}$ are analogous to
the latter. (See also the discussion after equation (2.12) of
\cite{Berkovits:2004jj}.) For recent developments in the
nonperturbative formulation of the topological string theory see
\cite{Aganagic:2003qj}\cite{Dijkgraaf:2004te}\cite{Aganagic:2004js}.

\subsection{$\Spin(7)$ in Berkovits's model}
\label{subsec:Spin7B}
It is easier, however,
to identify $T^{AB}$ in Berkovits's model.
As usual in a two-dimensional conformal field theory,
the symmetry generators correspond to holomorphic
and anti-holomorphic currents.
We will denote these currents by $\Jc^{AB}.$
Let us first identify
the worldsheet currents that correspond to the generators
${T^A}_B.$
They are easily constructed from the $SU(4)$ transformation
properties of $\Theta^A$ and $\Upsilon_A$, and we get
$$
{\Jc^A}_B\defineas\Upsilon_B\Theta^A
-\frac{1}{4}\delta^A_B\Upsilon_C\Theta^C.
$$
We claim that the worldsheet currents that correspond to $T^{AB}$
can be expressed, formally, as 
\be\label{eqn:TABrUrU}
\framebox{$\displaystyle{
\Jc^{AB}\defineas
\rU\Theta^A\Theta^B
+\tfrac{1}{2}\rU^{-1}\epsilon^{ABCD}\Upsilon_C\Upsilon_D.
}$}
\ee
We determined
the generators \eqref{eqn:TABrUrU} by looking for expressions with
total conformal dimension $1$, $GL(1)$ charge zero, and the
correct residues in their OPEs among themselves and with the
Poincar\'e currents \eqref{eqn:PgenCu} and  supersymmetry currents
\eqref{eqn:QgenCu}. These residues are determined by the known
commutation relations between the R-symmetry and super-Poincar\'e
generators. For example, using the residues of the simple poles in
the OPEs of the currents $\Jc^{AB}$ and ${\Jc^A}_B$, one can
verify the commutation relations \eqref{eqn:CommRels}.

The operator $\rU^{-1}$ that appears
in \eqref{eqn:TABrUrU} has to be defined carefully.
It is required to have the OPE
$$
\rU^{-1}(\zws)\rU(0) = 1 + O(\zws).
$$
This operator $\rU^{-1}$ can be handled by bosonization
as follows.
The $(\rU, W)$ system is very similar to the superconformal
ghosts $(\beta,\gamma)$ of superstring theory,
except that the conformal dimensions are shifted.
Thus, $(\rU, W)$ can be bosonized in much the same
way as the superconformal ghosts.
(See, e.g., \S10.4 of \cite{Polchinski:1998rq}.
Bosonization of the $(Y_I, Z^I)$ fields
has also been discussed in
\cite{Berkovits:2004tx}\cite{Polyakov:2005yv},
and see also \cite{Berkovits:1992bm}.)
Let $\phi$ be a chiral boson, and let $(\xi,\eta)$
be anti-commuting ghosts,
with OPEs
$$
\phi(\zws)\partial\phi(0)\sim \frac{1}{\zws},
\qquad
\eta(\zws)\xi(0)\sim\frac{1}{\zws}.
$$
(See \S10.4 of \cite{Polchinski:1998rq}.)
The bosonization formulas are then
\be\label{eqn:WUbos}
W \simeq -e^{-\phi}\partial\xi,\qquad
\rU\simeq e^{\phi}\eta.
\ee
We can then take
\be\label{eqn:defUinv}
\rU^{-1}\simeq e^{-\phi}\xi,
\ee
and this has all the properties required
of the inverse of $\rU.$

We can now use the R-symmetry operators \eqref{eqn:TABrUrU}
to complete partial $\Spin(7)$ multiplets
in mini-twistor string theory.
We will apply this procedure in \secref{subsec:ApplSpin7},
where we will again encounter our operators
$E_{AB}, \overline{E}^{AB}$ and ${{V_4}^A}_B$,
from \secref{subsec:Spin7}.
More details on bosonization can be found in \appref{app:Bosonization}.

\section{Mass terms}\label{sec:mass}
We come now to the main new point of our paper --- adding mass
terms. We can augment the $D=3$ super Yang-Mills Lagrangian with
mass terms for the scalars and fermions. The $D=3$ fermions are in
the spinor representation $\rep{8}$ of the R-symmetry group
$\Spin(7).$ Fermion mass terms are linear in the mass parameter
and correspond to operators of the form 
$M_{ab}\chi^a_\a \chi^{\a b}$, 
where $M_{ab}$ is the mass matrix, and we used the notation
of \eqref{eqn:Lagrangian}. These operators are in the symmetric
part of the tensor product representation $\rep{8}\otimes\rep{8}$
of $\Spin(7).$ (It decomposes into the irreducible representations
$\rep{1}+\rep{35}.$) The $D=3$ scalars are in the vector
representation $\rep{7}$ of $\Spin(7).$ Scalar mass terms are
quadratic in the mass parameter and correspond to operators in the
symmetric part of the tensor product representation
$\rep{7}\otimes\rep{7}.$ (It decomposes into irreducible
representations $\rep{1} + \rep{27}.$) The $D=3$
super-mini-twistor space formalism only exhibits a manifest
$SU(4)\subset \Spin(7)$ subgroup of the R-symmetry. This is the
subgroup inherited from $D=4$, and the anti-commuting variables
$\theta^A$ ($A=1,\dots, 4$) are in the $\rep{4}$ of this $SU(4),$
while the $\partial/\partial\theta^A$ derivatives are in the
$\rep{\overline{4}}$ conjugate representation. The operators that
correspond to $D=3$ fermion mass terms split into the following
$SU(4)$ irreducible representations 
\be\label{eqn:massreps}
(\rep{8}\otimes \rep{8})_S = (\rep{4}\otimes \rep{4})_S
+(\rep{\overline{4}}\otimes \rep{\overline{4}})_S
+\rep{4}\otimes\rep{\overline{4}} = \rep{10}+\rep{\overline{10}}
+[\rep{1}+\rep{15}]. 
\ee 
Here $(\cdots)_S$ denotes the symmetric
part of a tensor product, and $\rep{15}$ is the adjoint
representation of $SU(4).$ We will now study the deformations of
mini-twistor string theory that correspond to these mass
terms. Generating the mass terms in the adjoint representation
$\rep{15}$ is simpler, and we therefore start with those.

\subsection{Dimensional reduction with twists}
\label{subsec:drtwists}
A convenient way to achieve a massive theory with mass terms in
the $\rep{15}$ representation is to modify the procedure of
dimensional reduction by adding an R-symmetry twist. In
\secref{subsec:drstr} we obtained massless $D=3$ super Yang-Mills
theory from the massless $D=4$ theory by gauging a translation
generator $P_4$ (in the worldsheet theory). To obtain massive
$D=3$ super Yang-Mills theory from the massless $D=4$ theory, we
gauge a linear combination of translation and R-symmetry, $P_4 -
\tr{(M R)}$, where $R$ is the $SU(4)$ R-symmetry charge (realized
as a $4\times 4$ hermitian traceless matrix) and $M$ is a constant
$4\times 4$ hermitian traceless mass-matrix that will end up as
the mass-matrix of the $D=3$ fermions. The logic behind this
reduction is as follows. Let $\psi_\a^A$ ($A=1,\dots, 4$) be a
$D=4$ Weyl fermion field, which transforms in the fundamental
representation $\rep{4}$ of the R-symmetry group $SU(4).$ Gauging
the combination $P_4-\tr{(M R)}$ on the worldsheet corresponds to
setting
$$
i\px{3}\psi_\a^A = {M^A}_B\psi_\a^B
$$
in physical space. The term
$\bpsi^\dta_A\sigma^3_{\a\dta}\px{3}\psi^{A\a}$ in the $D=4$
Lagrangian will become a mass term in the $D=3$ dimensionally
reduced Lagrangian. The fermion mass matrix will be $M$, and the
scalar mass matrix squared will be the anti-symmetric part of
$M\otimes M$, which, in the $\rep{6}$ representation of $SU(4)$,
is the $6\times 6$ matrix representative of $M.$

In twistor space, we augment \eqref{eqn:P3ws} to get 
\bear
P_4-\tr{(M R)}&:&\delta Z_1 = \delta Z_2 = 0,\quad 
\delta Z_3 =
\epsilon Z_1, \quad 
\delta Z_4 = -\epsilon Z_2, 
\nn\\ &&
\delta\theta^A = \epsilon {M^A}_B\theta^B\quad 
(A,B =1,\dots, 4).
\label{eqn:P3wstw} 
\eear 
Gauging this translation symmetry can be
done along the lines that led to \eqref{eqn:strzw}. We get a
similar supermanifold that can be covered by two patches $U_1,
U_2$ with transition functions that are a slight modification of
\eqref{eqn:strzw}, 
\be\label{eqn:strzwtw}
\framebox{$\displaystyle{
z' = \frac{1}{z}, 
\qquad
w' = -\frac{w}{z^2},
\qquad 
\theta' = \frac{1}{z}e^{i\frac{w}{z}M}\theta
}$}
\ee
where $\theta$ represents a 4-component vector, and
$\exp[i(w/z)M]$ is a $4\times 4$ matrix [in the complexification
of $SU(4)$].
This is one of our main results.
Equation \eqref{eqn:strzwtw} describes 
the target space of a twistor string theory
that is, by conjecture, dual to a mass-deformed $D=3$ 
super Yang-Mills theory with a mass-matrix $M.$

Similarly to the discussion at the end of \secref{subsec:drstr},
we can think of \eqref{eqn:P3wstw} as the $r\rightarrow 0$ limit
of the discrete orbifold of $\CP^{3|4}$ by the group
$\Gamma_{r,M}\simeq\Z$ generated by 
\bear 
\gamma_{r,M}\equiv
e^{2\pi i r (P_4-\tr{(M R)})}&:& 
Z_1 \mapsto Z_1,\quad Z_2\mapsto
Z_2,\quad Z_3\mapsto Z_3 +2\pi r Z_1, 
\nn\\ && 
Z_4\mapsto Z_4 -2\pi r Z_2, 
\quad 
\theta\mapsto e^{-2\pi i r M}\theta,
\label{eqn:Orbptw} 
\eear 
where $\theta$ again denotes a 4-component vector. 
Physically, \eqref{eqn:Orbptw} corresponds to
a circle compactification of $D=4$ super Yang-Mills theory with an
R-symmetry twist. That is, the scalars and fermions have
nonperiodic boundary conditions on the circle. Such
compactifications have been studied, for example, in
\cite{Cheung:1998te}. R-symmetry orbifolds of the $D=4$ theory
(without the $P_4$ generator) have been recently studied in
\cite{Park:2004bw}.
For recent developments in worldsheet orbifolds see
\cite{Halpern}.

\subsection{Mini-twistor string theory mass operators}
\label{subsec:strop}
For any value of the mass matrix $M$, the manifold described by
\eqref{eqn:strzwtw} is, presumably, the target space of the
worldsheet (mini-)twistor string theory that describes the massive
deformation studied in \secref{subsec:drtwists}. In particular,
for an infinitesimally small $M$, \eqref{eqn:strzwtw} describes a
small deformation of the complex structure of the 
super mini-twistor
space $\Tw_3$ (defined as $T\CP^1$ with four extra anti-commuting
coordinates fibered over it).

In general, a small deformation of the complex structure
corresponds to an operator in the worldsheet theory, which is the
topological B-model with target space $\Tw_3.$ For example, an
interesting case was studied in \cite{Kulaxizi:2004pa} where
marginal deformations of the $D=4$ theory
\cite{Leigh:1995ep}\cite{Aharony:2002hx}\cite{Dorey:2004xm} were
associated with certain B-model closed string operators. The
operators of the topological B-model on a manifold $X$ correspond
to the sheaf cohomology classes $H^p(X,\wedge^m T^{(1,0)}X)$
\cite{Witten:1991zz}. Locally on $X$, with a choice of complex
coordinates $z_i$ ($i=1,\dots,\dim_\C X$) and their complex
conjugates $\bz_\oi,$ these sheaf cohomology classes can be
realized as tensors ${V_{\oi_1\cdots\oi_p}}^{j_1\cdots j_m}$ that
are anti-symmetric in the $\oi_s$ and $j_t$ indices. Deformation
by the operators with $p=1$ and $m=1$ corresponds to a complex
structure deformation. (The target space action that describes
these deformation is the Kodaira-Spencer action
\cite{Bershadsky:1993cx}.) This correspondence can be extended,
presumably, to supermanifolds and we get the cohomology classes
$H^{p|q}(X,\wedge^{m|n} T^{(1,0)}X).$ In \cite{Kulaxizi:2004pa},
B-model operators that correspond to elements of
$H^{0|2}(X,\wedge^{0|2} T^{(1,0)}X)$, with $X=\CP^{3|4}$, were
identified with cubic deformations of the $D=4$, $N=4$ super
Yang-Mills superpotential.
Furthermore,
general deformations of holomorphic vector bundles over 
weighted projective superspaces were recently 
studied in \cite{Wolf:2004hp}. 
The particular complex structure deformation that we need
for the mass term is a special case.

Let us now study the $D=3$ super-cohomology classes. 
First, let us
take the mass deformation in the limit of infinitesimal mass
matrix $M$. Equation \eqref{eqn:strzwtw} teaches us that the
infinitesimal change in coordinate transition functions on the
intersection $U_{12}\equiv U_1\cap U_2$ of the two patches
$U_1\equiv \{z<\infty\}$ and $U_2\equiv\{z'<\infty\}$ is
\be\label{eqn:delzwth}
\delta z' = 0,\quad 
\delta w' = 0,\quad
\delta\theta' = i\frac{w}{z^2}M\theta
= -i\frac{w'}{z'}M\theta' +O(M)^2. 
\ee 
Here, we think of $U_1, U_2$ as
two patches of the undeformed (${M^A}_B=0$) $D=3$ twistor
supermanifold $\Tw_3.$ Equation \eqref{eqn:delzwth} defines a
holomorphic vector field on $U_{12}$ whose components are given,
in local coordinates, by 
\be\label{eqn:vecwzM} 
\delta z'\ppx{z'}+\delta w'\ppx{w'}+{\delta\theta'}^T\ppx{\theta'}
=-i\frac{w'}{z'}{\theta'}^T M^T\ppx{\theta'} 
=i\frac{w}{z}\theta^T M^T\ppx{\theta}. 
\ee
This defines an element of the sheaf
cohomology $H^{1|0}(X, \wedge^{0|1}T^{(1,0)}X)$ for $X=\Tw_3.$ In
fact, $H^{1|0}(X, \wedge^{0|1}T^{(1,0)}X)$ is the space of
holomorphic vector fields of the form
$$
{V^A}_B(z, w, \theta_1,\dots,\theta_4) \theta_A\ppx{\theta_B}
$$
that are defined for $0<z<\infty$ up to the equivalence relation
\be\label{eqn:scexact} {V^A}_B \sim {V^A}_B + {(\Vpch{1})^A}_B -
{(\Vpch{2})^A}_B, \ee where $\Vpch{1}$ is a  holomorphic vector
field of a similar form that is defined for all $z<\infty$
(including $z=0$) and $\Vpch{2}$ is similarly defined for all
$0<z$ (including $z=\infty$). [Recall that in general, given a
cover of $X$ by contractible open patches $X=\cup_\a U_\a$, an
element of $H^p(X, T^{(1,0)}X)$ can be described by a collection
of local vector fields, one vector field for each nontrivial
intersection of $(p+1)$ patches $U_{\a_1}\cap\cdots\cap
U_{\a_{p+1}}$, such that a certain linear relation---the cocycle
condition---holds on intersections of $(p+2)$ patches. In our
case, there are only two patches, $U_1$ and $U_2$, and a vector
field that is defined on the intersection $U_1\cap U_2$ defines an
element of $H^1(X, T^{(1,0)}X).$]

The vector field corresponding to \eqref{eqn:vecwzM} is
$$
{V^A}_B = \frac{w}{z}{M^A}_B = -\frac{w'}{z'}{M^A}_B.
$$
This vector field therefore has poles both at $z=0$
and $z'=0.$

More generally, the field 
\be\label{eqn:fieldmn}
\frac{w^r}{z^s}{M^A}_B = \frac{(-w')^r}{{z'}^{2r-s}}{M^A}_B 
\ee
has poles both at $z=0$ and $z'=0$ only if $0<s<2r.$ For other
values of $s$, the vector field can be extended to either $z=0$ or
$z=\infty$, and therefore is trivial in cohomology. Let us check
for which values of $s,r$ this cohomology class is invariant under
translations of the physical $\R^3.$ The generators of
translations were written down in \eqref{eqn:Pgen}. Since the
field \eqref{eqn:fieldmn} is holomorphic, only the holomorphic
parts of $P_1, P_2, P_3$ in \eqref{eqn:Pgen} are relevant. We
therefore need to check that $z^k\partial/\partial w$, for
$k=0,1,2$, annihilates \eqref{eqn:fieldmn} in cohomology. By the
discussion above, this is equivalent to requiring that
$0<s-k<2(r-1)$ is not satisfied for any of the values $k=0,1,2.$
It is not hard to check that the only solution is
$s=r=1$, which is the field from \eqref{eqn:vecwzM}.
Similarly, using equations \eqref{eqn:Jgen}
for the rotation generators, it can  be checked that
\eqref{eqn:vecwzM} is rotationally invariant.
Thus, the mass deformation \eqref{eqn:vecwzM} corresponds
to the unique (up to multiplication by a constant)
translationally invariant element
of $H^{1|0}(X, \wedge^{0|1}T^{(1,0)}X).$

Alternatively, the cohomology $H^{1|0}(X, \wedge^{0|1}T^{(1,0)}X)$
can be represented by a {\it global} differential 1-form with
coefficients in $\wedge^{0|1}T^{(1,0)}X.$ The way to convert a
sheaf cohomology representative to a global form is to write
$$
{V^A}_B = {(\Vpch{1})^A}_B - {(\Vpch{2})^A}_B,
$$
where $\Vpch{1}$ and $\Vpch{2}$
are well-defined on $U_1$ and $U_2$ respectively,
but, unlike \eqref{eqn:scexact}, $\Vpch{1}$ and $\Vpch{2}$
are not required to be holomorphic
(otherwise ${V^A}_B$ would be trivial in cohomology).
Then, the 1-form $\bpar {(\Vpch{1})^A}_B =\bpar {(\Vpch{2})^A}_B$
(where $\bpar$ is the Dolbeault differential operator)
is globally defined and represents the cohomology class.
In our case, we can pick an arbitrary differentiable
function $f(\abs{z}^2)$ such that $f(0)=1$ and
$f(\infty)=0$ and set
$$
{(\Vpch{1})^A}_B = [1-f(\abs{z}^2)]\frac{w}{z}{M^A}_B,
\qquad
{(\Vpch{2})^A}_B = -f(\abs{z}^2)\frac{w}{z}{M^A}_B.
$$
Then, \be\label{eqn:MassForm} \bpar {(\Vpch{1})^A}_B = \bpar
{(\Vpch{2})^A}_B = -f'(\abs{z}^2){M^A}_B w d\bz \ee is a global
1-form that represents the cohomology class. The corresponding
B-model operator is easily constructed from this form. It is
\be\label{eqn:VtxMass15} V_{(\rep{15})} = -f'(\tZ\btZ) \oeta^\bz
\varth_A {M^A}_B\tW \Theta^B. \ee (The worldsheet fields of the
B-model are listed in Table~\ref{table:fields}.) Can the operator
corresponding to \eqref{eqn:MassForm} be related by dimensional
reduction to a $D=4$ operator? We will work in the coordinates
\eqref{eqn:coCP3} for $\CP^{3|4}$, and define the two patches
$$
U_1'\defineas \{Z_1\neq 0\},
\qquad
U_2'\defineas \{Z_2\neq 0\}.
$$
Together, $U_1'$ and $U_2'$ cover the $D=4$ twistor space
(given by the condition that $Z_1$ and $Z_2$ do not
vanish simultaneously).
Using \eqref{eqn:wlammu}, we calculate
$$
\frac{w}{z} =
  \frac{\mu^{\dot{1}}}{\lam_1}
 -\frac{\mu^{\dot{2}}}{\lam_2}
= \frac{Z_3}{Z_1}-\frac{Z_4}{Z_2}.
$$
The element \eqref{eqn:vecwzM} can
therefore be dimensionally ``oxidized'' (the inverse of
``dimensionally reduced'') to the following (super-)vector field
on $U_1'\cap U_2'$, 
\be\label{eqn:VmD4}
V_m^{(D=4)} =
\Bigl(\frac{Z_3}{Z_1} -\frac{Z_4}{Z_2}\Bigr) \theta^A
{M_A}^B\ppx{\theta^B} = 
\Bigl(\frac{Z_3}{Z_1}\theta^A {M_A}^B\ppx{\theta^B}\Bigr) 
-\Bigl(\frac{Z_4}{Z_2}\theta^A {M_A}^B\ppx{\theta^B}\Bigr). 
\ee
In the last equality, we have
written $V_m$ as the difference of two terms, the first of which
can be extended to a holomorphic function on the entire $U_1'$
patch, and the second can be extended to a holomorphic function on
the entire $U_2'$ patch. $V_m$ is therefore exact in sheaf
cohomology and corresponds to the zero operator. This was to be
expected, since the $D=3$ mass terms that we consider in this
subsection, which are in $\rep{15}$ of $SU(4)$, lift to terms of
the form ${M^A}_B A_3 \sigma^3_{\a\dta}\psi_A^\a\bpsi^{B\dta} +
O(M)^2$ in $D=4$, where $A_3$ is the $4^{th}$ component of the
gauge field (counting from $0$). These terms can be gauged away by
an $x_4$-dependent gauge transformation.

\subsection{The $\rep{10}$ and $\rep{\overline{10}}$ mass terms.}
\label{subsec:massterms10}
We still have to analyze the remaining mass terms from
\eqref{eqn:massreps}.
These are the terms
in the representations $\rep{10}$, their complex conjugates
in $\rep{\overline{10}},$ and the representation $\rep{1}.$
We will discuss the singlet $\rep{1}$ separately,
and concentrate on the $\rep{10}$ and $\rep{\overline{10}}$ first.

First, let us ask whether these $D=3$ mass terms can be derived by
dimensional reduction of $D=4$ mass terms. The answer is yes!
Looking back at \eqref{eqn:massreps}, we see that, unlike the mass
terms in $\rep{15}$ that were discussed in
\secref{subsec:drtwists}, the mass terms in $\rep{10}$ and
$\rep{\overline{10}}$ can indeed be gotten by dimensional
reduction of $D=4$ mass terms. This is because $D=4$ mass terms
are of the form $\epsilon^{\a\b}\psi^A_\a\psi^B_\b$ and
$\epsilon^{\dta\dtb}\bpsi_{\dta A}\bpsi_{\dtb B}$ but cannot
involve $\psi^A_\a\bpsi_{\dta B}$, since spinors of different
chirality in $D=4$ cannot be contracted in a Lorentz invariant
manner. After dimensional reduction to $D=3$, however, both
spinors become the same representation. Thus, the $D=3$ mass terms
in $\rep{10}$ and $\rep{\overline{10}}$ can be derived from
dimensionally reduced $D=4$ mass terms, but the mass terms in
$\rep{15}$ have to be derived by twisting as in
\secref{subsec:drtwists}.

So what are the $D=4$ operators that can give such mass terms upon
dimensional reduction? These are closed string vertex operators of
the B-model twistor string theory of $D=4$ super Yang-Mills. It
has been argued in
\cite{Witten:2003nn}\cite{Berkovits:2004jj}\cite{Ahn:2004yu} that
they correspond to fields of $D=4$ superconformal supergravity. In
particular, Table 1 in \S4.2 of \cite{Berkovits:2004jj} lists the
physical states of that theory. And indeed, among these fields we
find two that have the right quantum numbers. In the notation of
\cite{Berkovits:2004jj}, these are $E_{(AB)}$ ($A,B=1,\dots 4$ are
$SU(4)$ indices) in $\rep{\overline{10}},$ and
$\overline{E}^{(AB)}$ in $\rep{10}.$ The conformal dimensions of
these fields can be read off from their $U(1)_R$ charge, which is
$Q=-2$ (and is also listed in Table 1 of \cite{Berkovits:2004jj}).
Their conformal dimensions are $\Delta = D+(3/2)Q = 1$, and this
is consistent with a fermionic mass term.

Now let us write down the vertex operators for the
dimensionally reduced $D=3$ mass terms in the
$\rep{10}$ representation.
The $\overline{E}^{AB}$ operators correspond to
linear combinations of vector fields
of the form
\be\label{eqn:EdABwz}
\frac{w^r}{z^s}M^{AB}\epsilon_{ACDE}
\theta^C\theta^D\theta^E\ppx{\theta^B},
\ee
defined
on the intersection $U_1\cap U_2$ (i.e., $0<\abs{z}<\infty$),
where $r,s$ are integers.
By the same arguments that follow \eqref{eqn:fieldmn}
we conclude that only $r=0$ and $s=1$ describes a nonzero
Poincar\'e invariant operator. Converting from sheaf
to Dolbeault $\bpar$-cohomology, we find the operator
\be\label{eqn:VtxMass10}
V_{(\rep{10})}=
-\tfrac{1}{6}f'(\tZ\btZ)\oeta^\bz
M^{AB}\epsilon_{ACDE}
\Theta^C\Theta^D\Theta^E\varth_B.
\ee
Here, $f(\abs{z}^2)$ is, as in \eqref{eqn:MassForm}, an
arbitrary function that satisfies $f(0)=1$ and $f(\infty)=0.$

\subsection{Chiral $D=4$ mass terms}
\label{subsec:D4Mass}
Let us make a brief digression to
check that $\overline{E}^{AB}$ 
indeed corresponds to a mass term in $D=4.$
We hope to present a more complete analysis elsewhere
\cite{P}.

The mass-deformed $N=4$ super Yang-Mills theory that, we claim,
corresponds to turning on  an $\overline{E}^{AB}$ 
VEV is nonstandard. It is the CPT violating theory 
that one obtains by giving a mass term
only to the positive helicity spinors of $D=4, N=4$ super
Yang-Mills theory. This is a perturbation of the form
$M^{AB}\psi_{\a A}\psi^\a_B$, where $\psi_{\a A}$ denote, 
as before, the $D=4$ fermions in the 
$\overline{\rep{4}}$ of $SU(4).$

In momentum space, the free massive Dirac equation now reads
\be\label{eqn:fMD}
p_{\a\dta}\psi^{\a A} = M^{AB}\bpsi_{\dta B},
\qquad
p_{\a\dta}\bpsi^\dta_A = 0.
\ee
Thus, the 4-momentum $p_{\a\dta}$ is still lightlike,
and we can decompose it as in the massless case,
$$
p_{\a\dta}=\lam_\a\tlam_\dta.
$$
The general solution to \eqref{eqn:fMD} is of the form
\be\label{eqn:solfMD} 
\bpsi_{\dta A} = \tlam_\dta\wvrho_A, 
\qquad
\psi_\a^A = \lam_\a\varrho^A +M^{AB}\eta_\a\wvrho_B, 
\ee 
where
$\wvrho_A(\lam,\tlam)$ and $\varrho^A(\lam,\tlam)$ are arbitrary
functions, and $\eta_\a$ is any ($\lam$-dependent) solution to the
linear equation \be\label{eqn:etalam} \eta_\a\lam^\a = 1. \ee
There is, of course, a family of solutions to this equation. Given
a solution $\eta_\a$ and an arbitrary function
$\zeta(\lam,\tlam)$, the following is also a solution to
\eqref{eqn:etalam}: \be\label{eqn:etapr} \eta'_\a\defineas
\eta_\a+\lam_\a\zeta. \ee If we choose $\eta'_\a$ instead of
$\eta_\a$ as the solution to \eqref{eqn:etalam}, we can preserve
the physical wavefunctions \eqref{eqn:solfMD} by setting
\be\label{eqn:wvrzet} \wvrho_A'\defineas\wvrho_A,\qquad
\varrho^{\prime A}\defineas\varrho^A-M^{AB}\zeta\wvrho_B, \ee so
that
$$
\bpsi_{\dta A} = \tlam_\dta\wvrho'_A,
\qquad
\psi_\a^A = \lam_\a\varrho^{\prime A}
+M^{AB}\wvrho'_B\eta'_\a.
$$
An example of a solution
to \eqref{eqn:etalam}
is given by $\eta_1=1/\lam^1$ and $\eta_2=0$.
It is well-defined
on the patch of $\lam$-space where $\lam^1\neq 0.$
Similarly, on the patch where $\lam^2\neq 0$
we can take $\eta'_2=1/\lam^2$ and $\eta'_1=0$,
so that on the intersection of the two patches,
where both components of $\lam$ are nonzero,
we have $\eta'_\a = \eta_\a -\epsilon_{\a\b}\lam^\b/(\lam^1\lam^2)$.
If we choose to work with $\eta$ on the patch
$\lam^1\neq 0$ and with $\eta'$ on the patch $\lam^2\neq 0$,
we need the transition relations
$$
\varrho^{\prime A} =
   \varrho^A +\frac{1}{\lam^1\lam^2}M^{AB}\wvrho_B,
\qquad
\wvrho_A'=\wvrho_A.
$$
External fermions in scattering amplitudes are described by
wave-functions of the form \eqref{eqn:solfMD}. These
wave-functions can be twistor-transformed as usual,
\be\label{eqn:twhwv} 
\hvarrho^A(\lam,\mu) \defineas \int d^2\tlam
e^{i\mu_\dta\tlam^\dta} \varrho^A(\lam,\tlam), 
\qquad
\hwvrho_A(\lam,\mu) \defineas \int d^2\tlam
e^{i\mu_\dta\tlam^\dta} \wvrho_A(\lam,\tlam). 
\ee 
The freedom
\eqref{eqn:wvrzet} extends to the twistor transforms
\be\label{eqn:hwvrzet}
\hwvrho_A'\defineas\hwvrho_A,
\qquad
\hvarrho^{\prime A}\defineas
\hvarrho^A-M^{AB}\hzeta\cdot\hwvrho_B,
\ee 
where $\hzeta\cdot\hwvrho_A$ denotes the convolution of
$\hwvrho_A$ with the twistor transform of $\zeta$, and the
convolution is taken with respect to $\mu_\dta.$ Note, however,
that the equivalence \eqref{eqn:wvrzet} can be generalized to
$\zeta$'s that are operator functions of $\lam_\a,$ $\tlam_\dta$
and $\partial/\partial\tlam_\dta.$ For the special case that
$\zeta$ is of the form
$\zeta(\lam_\a,i\partial/\partial\tlam_\dta)$ the convolution
becomes an ordinary product, and \eqref{eqn:hwvrzet} becomes
\be\label{eqn:hwvrz} \hwvrho_A'\defineas\hwvrho_A,\qquad
\hvarrho^{\prime A}\defineas\hvarrho^A-M^{AB}\zeta\hwvrho_B, \ee
where $\zeta(\lam,\mu)$ is a function on twistor space.

Following similar steps as in the appendix of
\cite{Witten:2003nn}, we take the twistor transforms
\eqref{eqn:twhwv}, plug them into \eqref{eqn:solfMD}, and
integrate over $\lam$ and $\tlam$ to convert from momentum-space
back to coordinate space. 
We perform the $\lam_\a$-integrals by gauge-fixing
$\lam^1=1$ and integrating $z\equiv\lam^2$ over a path $C$ around
the origin. 
\bear 
\bpsi_{\dta A}(x) &=& \frac{1}{2\pi}\oint_C
d\lam^2 \int d^2\tlam \int d^2\mu e^{i
x_{\a\dta}\lam^\a\tlam^\dta-i\mu_\dta\tlam^\dta}
\tlam_\dta\hwvrho_A(\lam,\mu) = \frac{1}{2\pi i}\oint_C dz
\pypx{\hwvrho_A}{\mu_\dta} \bigr\rvert_{(\lam^\a,
x_{\a\dta}\lam^\a)},
\nn\\
\psi_\a^A(x) &=& \frac{1}{2\pi}\oint_C d\lam^2 \int d^2\tlam \int
d^2\mu e^{i x_{\a\dta}\lam^\a\tlam^\dta-i\mu_\dta\tlam^\dta}
\bigl\lbrack \lam_\a\hvarrho^A(\lam,\mu)
+M^{AB}\eta_\a(\lam)\hwvrho_B(\lam,\mu) \bigr\rbrack
\nn\\
&=& \frac{1}{2\pi}\oint_C dz \bigl\lbrack
\lam_\a\hvarrho^A(\lam^\a, x_{\a\dta}\lam^\a)
+M^{AB}\eta_\a(\lam)\hwvrho_B(\lam^\a, x_{\a\dta}\lam^\a)
\bigr\rbrack, 
\label{eqn:psix} 
\eear 
where
$$
(\lam^1,\lam^2)\equiv (1, z),
\qquad
(\eta_1, \eta_2)\equiv (1, 0).
$$
Now let us
analyze the coupling of the closed string B-model mode
$\overline{E}^{AB}$ to the open string modes.
In general, the coupling of closed string B-model modes to
open string modes has been extensively studied.
(See
\cite{Hofman:2000ce}\cite{Lazaroiu:2000rk}\cite{Hofman:2002cw}
and \cite{Kulaxizi:2004pa} for example.)
Turning on a VEV 
for $\overline{E}^{AB}$ corresponds to a
perturbation of the complex structure of the
super-manifold target space $\CP^{3|4}.$
We will denote by
$X^i$ ($i=1,\dots,4$) and $\Psi^A$ ($A=1,\dots,4$)
the homogeneous coordinates on $\CP^{3|4}.$
(In \cite{Witten:2003nn}, $X^i$ is denoted
by $Z^I$, but we use the symbols $Z^1, Z^2$
to denote some of the projective coordinates of $W\CP^{1,1|2}.$)
We will cover
the $D=4$ twistor space $\CP^{3|4}\setminus\CP^{1|4}$
with the two patches
$$
U_1 \defineas \{X^1\neq 0\},
\quad
\text{and}
\quad
U_2\defineas\{X^2\neq 0\}.
$$
On the patch $U_1$ where $X^1\neq 0$,
we can rescale the projective coordinates and set $X^1=1.$
Then $X^2, X^3, X^4$ and the $\Psi^A$ are
the independent coordinates.
The good coordinates on the patch $U_2$,
where $X^2\neq 0$, are then
\be\label{eqn:U2vars}
\frac{1}{X^2}, \frac{X^3}{X^2}, \frac{X^4}{X^2},
\frac{\Psi^1}{X^2},\dots,\frac{\Psi^4}{X^2}.
\ee
On the patch $X^1\neq 0$, 
the VEV $\langle\overline{E}^{AB}\rangle=M^{AB}$
corresponds to the local holomorphic vector field
\be\label{eqn:InvEAB}
\frac{1}{6X^2}M^{AB}\epsilon_{ACDE}
\Psi^C\Psi^D\Psi^E\ppx{\Psi^B},
\ee
[see \eqref{eqn:EdABwz} with $r=0$ and $s=1$].
This vector field, in turn,
corresponds to a (super-)complex structure
deformation of the B-model target space $\CP^{3|4}.$

Formula (4.16) of \cite{Witten:2003nn} gives us the
component expansion of the B-model super 1-form field
as follows,
\bear
\cA(X,\bX,\Psi) &=&
d\bX^\oi\Bigl(
A_\oi+\Psi^A\chi_{\oi A}
+\tfrac{1}{2}\Psi^A\Psi^B\phi_{\oi AB}
\nn\\ &&
+\tfrac{1}{6}\epsilon_{ABCD}\Psi^A\Psi^B\Psi^C\wchi_\oi^D
+\tfrac{1}{24}\epsilon_{ABCD}\Psi^A\Psi^B\Psi^C\Psi^DG_\oi
\Bigr)
\label{eqn:AXXPsi}
\eear
where $A_\oi, \chi_{\oi A}, \phi_{\oi AB}, \wchi_\oi^D$
and $G_\oi$ are functions of $X$ and $\bX.$
The classical action
is given by equation (4.18) of \cite{Witten:2003nn},
\be\label{eqn:I}
I = \frac{1}{2}\int\Omega\wedge
\bigl(\cA\wedge\bpar\cA+\tfrac{2}{3}\cA\wedge\cA\wedge\cA\bigr)
\ee
The equations of motion of this holomorphic Chern-Simons theory,
to first order, state that $\cA$ is an element
of $\bpar$-cohomology
$$
\bpar\cA = 0,\qquad
\cA\sim\cA+\bpar\Lambda,
$$
where $\Lambda$ is an arbitrary function.

For our purposes, however, it is more convenient to let $\cA$ be
an element of \Cech cohomology rather than $\bpar$-cohomology.
For ordinary manifolds these cohomologies are equivalent, 
but the advantage of \Cech cohomology is that we work
with holomorphic functions. 
For supermanifolds the situation is more complicated,
and we refer the reader to 
\cite{Movshev:2003ib}\cite{Samann:2004tt}\cite{Wolf:2004hp}
for further details.\footnote{We
are grateful to Alexander Popov, 
Christian S\"amann and Martin Wolf for
pointing this to us.}

In \Cech cohomology, which is a special
case of sheaf cohomology, the B-model field is represented by a
holomorphic function $\cA$ that is defined on the {\it
intersection} of patches $U_1\cap U_2.$ This is in contrast to
$\bpar$-cohomology for which $\cA$ was a $(1,0)$-form and was not
necessarily represented by holomorphic functions, but was defined
on the entire manifold $U_1\cup U_2.$
In \Cech cohomology, there is an equivalence relation
\be\label{eqn:CechEquiv} 
\cA\sim\cA+\Lambda_1+\Lambda_2, 
\ee 
where
$\Lambda_1$ and $\Lambda_2$ are holomorphic functions on $U_1$ and
$U_2$, respectively. 

How does the infinitesimal complex structure deformation that
corresponds to \eqref{eqn:InvEAB} change the \Cech cohomology class
$\cA$? The infinitesimal complex structure deformation can be
interpreted as a change in the holomorphic transition functions
between the patch $U_1$ and the patch $U_2.$ In the deformed
space, \eqref{eqn:U2vars} are no longer good coordinates on $U_2.$
Instead, a good set of coordinates is 
\be\label{eqn:U2varsM}
\frac{1}{X^2}, \frac{X^3}{X^2}, \frac{X^4}{X^2},
\frac{\Psi^A}{X^2}
+\frac{1}{6(X^2)^2}M^{AB}\epsilon_{BCDE}\Psi^C\Psi^D\Psi^E
\qquad
A=1,\dots,4, 
\ee 
where we used \eqref{eqn:InvEAB}. Thus, the \Cech
equivalence relation \eqref{eqn:CechEquiv} has to be modified to
\bear 
\cA\Bigl(X^2, X^3, X^4, \{\Psi^A\}\Bigr) &\sim&
\cA\Bigl(X^2, X^3, X^4, \{\Psi^A\}\Bigr) +\Lambda_1\Bigl(X^2, X^3,
X^4, \{\Psi^A\}\Bigr)
\label{eqn:CechEquivM}\\
&+& \Lambda_2\Bigl( \frac{1}{X^2}, \frac{X^3}{X^2},
\frac{X^4}{X^2}, \bigl\{\frac{\Psi^A}{X^2}
+\frac{1}{6(X^2)^2}M^{AB}\epsilon_{BCDE}\Psi^C\Psi^D\Psi^E \bigr\}
\Bigr), 
\nn 
\eear 
where $\Lambda_1$ and $\Lambda_2$ are
holomorphic functions of their variables. Let us expand $\cA$ in
components, similarly to \eqref{eqn:AXXPsi}. 
\bear
\cA &=& A(X^2,
X^3, X^4)+\Psi^A\chi_A(X^2, X^3, X^4)
+\tfrac{1}{2}\Psi^A\Psi^B\phi_{AB}(X^2, X^3, X^4) 
\nn\\ &&
+\tfrac{1}{6}\epsilon_{ABCD}\Psi^A\Psi^B\Psi^C\wchi^D(X^2, X^3,
X^4) +\tfrac{1}{24}\epsilon_{ABCD}\Psi^A\Psi^B\Psi^C\Psi^D G(X^2,
X^3, X^4), 
\label{eqn:AXXPsiC} 
\eear 
where $A, \chi_A, \phi_{AB},
\wchi_D$ and $G$ are holomorphic functions of $X^2, X^3, X^4$ and
are defined on $U_1\cap U_2$, that is, for $X^2\neq 0$ and
$X^2\neq\infty.$ (Since confusion is not likely to arise, we use
the same notation for the component fields in \Cech cohomology as
in $\bpar$-cohomology.) The freedom to add $\Lambda_1$ in
\eqref{eqn:CechEquivM} implies that each of the component fields
can be augmented by a holomorphic function of $X^2, X^3, X^4$ that
is nonsingular at $X^2=0.$ Thus, the \Cech cohomology classes of
the component fields are only sensitive to the singular
behavior of the fields at $X^2=0.$

The freedom to add $\Lambda_2$ in \eqref{eqn:CechEquivM}
is now more complicated.
Let the component expansion of $\Lambda_2$ be
\be\label{eqn:Lambda2C}
\Lambda_2\bigl(y^2, y^3, y^4, \{\theta^A\}\bigr)
=
\varpi
+\theta^A\zeta_A+
\tfrac{1}{2!}\theta^A\theta^B\varsigma_{AB}
+\tfrac{1}{3!}\epsilon_{ABCD}\theta^A\theta^B\theta^C\upsilon^D
+\tfrac{1}{4!}\epsilon_{ABCD}\theta^A\theta^B\theta^C\theta^D\kappa,
\ee
where $\varpi, \varsigma, \upsilon, \kappa$ are holomorphic
functions of their generic variables $y^2, y^3, y^4.$
Expanding \eqref{eqn:CechEquivM} in components, we now
find the equivalence relations
\be
\begin{split}
\chi_A(X^2, X^3, X^4) &\sim \chi_A(X^2, X^3, X^4)
 +\frac{1}{X^2}\zeta_A\bigl(\frac{1}{X^2},
         \frac{X^3}{X^2}, \frac{X^4}{X^2}\bigr),
\\
\wchi^A(X^2, X^3, X^4) &\sim \wchi^A(X^2, X^3, X^4)
 +\frac{1}{(X^2)^3}\upsilon^A
   \bigl(\frac{1}{X^2}, \frac{X^3}{X^2}, \frac{X^4}{X^2}\bigr)
-\frac{1}{(X^2)^2}M^{AB}
\zeta_B\bigl(\frac{1}{X^2}, \frac{X^3}{X^2}, \frac{X^4}{X^2}\bigr)
\end{split}
\label{eqn:Equiv2MC}
\ee
for the fermions,
and
\be
\begin{split}
A(X^2, X^3, X^4) &\sim A(X^2, X^3, X^4)
 +\varpi\bigl(\frac{1}{X^2}, \frac{X^3}{X^2}, \frac{X^4}{X^2}\bigr),
\\
\phi_{AB}(X^2, X^3, X^4) &\sim \phi_{AB}(X^2, X^3, X^4)
 +\frac{1}{(X^2)^2}\varsigma_{AB}
   \bigl(\frac{1}{X^2}, \frac{X^3}{X^2}, \frac{X^4}{X^2}\bigr),
\\
G(X^2, X^3, X^4) &\sim G(X^2, X^3, X^4)
 +\frac{1}{(X^2)^4}\kappa
\bigl(\frac{1}{X^2}, \frac{X^3}{X^2}, \frac{X^4}{X^2}\bigr)
\end{split}
\label{eqn:Equiv2MCsc}
\ee
for the scalars,
where we have used the symmetry of $M^{AB}$ in the $A,B$ indices.
We see that only the equivalence relation for
the field $\wchi^A$ is modified.
This field is the twistor transform of the $h=-1/2$
helicity spinor $\psi_\a^A.$
We would now like to relate the modified equivalence
relation to the solution to the massive Dirac equation
\eqref{eqn:solfMD}.

Let us first recall the origin of the equivalence relations
\eqref{eqn:Equiv2MC} in the massless case ($M^{AB}=0$).
It was explained in the appendix of \cite{Witten:2003nn}
that the physical wave-functions are recovered
from $\wchi^A$ and $\chi_A$ by a Cauchy integral
along a path $C$ that encircles the origin,
\be\label{eqn:psixC}
\begin{split}
\psi_\a^A(x) &= \frac{1}{2\pi i}\oint_C \lam_\a
\wchi^A(z, x_{1\dot{1}}+x_{2\dot{1}}z,
           x_{1\dot{2}}+x_{2\dot{2}}z) dz,
\\
\psi^\dta_A(x) &= \frac{1}{2\pi i}\oint_C
\ppx{x_{1\dta}}\chi_A(z, x_{1\dot{1}}+x_{2\dot{1}}z,
       x_{1\dot{2}}+x_{2\dot{2}}z)dz,
\end{split}
\ee
where we have fixed the rescaling freedom by
\be
\lam^1\equiv 1,
\quad
\lam^2\equiv z.
\ee
[Equations \eqref{eqn:psixC}
are the analogs of Whittaker's formula
\eqref{eqn:Whit} for spinor fields in $D=4.$]
For $M^{AB}=0$, the equivalence relation
\eqref{eqn:Equiv2MC} can be written in this patch as
\be\label{eqn:chiequiv}
\begin{split}
\wchi^A(z, x_{1\dot{1}}+x_{2\dot{1}}z,
           x_{1\dot{2}}+x_{2\dot{2}}z)
&\sim
\wchi^A(z, x_{1\dot{1}}+x_{2\dot{1}}z,
           x_{1\dot{2}}+x_{2\dot{2}}z)
+   \frac{1}{z^3}\upsilon^A\bigl(\frac{1}{z},
     \frac{x_{1\dot{1}}}{z}+x_{2\dot{1}},
     \frac{x_{1\dot{2}}}{z}+x_{2\dot{2}}\bigr),
\\
\chi_A(z, x_{1\dot{1}}+x_{2\dot{1}}z,
       x_{1\dot{2}}+x_{2\dot{2}}z)
&\sim
\chi_A(z, x_{1\dot{1}}+x_{2\dot{1}}z,
       x_{1\dot{2}}+x_{2\dot{2}}z)
+   \frac{1}{z}\zeta_A\bigl(\frac{1}{z},
     \frac{x_{1\dot{1}}}{z}+x_{2\dot{1}},
     \frac{x_{1\dot{2}}}{z}+x_{2\dot{2}}\bigr).
\end{split}
\ee
These equivalences hold because of the identities
\be\label{eqn:upsintz}
\begin{split}
0 &=
\frac{1}{2\pi i}\oint_C \lam_\a
\frac{1}{z^3}\upsilon^A
   \bigl(\frac{1}{z},
     \frac{x_{1\dot{1}}}{z}+x_{2\dot{1}},
     \frac{x_{1\dot{2}}}{z}+x_{2\dot{2}}\bigr) dz,
\\
0 &=
\frac{1}{2\pi i}\oint_C
\frac{1}{z^2}
\ppx{x_{2\dta}}\zeta_A\bigl(\frac{1}{z},
     \frac{x_{1\dot{1}}}{z}+x_{2\dot{1}},
     \frac{x_{1\dot{2}}}{z}+x_{2\dot{2}}\bigr)dz,
\end{split}
\ee
that can be derived by deforming the contour of integration
into a circle of radius $\abs{z}\rightarrow\infty.$
(Note that $\lam_\a$ behaves at worst like $z$,
and $\upsilon^A$ at worst like a constant.)
In the second identity in \eqref{eqn:upsintz}
we have set
$\partial\zeta_A/\partial x_{1\dta} =
(\partial\zeta_A/\partial x_{2\dta})/z.$

For the massless case $M^{AB}=0$,
equation \eqref{eqn:psixC} is the same as
\eqref{eqn:psix} with
\be\label{eqn:hvchi}
\hwvrho^A\equiv\wchi^A,
\qquad
\hvarrho_A\equiv\chi_A.
\ee
Now let us turn to the massive case.
Rewriting the Dirac wave-functions \eqref{eqn:psix}
in terms of the B-model fields \eqref{eqn:hvchi},
we get
\be\label{eqn:psixCM}
\begin{split}
\psi_\a^A(x) &= \frac{1}{2\pi i}\oint_C \
\bigl\lbrack
\lam_\a
\wchi^A(z, x_{1\dot{1}}+x_{2\dot{1}}z,
           x_{1\dot{2}}+x_{2\dot{2}}z)
+M^{AB}\eta_\a\chi_B(z,
       x_{1\dot{1}}+x_{2\dot{1}}z,
       x_{1\dot{2}}+x_{2\dot{2}}z)
\bigr\rbrack dz,
\\
\psi^\dta_A(x) &= \frac{1}{2\pi i}\oint_C
\ppx{x_{1\dta}}\chi_A(z, x_{1\dot{1}}+x_{2\dot{1}}z,
       x_{1\dot{2}}+x_{2\dot{2}}z) dz,
\end{split}
\ee
where
$$
(\lam^1,\lam^2)\equiv (1, z),
\qquad
(\eta_1, \eta_2)\equiv (1, 0).
$$
The equivalences \eqref{eqn:chiequiv} now have to be
modified to
\bear
\lefteqn{
\wchi^A(z, x_{1\dot{1}}+x_{2\dot{1}}z,
           x_{1\dot{2}}+x_{2\dot{2}}z)
\sim
\wchi^A(z, x_{1\dot{1}}+x_{2\dot{1}}z,
           x_{1\dot{2}}+x_{2\dot{2}}z)
}\nn\\ &+&
   \frac{1}{z^3}\upsilon^A\bigl(\frac{1}{z},
     \frac{x_{1\dot{1}}}{z}+x_{2\dot{1}},
     \frac{x_{1\dot{2}}}{z}+x_{2\dot{2}}\bigr)
+   \frac{1}{z^2}M^{AB}\zeta_B\bigl(\frac{1}{z},
     \frac{x_{1\dot{1}}}{z}+x_{2\dot{1}},
     \frac{x_{1\dot{2}}}{z}+x_{2\dot{2}}\bigr),
\nn\\
\lefteqn{
\chi_A(z, x_{1\dot{1}}+x_{2\dot{1}}z,
       x_{1\dot{2}}+x_{2\dot{2}}z)
}\nn\\ &\sim&
\chi_A(z, x_{1\dot{1}}+x_{2\dot{1}}z,
       x_{1\dot{2}}+x_{2\dot{2}}z)
 +\frac{1}{z}\zeta_A\bigl(\frac{1}{z},
     \frac{x_{1\dot{1}}}{z}+x_{2\dot{1}},
     \frac{x_{1\dot{2}}}{z}+x_{2\dot{2}}\bigr).
\label{eqn:chiequivM}
\eear
Indeed, these transformations leave the
Cauchy integrals \eqref{eqn:psixCM} invariant.
Moreover, the \Cech equivalence relations
\eqref{eqn:Equiv2MC}, in the form
\eqref{eqn:chiequivM},
determine the form of the Cauchy integrals
\eqref{eqn:psixCM}, with $\eta_\a$ determined up
to the ambiguity \eqref{eqn:etapr}.
Thus we have shown that a VEV for $\overline{E}^{AB}$
corresponds to a chiral mass term.

\subsection{Mass terms in Berkovits's model}
\label{subsec:Berkmt10}
We will now study the counterparts of the  mass operators
\eqref{eqn:VtxMass15} and \eqref{eqn:VtxMass10} in Berkovits's
twistor string theory.
We will work with the worldsheet
fields from \secref{subsec:drstrBerk}.
Berkovits and Witten explained in \cite{Berkovits:2004jj}
how to convert B-model operators to operators
in Berkovits's model.
Adapted to our $D=3$ setting, the procedure is as follows.
We start with a holomorphic vector field on mini-twistor
space $\Tw_3.$ It can be represented in homogeneous
coordinates as
\be\label{eqn:fifA}
f'\defineas
f^i(Z^1,Z^2, W,\Theta)\ppx{Z^i}
+f^W(Z^1,Z^2, W,\Theta)\ppx{W}
+f^A(Z^1,Z^2, W,\Theta)\ppx{\Theta^A},
\ee
subject to the equivalence relation
$$
f'\simeq f' +
\Lambda(Z^1,Z^2, W,\Theta)\Bigl(
Z^i\ppx{Z^i}+ 2W\ppx{W}+\Theta^A\ppx{\Theta^A}
\Bigr).
$$
This vector field is converted to a Berkovits vertex operator
$$
\Vo =
Y_i f^i + \rU f^W +\Upsilon_A f^A.
$$
In particular,
starting with a B-model operator that is represented
as a holomorphic vector field
\be\label{eqn:fzfwfA}
f\defineas
f^z(z,w,\theta)\ppx{z}
+f^w(z,w,\theta)\ppx{w}
+f^A(z,w,\theta)\ppx{\theta^A},
\ee
we can get the worldsheet vertex operator
by lifting \eqref{eqn:fzfwfA} to $\C^{4|4}$
by setting $Z^1=1$, $Z^2=z$ and $W=w.$
$$
\Vo=
Y_2 f^z\Bigl
(\frac{Z^2}{Z^1},\frac{W}{(Z^1)^2},\frac{\Theta}{Z^1}\Bigr)
+\rU f^w\Bigl
(\frac{Z^2}{Z^1},\frac{W}{(Z^1)^2},\frac{\Theta}{Z^1}\Bigr)
+\Upsilon_A f^A\Bigl
(\frac{Z^2}{Z^1},\frac{W}{(Z^1)^2},\frac{\Theta}{Z^1}\Bigr).
$$
This is an {\it open} string vertex operator
in Berkovits's model.
The vector field \eqref{eqn:fzfwfA} is required to
preserve the holomorphic
superform on mini-twistor space
$$
\Omega = dw\wedge dz\wedge d\theta^1\cdots d\theta^4.
$$
This is a condition that descends from a similar condition
in $D=4$ \cite{Berkovits:2004jj}.

Applying this Berkovits-Witten
prescription to \eqref{eqn:VtxMass15} we
get our first ansatz for the Berkovits-model mass operators
\be\label{eqn:VtxBMass15}
V_{(\rep{15})} \rightarrow
\Vo_{(\rep{15})}
=\frac{W}{Z^1 Z^2}\Upsilon_A {M^A}_B\Theta^B.
\ee
The factor of $Z^1 Z^2$ in the denominator
of \eqref{eqn:VtxBMass15} might appear strange at first,
but it can be handled by bosonization.
We have discussed a similar issue
at the end of \secref{subsec:Spin7},
and more details can be found in \appref{app:Bosonization}.

To gain more insight,
we will now derive \eqref{eqn:VtxBMass15}
by directly gauging translations
with an R-symmetry twist
\eqref{eqn:P3wstw} in the world-sheet action \eqref{eqn:BerkA}.
We therefore
augment the covariant derivative \eqref{eqn:NabBr}
according to the modified gauge transformation \eqref{eqn:P3wstw},
\be\label{eqn:NabBrM}
\nabla_\bzws Z_L^3 = \px{\bzws}Z_L^3 -A_\bzws Z_L^3 -\Br_\bzws Z_L^1,
\qquad
\nabla_\bzws Z_L^4 = \px{\bzws}Z_L^4 -A_\bzws Z_L^4 +\Br_\bzws Z_L^2,
\qquad
\nabla_\bzws \Theta^A_L = \px{\bzws}\Theta^A_L
 -i\Br_\bzws {M^A}_B\Theta^B,
\ee 
and similarly for the right-moving fields. Inserting these
covariant derivatives into the action \eqref{eqn:BerkA} and
integrating out the nondynamical gauge fields $\Br_\zws,
\Br_\bzws$, we get the modified constraint 
\be\label{eqn:YZYZM}
j_m(\zws)\defineas 
Y_3 Z^1 - Y_4 Z^2 - i{M^A}_B\Upsilon_A\Theta^B = 0. 
\ee 
(Again, we suppress the $L,R$ subscripts on fields.) Note
that for  ${M^A}_B\neq 0$, the left-moving gauge field $\Br_\bzws$
and the right-moving gauge field $\Br_\zws$ have to be related by
complex conjugation, because the left-moving gauge transformation
by itself is anomalous, with an anomaly proportional to ${M^A}_B
{M^B}_A.$

According to the previous discussion, and in particular
\eqref{eqn:VtxBMass15}, the mass deformation corresponds
to an open string vertex operator and therefore should
manifest itself as a change in the worldsheet boundary conditions.
How can we convert \eqref{eqn:YZYZM} to a change of boundary conditions?

Let us first recall, on the classical level, what the equations of
motion of the Berkovits action are. The charged fields have
equations of motion \be\label{eqn:Beoms} \nabla_\bzws Z_L^i =
0,\qquad \nabla_\bzws Y_{L i} = 0,\qquad \nabla_\zws Z_{R i}
=0,\qquad \nabla_\zws Y_R^i = 0,\qquad i=1,\dots,4, \ee and
similarly for $\Upsilon_A$ and $\Theta^A.$ The gauge fields
$A_\zws, A_\bzws$ and $\Br_\zws, \Br_\bzws$ do not have dynamical
equations of motion, and they can be arbitrary. The equations of
motion \eqref{eqn:Beoms} then imply that gauge invariant
combinations of left-moving fields are holomorphic, and gauge
invariant combinations of right-moving fields are
anti-holomorphic. This (anti-)analyticity and the constraints
\be\label{eqn:YZA} \sum_{i=1}^4 Y_{L i} Z_L^i + \Upsilon_{L
A}\Theta_L^A=0, \qquad \sum_{i=1}^4 Y^i_R Z_{R i} +
\Upsilon^A_R\Theta_{R A}=0, \ee together with \eqref{eqn:YZYZM},
are all the restrictions on gauge invariant combinations. For
${M^A}_B=0$, a full set of $\Br$-gauge invariant combinations is
given by \eqref{eqn:rU}-\eqref{eqn:rY}. For ${M^A}_B\neq 0$, we
need to modify these formulas. We define the following field
combinations, \be\label{eqn:MgInv}
\begin{split}
&Z^1,\qquad
Z^2,\qquad
W\defineas Z^1 Z^4 + Z^2 Z^3,
\qquad
\rU\defineas \frac{Y_4}{Z^1}
= \frac{Y_3}{Z^2} -\frac{i}{Z^1 Z^2}{M^A}_B\Upsilon_A\Theta^B,
\\
&\rY_1\defineas
Y_1-\rU Z^4
+i\frac{Z^3}{(Z^1)^2}{M^A}_B\Upsilon_A\Theta^B,
\qquad
\rY_2 \defineas Y_2 -\rU Z^3,
\\
&\rTheta\defineas
\exp\bigl(i\frac{Z^3}{Z^1}M\bigr)\Theta,
\qquad
\rUpsilon\defineas
\exp\bigl(-i\frac{Z^3}{Z^1}M\bigr)\Upsilon.
\end{split}
\ee 
They are invariant under the gauge transformations that
correspond to $\Br$, and using the constraint \eqref{eqn:YZA}, we
see that they satisfy \be\label{eqn:QuadCoM} 0 = \rY_1 Z^1+\rY_2
Z^2 + \rUpsilon_A\rTheta^A + 2\rU W. \ee Thus, the equations of
motion for $Z^1, Z^2, \rY_1, \rY_2, W, \rU, \rTheta, \rUpsilon$
are independent of the mass matrix ${M^A}_B.$ The mass matrix
must, therefore, enter into the boundary conditions.

Let us consider a worldsheet with the topology of a disk. It can
be represented as the upper half plane $\Imx\zws>0.$ The boundary
conditions at $\Imx\zws=0$ are \cite{Berkovits:2004jj}
\be\label{eqn:bc} Z_L^i = Z_{R i}^*,\qquad Y_{L i} =
(Y_R^i)^*,\qquad \Theta_L^A = \Theta_{R A}^*,\qquad \Upsilon_{L A}
= (\Upsilon_R^A)^*. \ee The ``doubling trick'' (see, e.g., \S2.6
of \cite{Polchinski:1998rq}) is a standard way of treating a
conformal field theory on a disk. We extend the definition of the
left-moving fields to the full $\zws$-plane by setting
\be\label{eqn:DblTrick} Z_L^i(\zws)\defineas
(Z_R^i(\bzws))^*,\quad Y_{L i}(\zws) = (Y_R^i(\bzws))^*,\quad
\Theta_L^A(\zws) = \Theta_{R A}(\bzws)^*,\quad \Upsilon_{L
A}(\zws) = (\Upsilon_R^A(\bzws))^*, \qquad\Imx\zws < 0. \ee
Including the point $\zws=\infty$, the fields are now defined on a
sphere.

Now let us look at the fields from \eqref{eqn:MgInv}. They are
well-defined only if $Z^1\neq 0.$ When ${M^A}_B=0,$ we can use the
alternative definition $\rU=Y_3/Z^2$, which makes sense when
$Z^2\neq 0.$ Thus, when ${M^A}_B=0$, the fields from
\eqref{eqn:MgInv} make sense whenever either $Z^1$ or $Z^2$ is
nonzero. But for ${M^A}_B\neq 0$ we have to use a different set of
fields in patches of the worldsheet that contain zeroes of $Z^1.$
Such an alternative set of fields is given by
\be\label{eqn:MgInvAlt}
\begin{split}
&Z^1,\qquad
Z^2,\qquad
W\defineas Z^1 Z^4 + Z^2 Z^3,
\qquad
\rU\defineas \frac{Y_3}{Z^2}
= \frac{Y_4}{Z^1} +\frac{i}{Z^1 Z^2}{M^A}_B\Upsilon_A\Theta^B,
\\
&\rY_1\defineas
Y_1-\rU Z^4,
\qquad
\rY_2 \defineas Y_2 -\rU Z^3
-i\frac{Z^4}{(Z^2)^2}{M^A}_B\Upsilon_A\Theta^B,
\\
&\rTheta\defineas
\exp\bigl(-i\frac{Z^4}{Z^2}M\bigr)\Theta,
\qquad
\rUpsilon\defineas
\exp\bigl(i\frac{Z^4}{Z^2}M\bigr)\Upsilon.
\end{split}
\ee These fields were determined by the requirement of invariance
under $\Br$-gauge transformations and the requirement that
\eqref{eqn:QuadCoM} is satisfied.

We still need to specify which of the two sets of formulas,
\eqref{eqn:MgInv} or \eqref{eqn:MgInvAlt}, to choose for the
left-moving and the right-moving sectors. We do not have any
compelling reason to prefer one choice over the other, but if we
choose \eqref{eqn:MgInvAlt} for the right-movers and
\eqref{eqn:MgInv} for the left-movers we will soon see that we
recover the mass operator \eqref{eqn:VtxBMass15}. It is not clear
to us why we cannot choose the same set of formulas, say
\eqref{eqn:MgInv}, for both left-movers and right-movers, but we
note that if we do that we need a certain condition, say
$Z_L^1\neq 0$, to hold throughout the doubled worldsheet, and this
could be too restrictive. 

With \eqref{eqn:MgInv} for the left-movers and \eqref{eqn:MgInvAlt}
for the right-movers, we get the boundary conditions
\be\label{eqn:bcM}
\begin{split}
&Z_L^1 = Z_{R 1}^*,\quad
Z_L^2 = Z_{R 2}^*,\quad
W_L = W_R^*,\quad
U_L = U_R^* + \frac{i}{Z_L^1 Z_L^2}{M^A}_B\Upsilon_{L A}\Theta^B_L,
\\
&\rY_{1 L} = \rY_{1 R}^*
+\frac{i W_L}{(Z^1_L)^2 Z^2_L}{M^A}_B\Upsilon_{L A}\Theta^B_L,
\qquad
\rY_{2 L} = \rY_{2 R}^*
-\frac{i W_L}{Z^1_L (Z^2_L)^2}{M^A}_B\Upsilon_{L A}\Theta^B_L,
\\
&\Theta_L^A = \exp\bigl(i\frac{W_L}{Z^1_L Z^2_L}M\bigr)
               \Theta_{R A}^*,\qquad
\Upsilon_{L A} = \exp\bigl(-i\frac{W_L}{Z^1_L Z^2_L}M\bigr)
(\Upsilon_R^A)^*.
\end{split}
\ee
These boundary conditions can be succinctly described
by adding an extra boundary term to the worldsheet action,
\be\label{eqn:deltaS}
\delta S =
-\frac{i}{2}\int d\zws \frac{W_L}{Z_L^1 Z_L^2}\Upsilon_{L A} {M^A}_B\Theta_L^B
+\frac{i}{2}\int d\bzws \frac{W_R}{Z_R^1 Z_R^2}\Upsilon_R^B {M^A}_B\Theta_{R A}.
\ee
To first order in $M$, the full action is given by the bulk worldsheet action
\eqref{eqn:BerkA} plus the boundary action \eqref{eqn:deltaS},
supplemented with the boundary conditions
$$
Z_L^1 = Z_{R 1}^*,\quad
Z_L^2 = Z_{R 2}^*,\quad
W_L = W_R^*,\quad
\Theta_L^A = \exp\bigl(i\frac{W_L}{Z^1_L Z^2_L}M\bigr)
               \Theta_{R A}^*.
$$
The remaining boundary conditions \eqref{eqn:bcM}
follow (upto $O(M^2)$ corrections) by minimizing the action.
To conclude, we note that the boundary term
\eqref{eqn:deltaS} is consistent
with the mass operator \eqref{eqn:VtxBMass15} that was
predicted from the B-model.

Similarly to \eqref{eqn:VtxBMass15},
we can convert \eqref{eqn:VtxMass10}
to Berkovits's model.
Using the fields from Table~\ref{table:Berkovits},
we get the open string vertex operator
\be\label{eqn:VtxBMass10}
V_{(\rep{10})} \rightarrow
\frac{1}{Z^1 Z^2}
M^{AB}\epsilon_{BCDE}
\Upsilon_A\Theta^C\Theta^D\Theta^E.
\ee
The mass terms in $\rep{\overline{10}}$, which were difficult to
identify in the B-model mini-twistor string theory, can be readily
identified in Berkovits's model. Dimensionally reducing the
operators $E_{AB}$ from \cite{Berkovits:2004jj}, we
get the open string vertex operator 
\be\label{eqn:VtxBMassOv10}
V_{(\rep{\overline{10}})} \rightarrow \frac{1}{Z^1 Z^2}
M_{AB}\Theta^A\partial\Theta^B. 
\ee

\subsection{Application of $\Spin(7)$ R-symmetry}
\label{subsec:ApplSpin7}
In \secref{subsec:strop} and \secref{subsec:Berkmt10}
we found three kinds of
$D=3$ mass operators. We found the operators
\eqref{eqn:VtxBMass10} in the \rep{10} of $SU(4)$,
the operators \eqref{eqn:VtxBMassOv10} in the
$\rep{\overline{10}}$ of $SU(4)$, and
the operators \eqref{eqn:VtxBMass15} in the $\rep{15}$
of $SU(4).$
The latter were the easiest to analyze, since they
could be derived by a twisted dimensional reduction.
However, all these operators should be related
by the $\Spin(7)$ R-symmetry that we discussed in
\secref{subsec:Spin7}.
In this subsection we will apply the R-symmetry
generators \eqref{eqn:TABrUrU} to
the operators \eqref{eqn:VtxBMass15}.
We will discover a surprise:
the operators \eqref{eqn:VtxBMass15}, \eqref{eqn:VtxBMass10}
and \eqref{eqn:VtxBMassOv10} do not fit
into an irreducible representation of $\Spin(7)$!

Nevertheless, starting with the operators \eqref{eqn:VtxBMassOv10}
(in the $\rep{\overline{10}}$), we can reconstruct a good
$\Spin(7)$ multiplet of operators by successively applying the
$\Spin(7)$ generators \eqref{eqn:TABrUrU}. These operators will be
different from our previous results \eqref{eqn:VtxBMass15},
\eqref{eqn:VtxBMass10}. We can ``distill'' out of
\eqref{eqn:VtxBMass15} the terms that do fall into the irreducible
representation $\rep{35}$ and use them to complete the $\Spin(7)$
multiplet.

Let us now do this in detail. Set 
\be\label{eqn:defZZo}
\ZZo\defineas \frac{1}{Z^1 Z^2}. 
\ee
The various mass terms that
we found in equations \eqref{eqn:VtxBMass15},
\eqref{eqn:VtxBMass10} and \eqref{eqn:VtxBMassOv10} are \bear
V_{(\rep{\overline{10}})} &\rightarrow& \ZZo
M_{AB}\Theta^A\partial\Theta^B,
\nn\\
V_{(\rep{15})} &\rightarrow&
\ZZo W\Upsilon_A {M^A}_B\Theta^B,
\nn\\
V_{(\rep{10})} &\rightarrow&
\ZZo
M^{AB}\epsilon_{BCDE}
\Upsilon_A\Theta^C\Theta^D\Theta^E.
\nn
\eear
These terms are linear combinations
of the vertex operators
\be\label{eqn:wVo}
\begin{split}
\Vo^{AB} &\defineas
\ZZo\Theta^{(A}\partial\Theta^{B)},
\qquad
{\wVo^A}_B\defineas
\ZZo W (\Upsilon_B\Theta^A
-\tfrac{1}{4}\delta_B^A\Upsilon_C\Theta^C),
\\
\wVo_{AB} &\defineas
\ZZo\epsilon_{CDE(B}
\Upsilon_{A)}\Theta^C\Theta^D\Theta^E.
\end{split}
\ee (Here and in the equations below, all operators are understood
to be normal ordered.) As we will see below, some of these vertex
operators are not quite what we are looking for, and the tilde
over $\Vo$ will remind us that they are about to be modified. We
will denote the modified vertex operators by ${\Vo^A}_B$  and
$\Vo_{AB}.$ These vertex operators should form an irreducible
representation of $\Spin(7)$, isomorphic to $\rep{35}.$ We can
therefore write down immediately their expected commutation
relations with the $\Spin(7)$ generators \eqref{eqn:CommRels}:
\be\label{eqn:TVoCom1}
\begin{split}
\lbrack{T^A}_B, \Vo_{CD}\rbrack &=
 \delta^A_C \Vo_{BD}
 +\delta^A_D \Vo_{BC},
\\
\lbrack{T^A}_B, \Vo^{CD}\rbrack &=
  -\delta^C_B\Vo^{AD}
 -\delta^D_B\Vo^{AC}
\\
\lbrack{T^A}_B, {\Vo^C}_D\rbrack &=\delta^A_D{\Vo^C}_B
  -\delta^C_B{\Vo^A}_D,
\end{split}
\ee and \be\label{eqn:TVoCom2}
\begin{split}
\lbrack T^{AB}, \Vo_{CD}\rbrack &=
  \delta^A_C {\Vo^B}_D
 -\delta^B_C {\Vo^A}_D
 +\delta^A_D {\Vo^B}_C
 -\delta^B_D {\Vo^A}_C,
\\
\lbrack T^{AB}, \Vo^{CD}\rbrack &=
  \epsilon^{ABCE}{\Vo^D}_E
 +\epsilon^{ABDE}{\Vo^C}_E,
\\
\lbrack T^{AB}, {\Vo^C}_D\rbrack &=
  \epsilon^{ABCE}\Vo_{ED}
 +\delta^A_D\Vo^{BC}
 -\delta^B_D\Vo^{AC},
\end{split}
\ee Thus, we can calculate the operators ${\Vo^A}_B$ from the
commutation relations of $T^{AB}$ with $\Vo^{CD}.$

We can read off these commutation relations from the coefficient
of the simple pole in the OPE of $\Jc^{AB}$ and $\Vo^{CD}.$ The
result is 
\bear 
\bigl\lbrack T^{AB}, \Vo^{CD}\bigr\rbrack &=&
\tfrac{1}{2}\epsilon^{ABCE}\ZZo (
 \rU^{-1}\partial\Upsilon_E\Theta^D
+\partial\rU^{-1}\Upsilon_E\Theta^D
-\rU^{-1}\Upsilon_E\partial\Theta^D)
\nn\\
&&
+\tfrac{1}{2}\epsilon^{ABDE}\ZZo (
 \rU^{-1}\partial\Upsilon_E\Theta^C
+\partial\rU^{-1}\Upsilon_E\Theta^C
-\rU^{-1}\Upsilon_E\partial\Theta^C)
\nn
\eear
A comparison with \eqref{eqn:TVoCom2} teaches us that
\be\label{eqn:VuAdB}
{\Vo^A}_B =
\tfrac{1}{2}\ZZo (
 \rU^{-1}\partial\Upsilon_B\Theta^A
+\partial\rU^{-1}\Upsilon_B\Theta^A
-\rU^{-1}\Upsilon_B\partial\Theta^A)
-\tfrac{1}{8}\delta^A_B
\ZZo (
 \rU^{-1}\partial\Upsilon_C\Theta^C
+\partial\rU^{-1}\Upsilon_C\Theta^C
-\rU^{-1}\Upsilon_C\partial\Theta^C). 
\nn 
\ee 
We have subtracted the trace to make ${\Vo^A}_B$ traceless, 
as is required of the irreducible representation $\rep{15}.$

Next, we calculate the commutation relations between the newly
found ${\Vo^A}_B$ and the $\Spin(7)$ generators $T^{AB}.$ A long
but straightforward calculation gives 
\be\label{eqn:TABVuCdD} 
\bigl\lbrack
T^{AB}, {\Vo^C}_D\bigr\rbrack =
-\delta^{[A}_D\ZZo\partial\Theta^{B]}\Theta^C
+\delta^{[A}_D\ZZo\Theta^{B]}\partial\Theta^C
-\epsilon^{ABCE}\ZZo\rU^{-2}\partial\Upsilon_{(E}\Upsilon_{D)}
\ee 
A comparison with \eqref{eqn:TVoCom2} therefore teaches us
that 
\be\label{eqn:VdAB}
\Vo_{AB} = \rU^{-2}\ZZo\Upsilon_{(A}\partial\Upsilon_{B)},
\ee 
where $\ZZo$ was defined in \eqref{eqn:defZZo}, and using the
bosonized fields from \secref{subsec:Spin7B}, we have defined
\be\label{eqn:defUinv2} 
\rU^{-2}\defineas -e^{-2\phi}\xi\partial\xi. 
\ee 
This definition is natural, given
the definition of $\rU^{-1}$ in \eqref{eqn:defUinv} and the
following OPE
$$
\rU^{-1}(\zws)\rU^{-1}(0) =
-e^{-2\phi(0)}\xi(0)\partial\xi(0) +O(\zws).
$$

We can now write down our final result for the three types of mass
operators:
\bear
\Vo^{AB} &=& \ZZo\Theta^{(A}\partial\Theta^{B)},
\label{eqn:VtxBMassOv10Cor}\\
\Vo_{AB} &=& \ZZo\rU^{-2}\Upsilon_{(A}\partial\Upsilon_{B)},
\label{eqn:VtxBMass10Cor}\\
{\Vo^A}_B &=& 
\tfrac{1}{2}\ZZo\bigl(
\rU^{-1}\partial\Upsilon_B\Theta^A
-\rU^{-1}\Upsilon_B\partial\Theta^A
+\partial\rU^{-1}\Upsilon_B\Theta^A
\bigr)
\nn\\
&& -\tfrac{1}{8}\ZZo\delta^A_B
\bigl(\rU^{-1}\partial\Upsilon_C\Theta^C
-\rU^{-1}\Upsilon_C\partial\Theta^C
+\partial\rU^{-1}\Upsilon_C\Theta^C\bigr).
\label{eqn:VtxBMass15Cor} 
\eear
We have checked that these
operators now satisfy all the relations
\eqref{eqn:TVoCom1}-\eqref{eqn:TVoCom2} and hence constitute an
irreducible representation of the R-symmetry group $\Spin(7)$ that
is isomorphic to $\rep{35}.$

Our results are summarized in Table~\ref{table:massterms}.
The operator \eqref{eqn:VtxBMassOv10Cor} agrees
with the previous result \eqref{eqn:VtxBMassOv10},
but the other operators
\eqref{eqn:VtxBMass10Cor}-\eqref{eqn:VtxBMass15Cor}
do not agree with the operators
\eqref{eqn:VtxBMass10} and
\eqref{eqn:VtxBMass15}
that we found in \secref{subsec:Berkmt10}.
The reason for this apparent discrepancy is
not completely clear to us, but we expect
the difference between \eqref{eqn:VtxBMass15} to
\eqref{eqn:VtxBMass15Cor} to decouple from physical
amplitudes. As for the relation between  
\eqref{eqn:VtxBMass10} and \eqref{eqn:VtxBMass10Cor},
we suspect that these operators are in a different
``picture,'' as we will now explain.

In \cite{Berkovits:2004tx}
it was explained that vertex operators
in Berkovits's model come in different ``pictures.''
A disk amplitude
with $(d+1)$ negative-helicity gluons and $(n-d-1)$ 
positive-helicity gluons requires an insertion
of $d$ ``instanton-changing operators,'' 
in addition to the $n$ vertex operators that correspond
to the physical states. These instanton-changing operators
are analogous to the picture-changing operators of
superstring theory. They contain $\delta$-functions
of the $Y_I$-fields, and are necessary to absorb zero-modes
of those fields. Similarly to superstring theory,
one can get rid of a picture-changing operator by
absorbing it in a physical vertex operator, thereby
changing the ``picture'' of that vertex operator.
This observation played a crucial role in identifying
parity symmetry \cite{Berkovits:2004tx}.

Let us now look at the zero-modes of $\rU.$
Using the bosonization formulas \eqref{eqn:WUbos},
we can define the $\rU$-picture of an operator to be its 
$\phi$-momentum, so that an operator that contains 
an exponent $\exp (p\phi)$ will be in the $p$-picture.
It follows from \eqref{eqn:TABrUrU} that when
$T^{AB}$ acts on an operator in the $p$-picture,
it produces a sum of operators in the $(p-1)$-picture
and $(p+1)$-picture.

To see why picture-changing is necessary,
recall our comment at the end of \secref{subsec:ft}.
There we saw that the notion of helicity in $D=3$ is not
invariant under a general R-symmetry transformation.
For example, we saw  that a positive-helicity gluon
satisfies $\Phi^7 = i \Phi^8$, but one can find an 
infinitesimal R-symmetry
transformation that acts as $\delta\Phi^7=\epsilon\Phi^6$, say.
After this transformation, the gluon state will acquire
a $0$-helicity component.
But replacing a $(+1)$-helicity state with a $0$-helicity
state in a scattering amplitude requires,
according to \cite{Berkovits:2004tx},
an extra instanton-changing operator.

Now take a positive-helicity gluon vertex operator in Berkovits's
model. In $D=3$ it is of the form $\Phi(Z^1, Z^2, W)J_C$,
where $J_C$ is a holomorphic current from the chiral-current
algebra component of the worldsheet theory, and $\Phi$ is 
a meromorphic function.
Acting with $T^{AB}$ gives
\be\label{eqn:TABPhi}
[T^{AB}, \Phi] = 
\pypx{\Phi}{W}\Theta^A\Theta^B 
+\tfrac{1}{2}\widetilde{\Phi}(Z^1, Z^2, \rU)
\epsilon^{ABCD}\Upsilon_C\Upsilon_D,
\ee
where $\widetilde{\Phi}$ is the residue of the simple pole
in the OPE of $\Phi$ with $\rU^{-1}$ and
can be calculated from formula \eqref{eqn:rUinvWn}.
The first term on the righthand side of \eqref{eqn:TABPhi}
describes the above-mentioned $0$-helicity state.
The second term, however, is in a different $\rU$-picture
and would formally correspond to a $(+2)$-helicity state.
Since no such term exists,
it must decouple from physical amplitudes.
Similarly, the mass operators
\eqref{eqn:VtxBMass10Cor} and
\eqref{eqn:VtxBMass10} are obviously in a different picture, 
since one contains $\rU^{-2}$ while the other does not.

\begin{table}[t]
\begin{tabular}{|l|c|c|c|}
\hline\hline
Operator
  & $M_{AB}\Vo^{AB}$ & ${M^A}_B {\Vo^B}_A$ & $M^{AB}\Vo_{AB}$ \\
\hline
$SU(4)$ representation
 & $\rep{\overline{10}}$ & $\rep{15}$ &  $\rep{10}$ \\
\hline
Fermion coupling  & $M_{AB}\chi^A\chi^B$ & ${M^A}_B\chi^A\chi_B$
          & $M^{AB}\chi_A\chi_B$ \\
\hline
Conformal SUGRA field
& $E_{AB}$ & ${{V_4}^A}_B$ & $\overline{E}^{AB}$ \\
\hline
B-model deformation: $\delta\theta^A=$
          & -
          & $z^{-1}{M^A}_B w\theta^B$
  &  $\tfrac{1}{6}
     z^{-1}M^{AB}\epsilon_{BCDE}\theta^C\theta^D\theta^E$ \\
\hline
Berkovits model
& $\ZZo\Theta^{(A}\partial\Theta^{B)}$
& ${1 \over 2} \ZZo\rU^{-1}\partial\Upsilon_B\Theta^A+\cdots$
& $\ZZo\rU^{-2}\Upsilon_{(A}\partial\Upsilon_{B)}$
\\
\hline\hline
\end{tabular}
\caption{The three types of mass operators with
their $SU(4)$ irreducible representations,
their coupling to the fermions,
the corresponding conformal supergravity fields,
the B-model complex structure deformations
($\delta\theta^A$ is the extra term in the change
of holomorphic variables from the patch $z\neq 0$
to the patch $z\neq\infty$)
and the Berkovits-model operators from
\eqref{eqn:VtxBMassOv10Cor}-\eqref{eqn:VtxBMass15Cor}.
The operator $\ZZo\equiv 1/Z^1 Z^2$ was defined in \eqref{eqn:defZZo}.
}
\label{table:massterms}
\end{table}

We conclude this subsection by demonstrating explicitly that
the operators ${\wVo^A}_B$
are not in the irreducible representation $\rep{35}$,
as we claimed at the beginning of this subsection.
If they were, we could complete them to a $\Spin(7)$
multiplet using the commutation relations
\eqref{eqn:TVoCom2}, just like we did above with $\Vo^{AB}.$
But an explicit computation gives
\bear
\bigl\lbrack T^{AB},
{\wVo^D}_C
\bigr\rbrack &=&
2\delta^{[A}_C\ZZo\partial\Theta^{B]}\Theta^D
+2\ZZo\partial\phi\delta^{[A}_C\Theta^{B]}\Theta^D
\nn\\ &&
+\epsilon^{ABDE}\ZZo\rU^{-2}\partial\Upsilon_E\Upsilon_C
-\epsilon^{ABDE}\ZZo\rU^{-2}\partial\phi\Upsilon_E\Upsilon_C
\nn\\ &&
+\ZZo\Theta^A\Theta^B\Theta^D\Upsilon_C
+{1 \over 2} \epsilon^{ABEF}\ZZo\rU^{-2}\Upsilon_E\Upsilon_F\Upsilon_C\Theta^D.
\label{eqn:TABW}
\eear
Here $\partial\phi = -:\rU W:$ is a bosonized current.
If \eqref{eqn:TVoCom2} were satisfied, we could set
\bear
\wVo_{AB} &=& \tfrac{1}{6}\epsilon_{CDEA}
\lbrack T^{CD}, {\wVo^E}_B\rbrack
\nn\\
&=&
\rU^{-2}\ZZo\partial\Upsilon_{(A}\Upsilon_{B)}
+\tfrac{1}{12}\epsilon_{CDE(A}\Theta^C\Theta^D\Theta^E\Upsilon_{B)}
\nn\\ &&
+{1 \over 2} \rU^{-2}\ZZo\partial (\Upsilon_A\Upsilon_B)
-\ZZo\Upsilon_A\Upsilon_B\rU^{-2}\partial\phi
-\tfrac{1}{3}\ZZo\epsilon_{ABCD}\partial\Theta^C\Theta^D
-\tfrac{1}{3}\ZZo\epsilon_{ABCD}\partial\phi\Theta^C\Theta^D
\nn\\ &&
+\tfrac{1}{12}\epsilon_{CDE[A}\ZZo
               \Theta^C\Theta^D\Theta^E\Upsilon_{B]}
+\tfrac{1}{3}\ZZo\rU^{-2}\Upsilon_A\Upsilon_B\Upsilon_C\Theta^C.
\label{eqn:wVoAB} \eear But $\wVo_{AB}$ has to be symmetric in the
indices $A,B$, and the last two lines of \eqref{eqn:wVoAB} are
antisymmetric in $A,B.$ This means that $\wVo_{AB}$ is a mixture
of the $SU(4)$ irreducible representations $\rep{10}$ and
$\rep{15}$. It suggests that our starting point, \eqref{eqn:wVo}
for ${\wVo^A}_B$, is a mixture of the irreducible $\Spin(7)$
representations $\rep{35}$ and the adjoint $\rep{21}.$ One can
similarly check that \bear \wVo^{BC} &=& \tfrac{1}{3}\lbrack
T^{AB}, {\wVo^C}_A\rbrack
\nn\\
&=&
 \ZZo\partial\Theta^{(B}\Theta^{C)}
+\tfrac{1}{6}\epsilon^{CDE(B}
  \ZZo\rU^{-2}\Upsilon_C\Upsilon_D\Upsilon_E\Theta^{C)}
\nn\\ &&
-\tfrac{1}{6}\epsilon^{BCDE}
  \ZZo\rU^{-2}\partial(\Upsilon_D\Upsilon_E)
+\tfrac{1}{3}\epsilon^{BCDE}
  \ZZo\rU^{-2}\partial\phi\Upsilon_D\Upsilon_E
-\tfrac{1}{3}\ZZo\Theta^B\Theta^C\Upsilon_A\Theta^A \nn\\ &&
+\tfrac{1}{6}\epsilon^{DEF[B}
  \ZZo\rU^{-2}\Upsilon_D\Upsilon_E\Upsilon_F\Theta^{C]}
+\tfrac{1}{2}\ZZo\partial(\Theta^B\Theta^C)
+\ZZo\partial\phi\Theta^B\Theta^C \nn \eear This is again not
symmetric in $A,B$ and thus contains a mixture of the $SU(4)$
irreducible representations $\rep{\overline{10}}$ and $\rep{6}.$

\subsection{The singlet mass term}
\label{subsec:masssingle}
The last remaining mass term in \eqref{eqn:massreps}
is the singlet $\rep{1}.$
A naive guess for the corresponding B-model operator
is to set $M$ in \eqref{eqn:vecwzM} to be proportional
to the identity matrix.
This means that we cannot obtain \eqref{eqn:vecwzM}
by dimensional reduction with an $SU(4)$ twist,
because the identity matrix is not traceless.
An even bigger problem is that if $M$ is not traceless,
the complex structure deformation to which
\eqref{eqn:vecwzM} corresponds does not preserve the
holomorphic volume form $\Omega$ of $\CP^{3|4}.$
Indeed, unless $M$ is traceless,
$$
\Omega = \frac{1}{{4!}^2}
\epsilon_{IJKL}Z^I dZ^J\wedge dZ^L \wedge dZ^L
\epsilon_{ABCD} d\psi^A d\psi^B d\psi^C d\psi^D
$$
is not preserved by \eqref{eqn:strzwtw}.
We do not know how to turn on singlet mass terms.
(See also the related discussion after equation
(2.11) in \S2.2 of \cite{Berkovits:2004jj}
regarding complex structure deformations that do not
preserve $\Omega.$)

\subsection{Comments on the decoupling limit $M\rightarrow\infty.$}
\label{subsec:largeM}
The discussion in the previous subsections was concerned
mainly with infinitesimal mass terms.
To first order, these mass terms are related to deformations
of the B-model action by closed string vertex operators.
Our analysis in \secref{subsec:drtwists}, however,
allows us to ``integrate'' the infinitesimal mass deformations
and describe finite mass terms.
Specifically, \eqref{eqn:strzwtw} describes a super
complex structure deformation of $\CP^{3|4}$
that preserves the holomorphic volume super-form for any
traceless $M.$

Let $m_1,\dots, m_4$ be the eigenvalues of $M$, and set $m_i = c_i
m$ for some constants $c_i.$ For generic $c_i$, the limit
$m\rightarrow\infty$ is quite interesting from the physical
perspective. If all $c_i$ and all $c_i+c_j$ are nonzero, all
fermions and six out of the seven scalars get a large bare mass
and decouple. We can preserve $D=3$, $N=2$ supersymmetry if we
keep $c_1=0.$ In this case,  one scalar and two $D=3$ gluinos
remain with zero bare mass. We can also preserve $D=3$, $N=4$
supersymmetry if we keep $c_1 = c_2=0.$ In this case, three
scalars and four gluinos remain with zero bare mass.

Let us now comment on the $m\rightarrow\infty$ limit
from the perspective of Berkovits's open
twistor string theory.
The mass appears only in the boundary term \eqref{eqn:deltaS}. In
the limit $m\rightarrow\infty$, the boundary term
\eqref{eqn:deltaS} becomes very big. This suggests that the bulk
action of $\Upsilon$ and $\Theta$ can be neglected. We end up with
a worldsheet theory that has bulk modes $Z^1, Z^2, W$ and their
conjugates $Y_1, Y_2, \rU.$ The $\Upsilon$ and $\Theta$ fields
live only on the boundary and couple to the bulk fields via the
boundary action \eqref{eqn:deltaS}. This state of affairs is
somewhat reminiscent of the Seiberg-Witten limit of large NS-NS
2-form $B$ field that leads to noncommutative geometry
\cite{Connes:1997cr}\cite{Douglas:1997fm}\cite{Seiberg:1999vs}. (A
connection between deformations of twistor string theory and
noncommutative geometry, and in particular its extension to
superspace \cite{Seiberg:2003yz} was also suggested in
\cite{Kulaxizi:2004pa}.) This limit in the present context will be
studied elsewhere \cite{P}.

\section{Conclusions and discussion}\label{sec:discussion}
We reviewed Hithcin's construction of mini-twistor space, which
relates $D=3$ mini-twistor space to $D=4$ twistor 
space by dimensional
reduction. The key point is that the 3(complex)-dimensional $D=4$
twistor space $\CP^3\setminus\CP^1$ can be written as a fiber
bundle with the 2(complex)-dimensional $D=3$ twistor space
$T\CP^1$ as the base, and the fiber is $\C.$ The structure group
is the additive translation group $\sim\C.$ 
 We used this fibration to relate tree-level
twistor amplitudes of $D=3$ Yang-Mills theory to $D=4$ amplitudes.
We calculated $D=3$ tree-level amplitudes from the $D=4$ ones by
integrating over the $\C$ fibers of the above fibration. This
immediately implies that Witten's observations
\cite{Witten:2003nn} regarding scattering amplitudes and
holomorphic curves in twistor space are valid in $D=3$, at least
at tree-level. In $D=3$ there is a known relation between
holomorphic curves on mini-twistor space and holomorphic curves in
complexified Minkowski space. We used this relation to give a
direct physical interpretation to the holomorphic curves in
mini-twistor space. At one-loop level and higher, the $D=3$ and
$D=4$ amplitudes are not directly related by dimensional
reduction, and whether or not the conjectures of
\cite{Witten:2003nn} extend to $D=3$ remains to be seen. For
developments regarding loop amplitudes in $D=4$ see
\cite{Bern:2004ky, Cachazo:2004zb, Cachazo:2004dr, Britto:2004nc,
Bedford:2004py, Bern:2005hs, Bidder:2005ri}.

Our main new results are related to
deformations of the $N=8$ supersymmetric
$D=3$ Yang-Mills theory by mass terms.
We proposed a variant of the topological B-model
that describes massive $D=3$ Yang-Mills theory.
This model can describe $15$ out of the $35$ different
possible mass terms.
These mass-deformed
$D=3$ theories can be obtained by twisting the
dimensional reduction
of massless $D=4$ theories.
As for the other mass terms,
we only discussed infinitesimal deformations and
conjectured that these deformations
correspond to VEVs of certain conformal
supergravity fields. In this paper, we gave circumstantial evidence in
support of the previous statement. It would be very interesting to
show this convincingly by examining the three point functions of two
supergravity fields and one gauge field operator. This should be
possible by performing a computation similar to the one in
\cite{Berkovits:2004jj} [see equation (5.5) of that paper], where the
correlation functions of a
conformal supergravity vertex operator with an arbitrary number of
Yang-Mills vertex operators were calculated in Berkovits's model.
We hope to report on this in another paper \cite{P}.

In $D=3$ we constructed the $\Spin(7)$ R-symmetry currents
that transform one type of infinitesimal mass term to another.
This allowed us to construct vertex operators
for all $35$ mass terms in Berkovits's model. We encountered a
surprise when we tried to compare the vertex operators in Berkovits's
model with the topological B-model. 
In Berkovits's model we found that the $35$ mass operators
do not quite fit into an irreducible representation of
the R-symmetry group $\Spin(7).$
We suggested a different set of operators that do fit 
into an R-symmetry multiplet.
It would be interesting to understand the meaning of 
this in more detail.

Our results provide a way to break the supersymmetry
from $N=8$ to
$N=4,2,1,0$ in $D=3$ by arbitrary mass terms, in
mini-twistor string theory.
One direction for further research is to analyze the
limit of infinite mass. In this limit we should obtain
pure $N=4,2,1$ Yang-Mills theories, and also $N=0$ with a scalar.
Our results also suggest a way to break $N=4$
in $D=4$ by infinitesimal mass terms.
It would be interesting to try to ``integrate'' these
infinitesimal deformations to get large mass terms.
Understanding R-symmetry better may enable us to
convert a non-infinitesimal mass term in the
representation $\rep{15}$ 
[described by \eqref{eqn:strzwtw} in the B-model
or \eqref{eqn:YZYZM} in Berkovits's model]
into a mass term in the $\rep{10}+\rep{\overline{10}}$,
and then it might be possible to lift it to $D=4.$
Alternatively, a better understanding of the nonperturbative
topological B-model,
perhaps along the lines suggested
recently in 
\cite{Aganagic:2003qj}\cite{Nekrasov:2004js}\cite{Dijkgraaf:2004te},
 may shed light on how to turn on
non-infinitesimal VEVs for the closed string modes
$E_{AB}$ and $\overline{E}^{AB}$ simultaneously.

Another possible direction for further research
is to explore the mirror manifold of mini-twistor space.
The same techniques of \cite{Aganagic:2004yh},
where the mirror manifold of the twistor space $\CP^{3|4}$
has been constructed, can
be applied to mini-twistor space.
It was shown in \cite{Aganagic:2004yh}
that the mirror of $\CP^{3|4}$ is a quadric in
$\CP^{3|3}\times\CP^{3|3}.$ One might then be tempted to examine the
dimensionally reduced version of the argument in \cite{Neitzke:2004pf}, where the
authors use S-duality
in conjunction with mirror symmetry to explain why the amplitudes of
${\cal N}= 4$ Yang-Mills are supported on holomorphic curves.
Moreover, certain symmetries (e.g. parity) might be more
manifest in the mirror. 
One would expect to lose some of the manifest
$SU(4|4)$ generators.

\acknowledgments
It is a pleasure to thank
Mina Aganagic, Korkut Bardakci, Itzhak Bars, Iosif Bena,
Eric Gimon, Nick Halmagyi,
Hitoshi Murayama, Anthony Ndirango, Eliezer Rabinovici
and Radu Tatar
for helpful discussions.
We also wish to thank 
Alexander Popov, Christian S\"amann and Martin Wolf
for comments on the first version of our paper.
This work was supported in part by the Director, Office of Science,
Office of High Energy and Nuclear Physics, of the U.S. Department of
Energy under Contract DE-AC03-76SF00098, and in part by
the NSF under grant PHY-0098840.

\appendix
\section{Dealing with inverses of fields}
\label{app:Bosonization}
The commuting worldsheet fields of Berkovits's model,
$(Y_1, Z^1)$,
$(Y_2, Z^2)$
and $(\rU, W)$,
have the same OPEs as the superconformal ghosts $(\beta,\gamma)$
in superstring theory.
Based on this, we bosonized the fields $(\rU, W)$
in \secref{subsec:Spin7B}.
Actually, we will need to bosonize the three pairs
$(Y_1, Z^1)$,
$(Y_2, Z^2)$
and $(\rU, W)$, simultaneously.
To ensure that fields from different pairs are commuting
(rather than anti-commuting) it is more convenient
to use a bosonization scheme with no anti-commuting fields.
Recall that any $(\beta,\gamma)$ pair
of worldsheet fields with OPE
$$
\beta(\zws)\gamma(0)\sim -\frac{1}{\zws}
$$
can be bosonized in one of two ways: 
\be\label{eqn:bgbos} 
\text{Either} \quad
\bigl\{\text{(i)}\quad
\beta=e^{-\phi+\chi}\partial\chi,\quad\gamma=e^{\phi-\chi}
\bigr\}\quad 
\text{or} 
\quad 
\bigl\{\text{(ii)}\quad
\beta=e^{-\phi+\chi},\quad
\gamma=e^{\phi-\chi}\partial\chi
\bigr\}, 
\ee 
where $\chi(\zws)$ and $\phi(\zws)$ are chiral
bosons. The current in both cases is
$$
\beta\gamma = \partial\phi.
$$
(See for example \S10.4 of \cite{Polchinski:1998rq}.)

At the outset, it seems that
for each pair of fields $(Y_i, Z^i)$ or $(\rU, W)$,
we can choose either the bosonization scheme (i) or (ii).
How can we decide which scheme to choose?

Before we pick the bosonization scheme, we
have to decide which of the $\beta, \gamma$ fields
is to have an inverse. If it is $\gamma$, we should pick
scheme (i), because we could then define
$$
\gamma^{-1}\defineas e^{-\phi+\chi}\qquad
\text{in bosonization scheme (i).}
$$
If it is $\beta$, we should pick scheme (ii) and set
$$
\beta^{-1}\defineas e^{\phi-\chi}\qquad
\text{in bosonization scheme (ii).}
$$
The field $\gamma$ does not have an inverse in scheme (i),
and $\beta$ does not have an inverse in scheme (ii).

For the application to  mini-twistor space we would like $Z^1,
Z^2$ and $\rU$ to have inverses. The inverse of $\rU$ was needed
in \eqref{eqn:TABrUrU} for the $\Spin(7)$ R-symmetry current. The
inverses $(Z^1)^{-1}$ and $(Z^2)^{-1}$ were needed for $\ZZo$
[defined in \eqref{eqn:defZZo}] and for all the mass operators
that contained $\ZZo.$ The inverses of the other fields,
$Y_1^{-1}, Y_2^{-1}, W^{-1}$ were never needed.

More basically, if we allow $(Z^i)^{-1}$ ($i=1,2$) it means that
we are restricting to the patch of mini-twistor space where
$Z^i\neq 0.$ If we proscribe $W^{-1}$ it means that we are
including points where $w=0.$ The discussion above compels us to
choose the following bosonization scheme: \be\label{eqn:bosYZUW}
\begin{split}
Y_i &= e^{-\phi_i+\chi_i}\partial\chi_i,\quad
Z^i = e^{\phi_i-\chi_i},\quad
(Z^i)^{-1} = e^{-\phi_i+\chi_i},\quad (i=1,2),
\\
\rU &= e^{-\phi_3+\chi_3},\quad
W = e^{\phi_3-\chi_3}\partial\chi_3,\quad
\rU^{-1} = e^{\phi_3-\chi_3},
\\
j &= Y_1 Z^1 + Y_2 Z^2 + 2\rU W =
\partial\phi_1 + \partial\phi_2 + 2\partial\phi_3.
\end{split}
\ee
We conclude with a few useful OPEs:
\bear
\rU(\zws)\rU^{-1}(0) &\sim&  1
+\zws \bigl\lbrack\partial\phi(0)+\eta(0)\xi(0)\bigr\rbrack
+ O(\zws^2),
\nn\\
\rU(\zws)W(0) &\sim& -\frac{1}{\zws}-\partial\phi(0)+O(\zws),
\nn\\
\rU^{-1}(\zws)W(0) &\sim& \frac{1}{\zws}\rU^{-2}(0)
-\rU^{-2}(0)\partial\phi(0)+O(\zws),
\nn\\
\rU^{-1}(\zws)\rU^{-1}(0) &\sim&
\rU^{-2}(0)
+\tfrac{1}{2}\zws\partial\rU^{-2}
+O(\zws)^2
\nn\\
\rU^{-1}(\zws)\rU^{-2}(0) &\sim&
\rU^{-3}(0) + O(\zws),
\nn\\
\rU(\zws)\partial\rU^{-1}(0) &\sim&
-\partial\phi(0)-\eta(0)\xi(0) + O(\zws).
\nn\\
\rU(\zws)\rU^{-2}(\zws) &\sim& \rU(0) -\zws\partial(\rU^{-1}(0))
+\zws^2 X(0)+O(\zws)^3,
\nn\\
\rU^{-1}(\zws)\partial\rU^{-1}(0) &\sim&
\tfrac{1}{2}\partial\rU^{-2}(0)
+O(\zws)
\nn
\eear
where we defined the operators
$$
X\defineas
e^{-\phi}\bigl(
\eta(\partial\xi)\xi
-\partial\phi\partial\xi
+\tfrac{1}{2}((\partial\phi)^2
+\partial^2\phi)\xi\bigr),
$$
and
$$
\rU^{-2}\defineas e^{-2\phi}(\partial\xi)\xi,
\qquad
\rU^{-3}\defineas
\tfrac{1}{2} e^{-3\phi}\partial^2\xi(\partial\xi)\xi.
$$
(All products on the righthand side
are assumed to be normal ordered.)
Finally, we need the OPEs
\be\label{eqn:rUinvWn}
\rU^{-1}(\zws)W^n(0) = \frac{1}{\zws}(-1)^n n! \rU^{-n-1}+O(1).
\ee

\begin{table}[p]
\begin{tabular}{lll}
\hline\hline
Symbol  & Subsection  & Meaning \\
\hline
$a, b, \dots$ & \secref{subsec:ft}
              & $\Spin(7)$ spinor R-symmetry indices ($1\dots 8$)\\
$A, B, C, \dots$ & \secref{subsec:supersp}
                 & $SU(4)$ R-symmetry indices ($1\dots 4$)\\
$A_\u$ & \secref{subsec:ft}
       & spacetime gauge field (either $D=3$ or $D=4$) \\
$A_\zws, A_\bzws$ & \secref{subsec:drstrBerk}
                  & worldsheet $GL(1)$ gauge field \\
$\Br_\zws, \Br_\bzws$ & \secref{subsec:drstrBerk}
                      & worldsheet gauge field,
                        used to gauge for translations\\
$E_{AB}, \overline{E}^{AB}$ & \secref{subsec:Spin7}
   & spacetime
     conformal supergravity fields from \cite{Berkovits:2004jj} \\
$f, f'$ & \secref{subsec:Berkmt10}
        & holomorphic vector field on mini-twistor space\\
$\Focal$ & \secref{subsec:evolute}
         & focal curve\\
$i,j,\dots$ & \secref{subsec:ft}
   & Euclidean spacetime indices $1,\dots,3$ or $1,\dots, 4$ \\
$I,\dots$ & \secref{subsec:ft}
   & $\Spin(7)$ vector index $1,\dots,7$ \\
$I$ & \secref{subsec:Spin7B}
  & a collective index for one of the eight
   $D=4$ Berkovits-model fields \\
$J_{\pm}, J_1$ & \secref{subsec:Poincare}
   & $D=3$ rotation generators \\
$\Lag$   & \secref{subsec:ft}
         & spacetime Lagrangian \\
${M^A}_B, M_{AB}, M^{AB}$ & \secref{subsec:drtwists}
                          & Mass matrices\\
$P_{\pm}, P_1$ & \secref{subsec:Poincare}
   & $D=3$ translation generators \\
$Q_{\pm}, \bQ_{\pm}$ & \secref{subsec:supersp}
   & $D=3$ supersymmetry generators \\
$T^{AB}, {T^A}_B$  & \secref{subsec:Spin7}
   & $\Spin(7)$ R-symmetry generators \\
$\Jc^{AB}, {\Jc^A}_B$  & \secref{subsec:Spin7B}
   & $\Spin(7)$ R-symmetry currents \\
$\twv$  & \secref{subsec:propagator}
        & $D=3$ mini-twistor \\
$\Tw_3$ & \secref{subsec:supersp}
        & Mini-twistor space \\
$U_1, U_2$ & \secref{subsec:dimsup}
      & patches of mini-twistor space \\
$\rU$ & \secref{subsec:drstrBerk}
    & worldsheet Berkovits-model field dual to $W$ \\
${\Vo^A}_B, \Vo^{AB}, \Vo_{AB}$ & \secref{subsec:ApplSpin7}
   & Berkovits-model mass operators \\
$\wVo{}^A_B, \wVo{}^{AB}, \wVo_{AB}$ & \secref{subsec:ApplSpin7}
    & Berkovits-model mass operators (converted from B-model) \\
$\tW, \btW$ & \secref{subsec:drstr}
         & B-model worldsheet field ($D=3$) \\
$W$ & \secref{subsec:drstrBerk}
         & Berkovits-model worldsheet fields ($D=3$) \\
$w, \bw$ & \secref{subsec:TCP1}
    & coordinates on mini-twistor space
      (for the fiber of $T\CP^1$)\\
$X^i$ & \secref{subsec:D4Mass}
      & worldsheet fields in $D=4$ B-model ($i=1,\dots,4$)\\
$x_{\a\dta}$ & \secref{subsec:dimred}
             & spacetime coordinate ($D=3$ or $D=4$)\\
$Y_i$ & \secref{subsec:drstrBerk}
         & Berkovits model dual field
 ($i=1,2$ in $D=3$) \\
$\tZ, \btZ$ & \secref{subsec:drstr}
         & B-model worldsheet fields ($D=3$) \\
$Z^i$ & \secref{subsec:drstrBerk}
      & worldsheet fields in $D=3$ Berkovits model ($i=1,2$)\\
$z, \bz$ & \secref{subsec:TCP1}
       &  coordinate on mini-twistor space
  (for the base of $T\CP^1$)\\
$\zws, \bzws$ & \secref{subsec:drstrBerk}
              & worldsheet coordinate \\
\hline\hline
\end{tabular}
\caption{Various symbols used in the text (Roman letters).
The ``subsection'' column is the place where the symbol
first appeared with its given meaning.}
\label{table:symbolsRoman}
\end{table}

\begin{table}[p]
\begin{tabular}{lll}
\hline\hline
Symbol  & Subsection  & Meaning \\
\hline
$\a, \b, \dots$ & \secref{subsec:Minkowski}
& chiral spinor indices ($D=3$ or $D=4$)\\
$\dta, \dtb, \dots$ & \secref{subsec:ft}
   & dotted spinors ($D=4$) \\
$\Gamma^\u$ & \secref{subsec:Minkowski}
  & $D=3$ Dirac matrices \\ 
$\zeta_A$ & \secref{subsec:D4Mass}
  & holomorphic component function on twistor space \\
$\zeta$ & \secref{subsec:D4Mass} 
  & arbitrary function of $\lam,\tlam$ \\
$\hzeta$ & \secref{subsec:D4Mass} 
  & twistor transform of $\zeta$ \\
$\eta$ & \secref{subsec:dimred}
    & fiber coordinate in the fibration $\CP^3\rightarrow T\CP^1$ \\
$\eta$ & \secref{subsec:Spin7B}
   & worldsheet field used for bosonization of $(\rU, W)$\\
$\eta_\a$ & \secref{subsec:D4Mass}
    &  solution to the equation $\eta_\a\lam^\a=1.$ \\
$\oeta^\bz, \oeta^\bw, \oeta_A$ & \secref{subsec:drstr}
         & B-model worldsheet fields \\
$\Theta^A, \oTheta^A$
&\secref{subsec:drstr}, \secref{subsec:drstrBerk}
         & worldsheet fields \\
$\varth_z, \varth_w, \varth_A$ & \secref{subsec:drstr}
         & B-model worldsheet anti-commuting fields \\
$\kappa$ & \secref{subsec:D4Mass}
  & holomorphic component function on twistor space \\
$\Lambda$ & \secref{subsec:D4Mass}
          & arbitrary function on twistor space \\
$\lam^\a$ & \secref{subsec:dimred} & twistor variable \\
$\tlam_\dta$ & \secref{subsec:Minkowski} 
     & appears in the decomposition of null vectors \\
$\mu_\dta$ & \secref{subsec:dimred} & twistor variable \\
$\mu, \nu,\dots$ & \secref{subsec:Minkowski}
   & spacetime indices $0,\dots,2$ or $0,\dots,3$\\
$\xi$ & \secref{subsec:Spin7B}
      & worldsheet field used for bosonization of $(\rU, W)$\\
$\xi_1, \xi_2, \xi_3$ & \secref{subsec:redtree}
        & projective coordinates on $W\CP^{1,1,2}$\\
$\ZZo$ & \secref{subsec:ApplSpin7}
       & Stands for the worldsheet field $1/(Z^1 Z^2)$ \\
$\pi, \pi'$ & \secref{subsec:dimred} 
   & projections from $D=4$ twistor space to
     $D=3$ mini-twistor space \\
$\varpi$ & \secref{subsec:D4Mass}
  & holomorphic component function on twistor space \\
$\rho^z, \rho^w, \rho^A$ & \secref{subsec:drstr}
         & B-model worldsheet fields \\
$\varrho, \wvrho$ & \secref{subsec:D4Mass}
  & functions of $\lam,\tlam$ appearing
   in the solution to Dirac's equation \\
$\hvarrho, \hwvrho$ & \secref{subsec:D4Mass}
  & twistor transforms of $\varrho, \wvrho$ \\
$\sigma^i_{\a\b}$ & \secref{subsec:ft}
   & $D=3$ Pauli matrices \\
$\varsigma_{AB}$ & \secref{subsec:D4Mass}
  & holomorphic component function on twistor space \\
$\wCurve$ & \secref{subsec:redtree}
         & curve in twistor space \\
$\Curve$ & \secref{subsec:redtree}
         & curve in mini-twistor space \\
$\upsilon^A$ & \secref{subsec:D4Mass}
  & holomorphic component function on twistor space \\
$\Upsilon_A$ & \secref{subsec:drstrBerk}
             & Berkovits model fields dual to $\Theta^A$ \\
$\phi$ & \secref{subsec:Spin7B}
       & worldsheet field used for bosonization of $(\rU, W)$\\
$\Phi^I$ & \secref{subsec:ft}
         & spacetime scalars \\
$\chi_A,\wchi^A$ & \secref{subsec:D4Mass}
  & component fields of the B-model 
    (functions of twistor space) \\
$\chi_\a^A, \chi_{\a A}, \chi^a_\a$ & \secref{subsec:ft}
   & $D=3$ fermions (functions of spacetime)\\
$\psi_\a^A, \bpsi^\dta_A$ & \secref{subsec:ft}
      & $D=4$ fermions (functions of spacetime) \\
$\Psi^A$ & \secref{subsec:D4Mass} & fermionic coordinates in the B-model \\
\hline\hline
\end{tabular}
\caption{Various symbols used in the text (Greek letters).
The ``subsection'' column is the place where the symbol
first appeared with its given meaning.
}
\label{table:symbolsGreek}
\end{table}

\end{document}